\documentclass[11pt, reqno, a4paper]{amsart}
\usepackage{amsmath,amssymb,amsthm}
\usepackage[margin=1in]{geometry}
\usepackage{graphicx}
\usepackage{tikz}
\usepackage{xr-hyper}
\usepackage{hyperref}
\usepackage{xcolor}
\usepackage{bbm}
\usepackage{enumitem}
\DeclareMathOperator{\Span}{span}
\allowdisplaybreaks
\newtheorem{theorem}{Theorem}[section]
\newtheorem{lemma}[theorem]{Lemma}
\DeclareMathOperator{\sinc}{sinc}
\usepackage{xcolor}
\newcommand\numberthis{\addtocounter{equation}{1}\tag{\theequation}}
\theoremstyle{definition}
\newtheorem{definition}[theorem]{Definition}

\newtheorem{prop}[theorem]{Proposition}
\DeclareMathOperator\supp{supp}

\tikzset{
  font={\fontsize{5pt}{12}\selectfont}}
\theoremstyle{remark}
\newtheorem{remark}[theorem]{Remark}

\numberwithin{equation}{section}

\begin{document}
	
	\title{On the reconstruction of bandlimited signals from random samples quantized via noise-shaping}
	
	\author{Rohan Joy}
	\address{Department of Mathematics, Indian Institute of Technology Madras, India}
	\email{rohanjoy96@gmail.com}
	
	\author{Felix Krahmer}
	\address{Center of Mathematics, Technical University of Munich, Germany}
	\email{felix.krahmer@tum.de}

        \author{Alessandro Lupoli}
	\address{Center of Mathematics, Technical University of Munich, Germany}
	\email{alessandro.lupoli@tum.de}
 
	\author{Radha Ramakrishnan}
	\address{Department of Mathematics, Indian Institute of Technology Madras, India}
	\email{radharam@iitm.ac.in}

	
	\maketitle

\begin{abstract}
Noise-shaping quantization techniques are widely used for converting bandlimited signals from the analog to the digital domain. They work by ``shaping" the quantization noise so that it falls close to the reconstruction operator's null space.
We investigate the compatibility of two such schemes, specifically $\Sigma\Delta$ quantization and distributed noise-shaping quantization, with random samples of bandlimited functions. Suppose $R>1$ is a real number and assume that $\{x_i\}_{i=1}^m$ is a sequence of i.i.d random variables uniformly distributed on $[-\tilde{R},\tilde{R}]$, where $\tilde{R}>R$ is appropriately chosen. We show that by using a noise-shaping quantizer to quantize the (randomly sign flipped) values of a real-valued $\pi$-bandlimited function $f$ at $\{x_i\}_{i=1}^m$, a function $f^{\sharp}$ can be reconstructed from these quantized values such that $\|f-f^{\sharp}\|_{L^2[-R, R]}$ decays with high probability as $m$ and $\tilde{R}$ increase. This decay holds uniformly over all bandlimited $f$. We emphasize that the sample points  $\{x_i\}_{i=1}^m$ are completely random, that is, they have no predefined structure, which makes our findings the first of their kind. 
\end{abstract}
	\section{Introduction} \label{paper179}
         In signal processing, one of the primary goals is to obtain a digital representation of a function in a signal space suitable for storage, transmission and recovery. This goal is usually attained through two steps, sampling and quantization. In the first step of sampling, we sample the function at carefully chosen data points—specifically, at a stable sampling set. This set is chosen to ensure that the function can be both stably and uniquely reconstructed from the obtained samples. For example, take any $\pi$-bandlimited function $f$. The Shannon-Nyquist sampling theorem tells us that if we sample $f$ at the integers, then it can be stably reconstructed from its samples, hence making the set of integers a stable set of sampling for the space of $\pi$-bandlimited functions. 
         The samples  $\{f(n)\}_{n \in \mathbb{Z}}$ however, still belong to the continuous space $\mathbb{R}$ and cannot be stored in full digitally; in the second step of quantization, we reduce these real or complex-valued function samples to a finite set. The finiteness of the set allows for the digital storage and processing of its elements. 

         \smallskip
Sampling schemes such as uniform, irregular, and random sampling have been explored extensively for the space of bandlimited functions and other signal spaces, including shift-invariant spaces and reproducing kernel Hilbert spaces. We refer the reader to  \cite{paper1aldroubisiamreview2001} for a comprehensive analysis of the sampling problem. Since in this paper, we work with random samples; we briefly review the relevant prior work done in this area. In  \cite{paper1basssiamjournalofmath2004}, Bass and Gr\"{o}chenig studied random sampling in the space of multivariate trigonometric polynomials. Later, Candes, Romberg and Tao  \cite{paper1candesieee2006,paper1candescpam2006} investigated the reconstruction of sparse trigonometric polynomials from a few random samples. In  \cite{paper1bassisrael2010,paper123}, Bass and Gr\"{o}chenig proved the sampling inequality for functions belonging to the space of bandlimited functions whose energy is concentrated on a cube $C_R=\big[-\frac{R}{2},\frac{R}{2}\big]^d$ in $\mathbb{R}^d$, with overwhelming probability. This result was then generalized for shift-invariant spaces  \cite{paper1aratiresultmath2022,paper1jiangjournalofcompmath2021,paper1jiangresultmath2023,paper1liinverseproblems2019,paper1yangaa2019,paper1yangjmaa2013}, signal spaces with finite rate of innovation  \cite{paper1Luataapplmath2020} and reproducing kernel Hilbert spaces  \cite{paper1liacha2021,paper14,paper1pateljmaa2020}.

\smallskip
         Given a signal $x$ and a sample vector $y$ containing linear measurements (function samples, frame measurements etc.) $Ax$ of x, the quantization process involves replacing the  vector $y$ with a vector $q$ with entries from a finite set $\mathcal{A}$, called the quantization alphabet, such that reconstruction of $x$ from $q$ is possible with minimal error. 

        One of the most intuitive approaches to quantization is memoryless scalar quantization (MSQ), which requires simply replacing each coefficient 
        of the measurement vector $y=Ax$ with its nearest element from $\mathcal{A}$.
        If the vector $y$ consists of frame coefficients, a simple method for reconstruction is to fix a dual frame and linearly reconstruct with MSQ quantized coefficients. 
        For a randomly sampled bandlimited signal, this reconstruction method is essentially identical to the one described in \cite{paper1aratiresultmath2022, paper1Luataapplmath2020}, applied to quantized samples. One could also apply  the reconstruction formula based on an iterative method  provided by the authors of  \cite{paper1liacha2021}, but as this method is specifically designed to cope with scenarios where the representation system is not a frame, we will use the first approach for our MSQ comparison.
        
        It should be noted, however, that just applying a reconstruction method to the quantized data and treating the quantization effect as noise is not an effective approach. Rather, each quantized sample defines a feasibility region, and one should ideally aim to find the intersection of these regions. This has been done successfully in finite-dimensional scenarios \cite{paper1plancpam2013,paper1powellfcm2016}, but we are not aware of such a method for bandlimited signals, which is why we use the aforementioned linear reconstruction approach for comparison in our numerical experiments.
        We note that the authors in  \cite{paper112} show that even when using an optimal reconstruction scheme to approximate $x$ from its MSQ quantized coefficients, the mean squared error cannot be better than $O(\lambda^{-1})$ with linear reconstruction methods. Here $\lambda$ denotes the oversampling ratio. Further, no such optimal analysis is available for random samples of bandlimited functions in the context of MSQ reconstruction. 

       One of the reasons why the MSQ approach falls short is because it naively quantizes each coefficient without regard for how the other coefficients in the vector are quantized. Quantization schemes such as $\Sigma \Delta$ quantization and distributed noise-shaping quantization were developed to address this issue. 
        
      $\Sigma \Delta$ quantization schemes were introduced in  \cite{paper113} for the quantization of oversampled bandlimited functions and have since seen widespread use in practice. Various studies  \cite{paper1gray1987}  have examined $\Sigma \Delta$ techniques from an engineering perspective. Nevertheless, their mathematical theory is relatively new. In their seminal work  \cite{paper120}, Daubechies and Devore showed that if the samples of a bandlimited function are quantized according to a stable $n^{\text{th}}$-order $\Sigma \Delta$ scheme, then an approximation error of the order $\lambda^{-n}$ can be achieved. G\"{u}nt\"{u}rk  \cite{paper121} subsequently proved that certain $1$-bit $\Sigma \Delta$ schemes could accomplish exponential error decay (with a constant $c \approx 0.07$). The authors of  \cite{paper122} improved the exponent to $ c \approx 0.102$. Lower bounds for the error decay incurred by coarse quantization schemes of real-valued bandlimited signals were given by the authors in \cite{paper1krahmeracha2012}. 
      
Inspired by the effectiveness of $\Sigma \Delta$ schemes in exploiting redundant measurements, Benedetto \textit{et al.} in  \cite{paper1benedettoieee2006} investigated $\Sigma \Delta$ quantization in the setting of finite frames. They showed that even the first-order $\Sigma \Delta$ scheme outperforms MSQ when the frames are adequately redundant and chosen from certain families. Further studies  \cite{paper1benedettoacha2006,paper118,paper1bodmannacha2007,paper1bodmannjfaa2007,paper119,paper1lammers2010} demonstrated that an $n^{\text{th}}$-order $\Sigma \Delta$ scheme could achieve an approximation error of order $\lambda^{-n}$. For instance, in  \cite{paper119} G\"{u}nt\"{u}rk \textit{et al.} proved that error bounds of the order $\lambda^{-(n-\frac{1}{2})\alpha}$ could be achieved using Sobolev duals for arbitrarily generated frames with $\alpha \in (0,1)$. In  \cite{paper18}, the authors established that it is possible to achieve root-exponential accuracy in the finite frame setting by selecting a family of tight frames that admit themselves as Sobolev duals. 
Recently, in  \cite{paper1ghao2021}, the authors constructed high-order, low-bit $\Sigma \Delta$ quantizers for the vector-valued configuration of fusion frames. In  \cite{paper1wang2018}, Wang examined $\Sigma \Delta$ quantization in relation to Fourier sensing matrices.

$\Sigma \Delta$ schemes have also been proven to be well-suited for quantizing compressed sensing measurements  \cite{paper1fengacha2019,paper119,paper1krahmerinfinference2014,paper1saabieee2018,paper1saabacha2018} and fast binary embeddings  \cite{paper15}.

In their work  \cite{paper16}, Chou and Güntürk introduced the concept of distributed noise-shaping quantization and obtained an exponentially small error bound in the quantization of Gaussian random finite frame expansions. They extended their findings to the setting of unitarily generated frames in  \cite{paper19}. This scheme has recently been applied to fast binary embeddings  \cite{paper15} and spectral super-resolution  \cite{paper114}. For a more in-depth look at the recent advancements in noise-shaping quantization methods, we refer the reader to   \cite{paper1chouanha2015}.

Although the quantization of oversampled bandlimited functions using uniform samples has been investigated significantly, the literature only contains a few papers that study quantized irregular samples of bandlimited functions  \cite{paper125,paper110}.
        In  \cite{paper125}, the authors first give a formula to reconstruct any bandlimited function $f$ from its samples $\{f(t_n)\}_{n \in \mathbb{Z}}$, where  $\{t_n\}_{n \in \mathbb{Z}}$ is a uniformly discrete sequence satisfying $\sup_{n \in \mathbb{Z}}|t_n-\frac{n}{\lambda}|<\infty$, and $\lambda>1$.
They then construct a dithered A/D converter and show that f can be accurately reconstructed from its quantized samples taken at $\{t_n\}_{n \in \mathbb{Z}}$. In  \cite{paper110}, the authors show that if a bandlimited function $f$ is sampled on an interleaved collection of $N$ uniform grids $\{kT+T_n\}_{k \in \mathbb{Z}}$ with $\{T_n\}_{n=1}^N$ chosen independently from $[0, T]$ ($T<1$), and the samples are quantized with a first order $\Sigma \Delta$ scheme, then with high probability the error $\|f-\tilde{f}\|_{L^\infty(\mathbb{R})}$ turns out to be atmost of the order $\frac{\log N}{N}$. Here  $\tilde{f}$ represents the function reconstructed from the quantized values. 

In contrast to  \cite{paper125}, where the specially constructed A/D converter necessitates a very specific sampling procedure, and  \cite{paper110}, where the sample points are interleaved randomly shifted grids, we do not impose such strict constraints on the sample point structure. This paper aims to investigate the effectiveness of popular noise-shaping quantization schemes when working with random samples. More precisely, we assume that the sample points $\{x_i\}_{i=1}^m$ are a sequence of i.i.d random variables uniformly distributed on a sufficiently large interval, that is, they are truly random and have no preconceived structure.

        Let $R>1$ be a real number and assume that $\{x_i\}_{i=1}^m$ is a sequence of i.i.d random variables uniformly distributed on $[-R-3m^{\frac{1}{16}},R+3m^{\frac{1}{16}}]$. Suppose $m$ satisfies certain divisibility conditions and is sufficiently large. We prove the following two results in this paper. Firstly, in Theorem \ref{paper16001}, we show that by applying a stable seventh-order $\Sigma\Delta$ quantization scheme to randomly sign flipped samples of bandlimited functions, we attain the bound
        \begin{equation}\label{paper115001}
            \|f-f^{\sharp}_{\Sigma \Delta}\|_{L^2[-R,R]}\leq \frac{d_1R}{m^{\frac{3}{8}}}
        \end{equation}
        uniformly over all  real-valued \(\pi\)-bandlimited function \(f\) satisfying \(\|f\|_{L^\infty(\mathbb{R})} \leq 1\) with probability greater than $1-17m^{\frac{15}{16}}\exp\left(-\frac{m^{\frac{5}{16}}}{d_2R}\right)$. Here $f^{\sharp}_{\Sigma\Delta}$ is the reconstructed function from the quantized (randomly sign flipped) samples of \(f\) at \(\{x_i\}_{i=1}^m\) and $d_1$ and $d_2$ are known positive constants. 
Secondly, in Theorem \ref{paper12001}, it is shown that  if a stable distributed noise-shaping quantization scheme is applied to randomly sign flipped samples of bandlimited functions, then the quantization error satisfies
\begin{equation}\label{paper115002}
    \|f-f^{\sharp}_{\beta}\|_{L^2[-R,R]}\leq \frac{d_1R}{m^{\frac{7}{16}}}
\end{equation}
uniformly over all  real-valued \(\pi\)-bandlimited function \(f\) satisfying \(\|f\|_{L^\infty(\mathbb{R})} \leq 1\) with probability greater than $1-17m^{\frac{15}{16}}\exp\left(-\frac{m^{\frac{3}{8}}}{d_2R}\right)$.

To illustrate our results, we discuss the following two cases. We assume that $m$ in both situations fulfils the requirements stated in the theorem assertions. In the first case, we obtain decay in $R$; this becomes particularly useful when $R$ is large. For this, select $m=R^{16}$ in the preceding configuration. Then, each $x_i$ is distributed uniformly on $[-4R, 4R]$. Further, the bound in \eqref{paper115001} simplifies to  $\|f-f^{\sharp}_{\Sigma \Delta}\|_{L^2[-R,R]}\leq \frac{d_1}{R^{5}}$ and holds with greater probability than $1-17R^{15}\exp\left(-\frac{R^{4}}{d_2}\right)$. Similarly, \eqref{paper115002} reduces to $\|f-f^{\sharp}_{\beta}\|_{L^2[-R,R]}\leq \frac{d_1}{R^6}$  and the probability bound changes to  $1-17R^{15}\exp\left(-\frac{R^5}{d_2}\right)$.

Given an $\epsilon >0$, in the second case, we estimate the sample size required to attain $\epsilon$ reconstruction accuracy, that is, $\|f-f^{\sharp}\|_{L^2[-R, R]}\leq \epsilon$. If $m=\left(\frac{d_1 R}{\epsilon}\right)^\frac{8}{3}$ is chosen, then each $x_i$ is uniformly distributed on $\left[-R-3\left(\frac{d_1 R}{\epsilon}\right)^\frac{1}{6}, R+3\left(\frac{d_1 R}{\epsilon}\right)^\frac{1}{6}\right]$ and the following estimate is obtained for the $\Sigma \Delta$ scheme.
\[\|f-f^{\sharp}_{\Sigma \Delta}\|_{L^2[-R,R]}\leq \epsilon \]
uniformly over all  real-valued \(\pi\)-bandlimited function \(f\) satisfying \(\|f\|_{L^\infty(\mathbb{R})} \leq 1\) with probability
greater than $ 1-17\left(\frac{d_1R}{\epsilon}\right)^{\frac{15}{6}}\exp\left(-\frac{d_1^{\frac{5}{6}}}{d_2\epsilon^{\frac{5}{6}}R^{\frac{1}{6}}}\right).$
Similarly, if $m=\left(\frac{d_1 R}{\epsilon}\right)^\frac{16}{7}$ is chosen, then we get 
\[\|f-f^{\sharp}_{\beta}\|_{L^2[-R,R]}\leq \epsilon\]
uniformly over all  real-valued \(\pi\)-bandlimited function \(f\) satisfying \(\|f\|_{L^\infty(\mathbb{R})} \leq 1\) with probability greater than $1-17\left(\frac{d_1R}{\epsilon}\right)^{\frac{15}{7}}\exp\left(-\frac{d_1^{\frac{6}{7}}}{d_2\epsilon^{\frac{6}{7}}R^{\frac{1}{7}}}\right)$,
where each sample $x_i$ is uniformly distributed on $\left[-R-3\left(\frac{d_1 R}{\epsilon}\right)^\frac{1}{7}, R+3\left(\frac{d_1 R}{\epsilon}\right)^\frac{1}{7}\right]$.

The estimates provided above  attain significance when $\epsilon$ is adequately small. It is worth noting that in both Theorem \ref{paper16001} and \ref{paper12001}, the sampling interval increases with increasing sample size. Also, the divisibility conditions on $m$ are due to the choices of certain variables we make and are not essential to the proof. The choices are made to arrive at the stated theoretical bounds and are not necessary for practical purposes, as shown in the numerical simulations towards the end of the paper.

Now we would like to discuss certain pertinent points regarding our results and the techniques that we use in our paper.

Our method is not adapted for online processing of temporal signals, as the samples are not quantized in their temporal order. Our primary focus is on signals that are not inherently temporal, such as spatial and optical signals, including images and 3D medical imaging. These types of signals are often higher-dimensional, where the issue of handling random samples is more intricate.   One possible approach to deal with random samples is to reorder them and treat them as irregular samples. However, this approach  proves less effective in higher dimensions, as the theory of irregular sampling in these settings remains underdeveloped. In fact, the issue of dealing with irregular samples in higher dimensions was what motivated Bass and Gröchenig \cite{paper1bassisrael2010} to introduce the concept of random sampling in the context of bandlimited functions. Additionally, if the samples arrive in a random order, reordering them prevents online processing, as it requires acquiring and storing all samples before quantization can begin. In the one-dimensional case (the case considered in this paper), this distinction is less pronounced because the theory of irregular sampling is well-established. However, our techniques do not rely on the dimensionality of the domain space and should be readily extensible to higher dimensions. We are currently working towards such an extension, leaving it for future work. As our work is the first in the literature to address the issue of random samples in this context, even in the one-dimensional case, we provide rigorous proofs for the one-dimensional scenario and discuss the theoretical challenges that arise.

\smallskip
In our approach, the values being quantized are not the function samples. Instead, the signs of the function samples are first randomly flipped by multiplying each sample with a symmetric Bernoulli random variable. These modified function samples are then quantized. Importantly, these random signs are added as an extra bit, as they are necessary for decoding the original signal from the quantized values. This use of random sign flips is a crucial component of our method and will be discussed later (see Remark \ref{paper1reasonforflips}). However, these additional flips give rise to a natural question. That is, would the random flipping of the function sample signs not destroy the fact that bandlimited signals vary more slowly than the Nyquist rate on the unit interval? And if so, would this not prevent the use of $\Sigma \Delta$ and $\beta$ quantization schemes, which typically exploit this property for accurate reconstruction from quantized samples?
The slow variation of bandlimited functions is exploited when the reconstruction is done using the method where the function samples are replaced by the quantized samples in the  (modified) Shannon sampling formula. 
That is, when for a given bandlimited function $f$ with representation
\begin{equation}\label{paper1samplingformulainitial}
f(t) = \frac{1}{\sqrt{\lambda}} \sum_{n \in \mathbb{Z}} f\left( \frac{n}{\lambda} \right) g\left( \cdot - \frac{n}{\lambda} \right),
\end{equation}
(see Section \ref{paper1500} for full details, especially \eqref{paper14}), 
the reconstructed function $ f^\sharp $ is computed from the quantized samples $ \{q_{\frac{n}{\lambda}}\}_{n \in \mathbb{Z}} $ as:
\begin{equation*}
f^\sharp(t) = \frac{1}{\sqrt{\lambda}} \sum_{n \in \mathbb{Z}} q_{\frac{n}{\lambda}} g\left( \cdot - \frac{n}{\lambda} \right) .
\end{equation*}

However, this standard technique is not used in our reconstruction method. Instead, we employ a frame-based reconstruction technique. The reason for this is that, since the samples are randomly picked and not sorted before reconstruction, the property that bandlimited functions vary slower than the Nyquist rate is lost and cannot be leveraged. Consequently, conventional noise-shaping techniques for bandlimited functions are not applicable. From this, we can also conclude that the addition of random flips does not cause us to lose any properties that were not already lost due to the randomness of the samples. We do not reorder the samples for two primary reasons. The first is, as mentioned earlier, that reordering the samples prevents us from handling them in an online manner. The second is that reordering the samples changes their distribution, and our method critically relies on the fact that the random samples are uniformly distributed.

The frame reconstruction method we use is highly general and we expect that it can be applied to many different function spaces. The bandlimitedness property is only used in the initial step. This step is used to obtain a sampling formula of the type \eqref{paper1samplingformulainitial}, which we then use to identify an appropriate approximation space. Therefore, we believe this technique can be extended to other spaces that admit a similar representation. Specifically, to spaces $\mathcal{H}$, where the following representation holds for all $f \in \mathcal{H}$:
\[
f = \sum_{k \in \mathbb{Z}} \left\langle f, h\left(\cdot - \frac{k}{\lambda}\right) \right\rangle g\left(\cdot - \frac{k}{\lambda}\right),
\]
for suitable functions $\lambda$, $g$, and $h$.

Examples of such spaces include the widely studied shift-invariant spaces, of which the bandlimited space is a particular case. However, in general shift-invariant spaces, both $g$ and $h$ are not explicitly known in closed form. For this reason, we focus on bandlimited functions in our work, where $h = g$ and leave the investigation of other function spaces to future work.

\smallskip
From our investigations, we were able to prove better theoretical bounds for the $\beta$ quantization scheme than for the $\Sigma \Delta$-quantization scheme. In our numerical experiments, the $\beta$ quantization scheme also performed much better than the $\Sigma \Delta$-quantization scheme. Furthermore, in order to achieve bounds weaker than those of the $\beta$ quantization scheme, we have to employ a seventh-order quantizer for the $\Sigma \Delta$-quantization scheme. This makes the $\Sigma \Delta$ scheme, compared to the $\beta$ scheme, much more computationally demanding and significantly slower. Additionally, in our method, we partition the random samples into three collections, which prevents the $\Sigma \Delta$ scheme from running in an online manner. However, the $\beta$ scheme continues to run in an online manner. For all these reasons, we conclude that the preferred quantization scheme for our method is the $\beta$ quantization scheme. In our numerical experiments, we also compare the performance of both the $\Sigma \Delta$ and the $\beta$ quantization schemes with MSQ. As seen from the numerical simulations, the MSQ scheme does not perform well. The error does not decrease with an increase in the number of samples, which is generally expected with the MSQ scheme when reconstruction is done by applying a dual frame.

\smallskip
The remainder of the paper is organized as follows: In Section \ref{paper175}, we introduce relevant concepts and results from frame, probability, quantization, and sampling theories. In Section \ref{paper1SQRpipeline}, we provide a detailed explanation of our sampling-quantization-reconstruction process. Our technique combines the theory of quantization of bandlimited functions with the theory of frames. The subsequent three sections are dedicated to expanding on the techniques mentioned in Section \ref{paper1SQRpipeline} and giving them mathematical rigor. We prove our main results towards the end of Section \ref{paper178}. In Section \ref{paper113001}, the effectiveness of our proposed reconstruction algorithm is demonstrated through numerical experiments. Finally, in Section \ref{paper140000}, we offer some concluding remarks.

\section{Preliminaries and Notations}\label{paper175}
\begin{itemize}
    \item For any Lebesgue measurable subset $I$ of $\mathbb{R}$ and $f\in L^2(\mathbb{R})$, we define $|I|$ to be the Lebesgue measure of $I$,
$\|f\|_{2,I}$:=($\int_{I}|f(x)|^2dx)^{\frac{1}{2}}$ and $\|f\|$:=($\int_{\mathbb{R}}|f(x)|^2dx)^{\frac{1}{2}}$. 
\item Let $X=(x_1, \ldots,x_{n_1}) \in \mathbb{C}^{n_1}$ and $Y=(y_1, \cdots,y_{n_2})\in \mathbb{C}^{n_2}$. Then
		\begin{enumerate}
			\item $X \frown Y:=(x_1, \ldots,x_{n_1},y_1, \ldots,y_{n_2}) \in \mathbb{C}^{n_1+n_2}$.
			\item For a positive integer $m$, \[\frown_{i=1}^mX_m:=X_1\frown X_2 \frown \cdots\frown X_m.\]
		\end{enumerate}
  \item For any positive real number $t$, define
		\[[t]:= \{-\lfloor t \rfloor, -(\lfloor t \rfloor-1),\cdots,-1, 0,1, \ldots,\lfloor t \rfloor-1, \lfloor t \rfloor\}.\]
  Here  $\lfloor t \rfloor$ denotes the greatest integer less than or equal to $t$.
  \item  For any $A \subset \mathbb{R}$, let $\mathbbm{1}_A$ denote the characteristic function on $A$. 
\end{itemize}
	\begin{definition}
	Let $\mathcal{H}$ be a separable Hilbert space. A sequence of functions $\{f_k: k \in \mathbb{Z}\}$ in $\mathcal{H}$ is said to be a frame for $\mathcal{H}$ if there exist constants $0< M \leq N< \infty$ such that 
	\begin{equation*}
		M\|f\|^2\leq \sum_{k \in \mathbb{Z}}|\left\langle f,f_k \right\rangle|^2 \leq 	N\|f\|^2
	\end{equation*}
for all $f \in \mathcal{H}$. The constants $M$ and $N$ are called the lower and upper frame bounds, respectively. 

Let $\{f_k: k \in \mathbb{Z}\}$ be a frame in a Hilbert space $\mathcal{H}$ with frame bounds $M$ and $N$. Define the frame operator $S:\mathcal{H} \longrightarrow \mathcal{H}$ as 
\[Sf=\sum_{k \in \mathbb{Z}}\left\langle f,f_k\right\rangle f_k.\]
Then, it is well known that the operator $S$ is bounded and invertible (see  \cite{paper12}). Moreover, $\{S^{-1}f_k: k \in \mathbb{Z}\}$ is also a frame for $\mathcal{H}$, it has bounds $\frac{1}{N}$ and $\frac{1}{M}$, and is called the canonical dual of $\{f_k:k \in \mathbb{Z}\}.$
\end{definition}
\begin{definition}
     A Hilbert space $V$ of functions defined on $\mathbb{R}$ is called a reproducing kernel Hilbert space (RKHS) if, for every $y \in \mathbb{R}$, the linear evaluational functional $E_y: V \longrightarrow \mathbb{R}$ defined by $E_y(f)=f(y)$ is bounded.
     
     Let $V$ be an RKHS on $\mathbb{R}$. Then, for every $y \in \mathbb{R}$, there exists a unique vector $k_y \in V$ such that \[f(y)=\left\langle f,k_y \right\rangle \hspace{0.2cm} \forall \hspace{0.1cm} f \in V.\] Let
     $K(x,y):=k_y(x) \hspace{0.2cm}\forall \hspace{0.2cm}x,y \in \mathbb{R}$. The multivariable function $K$ is called the reproducing kernel for $V$. Note that with this definition of $K$, we have 
    \begin{equation*} 
	   K(x,y)=k_y(x)=\left\langle k_y,k_x \right\rangle \text{ and }
	   \|k_y\|^2=\left\langle k_y,k_y \right\rangle = K(y,y).
    \end{equation*}
    \end{definition}
    	\begin{lemma}\label{paper116000}  \cite[Lemma 2]{paper124}
		Assume that $\mathcal{X} \subset \mathbb{R}$ is separated, that is, $\inf_{x,y\in \mathcal{X}:x\neq y}|x-y|>0$. Let $n$ and $m$ be distinct real numbers. If $r>1$ is a real number, then 
		\begin{align*}
			\sum_{x \in \mathcal{X}}\frac{1}{(1+|n-x|)^r(1+|m-x|)^r}&\leq \\ & \hspace{-1.5cm}\frac{1}{\left(1+\frac{|n-m|}{2}\right)^r}\left(\sum_{x \in \mathcal{X}}\frac{1}{(1+|n-x|)^r}+\sum_{x \in \mathcal{X}}\frac{1}{(1+|m-x|)^r}\right).
			\end{align*}
		\end{lemma}
		
    \begin{theorem}[Chernoff's inequality for small deviations]\label{paper13001}
        Let $\{X_i\}_{i=1}^N$ be independent Bernoulli random variables with parameters $\{p_i\}_{i=1}^N$. Consider the sum $S_N=\sum_{i=1}^NX_i$ and denote its mean by $\mu=\mathbb{E}S_N$. Then, for $\delta\in (0,1]$ we have 
        \begin{equation*}
            \mathbb{P}\{|S_N-\mu|\geq\delta \mu\}\leq 2e^{-\frac{\mu\delta^2}{3}}.
        \end{equation*}
    \end{theorem}
     \begin{definition}
      For a positive integer $L$ and real $\delta >0$, the quantization  alphabet $\mathcal{A}_L^\delta$ is defined as\begin{equation}\label{paper120001}
		    \mathcal{A}_L^\delta:=\{\pm (2l-1)\delta: 1\leq l \leq L, l \in \mathbb{Z}\}.
		\end{equation}
    The above-mentioned alphabet will be used by us throughout the paper.
 \end{definition}
	\subsection{Noise-shaping quantizers}
	Given a quantization alphabet $\mathcal{A}$, by a noise-shaping quantizer we mean any map $Q:\mathbb{R}^m\rightarrow \mathcal{A}^m$ that satisfies
	\begin{equation}
		\label{paper148}	y-q=Hu,
	\end{equation}
	where $q:=Q(y)$, $H$ is an $m \times m$ lower triangular Toeplitz matrix with unit diagonal and $u$ is a vector such that $\|u\|_\infty \leq c$ for some constant $c$ which is independent of $m$. The condition on $H$ may be relaxed, as done in the distributed noise-shaping quantizer, where we do not have $H$ to be Toeplitz. Noise-shaping quantizers do not exist unconditionally because of the restriction on $u$ and $H$.  However, under certain suitable assumptions on $H$ and $\mathcal{A}$, they exist and can be implemented via recursive algorithms.
	\begin{lemma}  \cite[Lemma 4.2]{paper15} \label{paper150}
		Let $\mathcal{A}:=\mathcal{A}^\delta_{L}$. Assume that $H=I-\tilde{H}$, where $\tilde{H}$ is strictly lower triangular, and $\mu\geq0$ is such that 
		\[\|\tilde{H}\|_{\infty\rightarrow\infty}+\frac{\mu}{\delta}\leq2L.\]
		Suppose that $\|y\|_{\infty}\leq\mu$. For each $s\geq 1$, let $w_s:=y_s+\sum_{j=1}^{s-1}\tilde{H}_{s,s-j}u_{s-j},$
		\begin{equation*}
			q_s:=(\mathcal{Q}(y))_s=\arg\min_{r \in \mathcal{A}}|w_s-r|,
		\end{equation*}
		and 
		\begin{equation*}
			u_s:=w_s-q_s.
		\end{equation*}
		Then the resulting $q$ satisfies the noise-shaping relation \eqref{paper148} with $\|u\|_\infty\leq \delta$.
	\end{lemma}
	\subsubsection{$\Sigma \Delta$ quantization}
	For a positive integer $n$, the standard $n^{\text{th}}$-order $\Sigma\Delta$ scheme with alphabet $\mathcal{A}_L^\delta$ runs the following iteration for $s=1, \ldots,m$.
	\begin{align}
		\nonumber q_s&=Q(u_{s-1}, u_{s-2},\ldots,u_{s-n},y_s),\\
		\label{paper146} (\Delta^nu)_s&=y_s-q_s.
	\end{align}
	Here  $Q: \mathbb{R}^{n+1} \rightarrow {\mathcal{A}_L^\delta}$ is the quantizer, $u_s$ are internal state variables in the algorithm and the operator $\Delta^n$ results from $n$ subsequent concatenations of the operator $(\Delta u)_s=u_{s}-u_{s-1}$. In the vector form \eqref{paper146} can be written as 
	\begin{equation}
		\label{paper167} y-q=D^nu,
	\end{equation}  
	where $D$ is the $m \times m$ first order difference matrix with entries given by
	\begin{equation}
		(D)_{ij}=
		\begin{cases}
			1 & \text{if } i=j,\\
			-1 & \text{if } i=j+1,\\
			0 & \text{otherwise}.\\
		\end{cases}
	\end{equation}
	An $n^{\text{th}}$-order $\Sigma \Delta$ quantization scheme is said to be stable if there exist positive constants $D_1$ and $D_2$, independent of $m$, such that for any positive integer $m$ and $y \in \mathbb{R}^m$ one has 
	\[\|y\|_{\infty}\leq D_1 \implies \|u\|_{\infty}\leq D_2.\]
	Although constructing such schemes for a given alphabet is a nontrivial task, such constructions exist. The following proposition from  \cite{paper18} will be used by us.
	\begin{prop} \cite{paper18} \label{paper139}
		There exists a universal constant $C >0$ such that for any alphabet $\mathcal{A}^\delta_L=\{\pm (2l-1)\delta: 1\leq l \leq L, l \in \mathbb{Z}\}$, there is a stable $n^{\text{th}}$-order $\Sigma\Delta
		$ scheme satisfying
		\begin{equation*}
			\|y\|_\infty \leq \mu \implies \|u\|_\infty \leq \delta C(n),
		\end{equation*}
		where $C(n):=C\left(\bigg\lceil \frac{\pi^2}{(\cosh^{-1}(2L- \frac{\mu}{\delta}))^2} \bigg\rceil \frac{e}{\pi}\cdot n \right)^n.$
	\end{prop} 
        Next, we define the $\Sigma\Delta$ condensation operator.
	\begin{definition}
		Let $m$ be a positive integer and $p\in \{1, \ldots,m\}$ be such that $\frac{m}{p}=\tilde{\lambda}n-n+1$ for some integer $\tilde{\lambda}$. Let ${\nu_{\Sigma \Delta}}$ be the row vector in $\mathbb{R}^{\frac{m}{p}}$ whose entry $\left({\nu_{\Sigma \Delta}}\right)_j$ is the $j$th coefficient of the polynomial $(1+x+\cdots+x^{\tilde{\lambda}-1})^n$. Then the $\Sigma\Delta$ condensation operator  \cite{paper15} $\tilde{V}_{\Sigma\Delta} \in \mathbb{R}^{p \times m}$ is defined as
		\begin{equation} \label{paper136} 
			\tilde{V}_{\Sigma\Delta}:=\frac{V_{\Sigma \Delta}}{\|\nu_{\Sigma \Delta}\|_1}, \text{ where }
			V_{\Sigma\Delta}= I_p \otimes {\nu_{\Sigma \Delta}}=
			\begin{pmatrix}
				{\nu_{\Sigma \Delta}} &  &  &  \\
				& {\nu_{\Sigma \Delta}}     &  &  \\ 	
				&   & \ddots &   \\
				&  &  & {\nu_{\Sigma \Delta}} 		\\
				&  &  & & {\nu_{\Sigma \Delta}} 
			\end{pmatrix}.
		\end{equation}
	\end{definition}
        In  \cite{paper15}, the $\Sigma\Delta$ condensation operator was $\ell_2$ normalized; here, we find that $\ell_1$ normalization is more convenient for our purposes. It can be shown \cite[Lemma 4.6]{paper15} that
	\begin{equation} \label{paper140}
		\|\tilde{V}_{\Sigma\Delta}D^n\|_{\infty \rightarrow 2} \leq \sqrt{p}(8n)^{n+1}\left({\frac{m}{p}}\right)^{-n}.
	\end{equation}
	\subsubsection[]{Distributed noise-shaping quantization}
	Given an alphabet $\mathcal{A}_L^\delta$, a distributed noise-shaping quantization (also called $\beta$ quantization) scheme with input $y$ computes a uniformly bounded solution $u$ to the equation
		\begin{equation*}
		y-q=H_\beta u,
	\end{equation*}  
    where the matrix $H_\beta$ is as defined below.
    
	Let $m$ and $p$ be positive integers such that $p$ divides $m$. Then, for any fixed $\beta >1$, the block diagonal distributed noise-shaping transfer operator \cite{paper16} $H_\beta$ is given by
	\begin{equation} \label{paper137}
		H_\beta= I_p \otimes \tilde{H}_\beta = 		\begin{pmatrix}
			\tilde{H}_\beta &  &  &  \\
			& \tilde{H}_\beta     &  &  \\
			&   & \ddots &   \\
			&  &  & \tilde{H}_\beta		\\
			&  &  & & \tilde{H}_\beta
		\end{pmatrix},
	\end{equation}
	where $\tilde{H}_\beta$ is the following $\frac{m}{p} \times \frac{m}{p}$ matrix 
	\begin{equation*}
		(\tilde{H}_\beta)_{ij}:=
		\begin{cases}
			1 & \text{if } i=j,\\
			-\beta & \text{if } i=j+1,\\
			0 & \text{otherwise}.
		\end{cases}
	\end{equation*} 
	\begin{definition}
		Let  the vector $\nu_\beta:=[\beta^{-1} \beta^{-2} \ldots \beta^{-\frac{m}{p}}]$. Then the $\beta$ condensation operator  \cite{paper16} $\tilde{V}_{\beta}$ is defined as
		\begin{equation} \label{paper138}
			\tilde{V}_{\beta}:=\frac{V_{\beta}}{\|\nu_\beta\|_1}, \text{ where }
			V_{\beta}= I_p \otimes \nu_\beta=
			\begin{pmatrix}
				\nu_\beta &  &  &  \\
				& \nu_\beta     &  &  \\
				&   & \ddots &   \\
				&  &  & \nu_\beta		\\
				&  &  & & \nu_\beta 
			\end{pmatrix}.
		\end{equation}
	\end{definition}
	\noindent It can be easily seen that $\|V_{\beta}H_\beta\|_{\infty \rightarrow \infty}= \beta^{-\frac{m}{p}}$, which along with $\|\nu_\beta\|_1\geq \frac{1}{\beta}$ implies that
	\begin{equation}\label{paper141}
		\|\tilde{V}_\beta H_\beta\|_{\infty \rightarrow 2}\leq \sqrt{p}\beta^{-\frac{m}{p}+1}.	
	\end{equation}	

 \subsection{Sampling in Bandlimited Functions} \label{paper1500}
 The Fourier transform of a function $f \in L^1(\mathbb{R})$ is defined as \[\hat{f}(\xi)=\frac{1}{\sqrt{2 \pi}}\int_{\mathbb{R}}f(t)e^{-it\xi}dt,\]
 with the usual extension via a unitary operator to functions in $L^2(\mathbb{R})$.
 Let $PW_{[-\pi,\pi]}$ denote the space of $\pi$-bandlimited functions, that is,
	\[PW_{[-\pi, \pi]}:=\{f \in L^2(\mathbb{R}): \supp(\hat{f})\subset [-\pi, \pi]\}.\]
 and
 \begin{equation}\label{paper12003}
     C_{[-\pi, \pi]}:=\{f \in PW_{[-\pi, \pi]}: \|f\|_{L^\infty(\mathbb{R})}\leq 1, f \text{ is real valued}\}.
 \end{equation}
	The celebrated Shannon-Nyquist  sampling theorem says that for any $f \in PW_{[-\pi, \pi]}$, we have 
	\begin{equation}\label{paper11}
		f(t)= \sum_{n \in \mathbb{Z}}f(n)\sinc(t-n) \hspace{0.6cm} \forall \hspace{0.2cm} t \in \mathbb{R},
	\end{equation}
	where $\sinc(t):= \frac{\sin\pi t}{\pi t}$. However, \eqref{paper11} is not useful in practice because $\sinc (x)$ decays too slowly. To circumvent this issue, it is useful to introduce oversampling. This amounts to viewing $PW_{[-\pi, \pi]}$ as a subspace of a shift-invariant space generated by a single smooth fast decaying function with orthonormal translates. This can easily be done as shown in  \cite{paper125}. We review the method over here.
 
	Let $\lambda > 1$ be a fixed real number. Choose a function $g_\lambda$ such that 
	\begin{enumerate}
		\item $\widehat{g_\lambda} \in C^\infty$.
		\item 
		\begin{equation}\label{paper12}
			\widehat{g}_\lambda(\xi)=
			\begin{array}{cc}
				\bigg\{ & 
				\begin{array}{cc}
					\frac{1}{\sqrt{2 \pi  \lambda}} & \xi \in [-\pi, \pi], \\
					0 & |\xi|> (2  \lambda-1) \pi.\\
				\end{array}
			\end{array}
		\end{equation}
		\item 
		\begin{equation}\label{paper13}
			\sum_{k \in \mathbb{Z}}\big|\widehat{g_\lambda}(\xi+2k \lambda\pi)\big|^2
			=\frac{1}{2 \pi \lambda} \hspace{0.2cm} \forall \hspace{0.1cm} \xi \in \mathbb{R}.
		\end{equation}
	\end{enumerate}
	\begin{definition}
		Define,
		\begin{equation*}
			V(g_\lambda):=\overline{\Span}\bigg\{g_\lambda\left(\cdot-\frac{k}{\lambda }\right): k \in \mathbb{Z}\bigg\}.
		\end{equation*}
	\end{definition}
	Using  \cite[Theorem 9.2.5]{paper12} and \eqref{paper13}, it can be seen that $\left\{g_\lambda\left(\cdot-\frac{k}{\lambda }\right): k \in \mathbb{Z}\right\}$ forms a orthonormal basis for $V(g)$. Now, consider any bandlimited function $f \in PW_{[-\pi, \pi]}$, then its Fourier transform can be viewed as an element in $L^2[-(2 \lambda -1)\pi, (2  \lambda-1) \pi]$. Using the Fourier series expansion of $\hat{f}$ on $[-(2  \lambda-1)\pi, (2  \lambda-1)\pi]$ and \eqref{paper12}, it can be shown that 
	\begin{equation}\label{paper14}
		f(t)=\frac{1}{\sqrt{ \lambda}}\sum_{n \in \mathbb{Z}}f\left(\frac{n}{\lambda}\right)g_\lambda\left(t- \frac{n}{ \lambda
		}\right) \hspace{0.2cm} \forall \hspace{0.1cm} t \in \mathbb{R}.    
	\end{equation}
    That is, $f \in V(g_\lambda)$ and $\left \langle f,g_\lambda\left(\cdot- \frac{n}{ \lambda
		}\right) \right \rangle= \frac{1}{\sqrt{ \lambda}}f\left(\frac{n}{\lambda}\right) \hspace{0.2cm} \forall \hspace{0.1cm} n \in \mathbb{Z}$.
  \begin{remark}\label{paper1localizationpropertyofmodifiedshannonformula}
	In comparison to \eqref{paper11}, the above reconstruction formula \eqref{paper14}, although requiring more samples, has the advantage that each sample is weighted in a well-localized way ($f\left(\frac{n}{ \lambda}\right)$ only contributes in a small neighbourhood of $\frac{n}{ \lambda}$). This property will be exploited by us when we find a finite-dimensional approximation space for $PW_{[-\pi, \pi]}$ in the next section.
  \end{remark}
  \begin{remark}
      Throughout this paper, we fix \(\lambda = 2\). Typically, \(\lambda\) serves as the oversampling factor; however, in our case, it is introduced solely to derive formulation \eqref{paper14} and does not function as the oversampling factor. Let $g:=g_2$. Hence, from \eqref{paper14}, we can conclude that for all $f \in PW_{[-\pi,\pi]} \subset V(g)$,
      \begin{equation}\label{paper14new}
		f(t)=\frac{1}{\sqrt{2}}\sum_{n \in \mathbb{Z}}f\left(\frac{n}{2}\right)g\left(t- \frac{n}{2
		}\right) \hspace{0.2cm} \forall \hspace{0.1cm} t \in \mathbb{R}.    
	\end{equation}
  \end{remark}
\section{Sampling-Quantization-Reconstruction Pipeline}\label{paper1SQRpipeline}
This section provides an overview of our entire sampling-quantization-reconstruction process. Our method consists of three components, and a summary of each component is presented as a subsection below. Each of the subsections will be discussed in detail, with the proofs provided in separate sections following this one. Thus, this section serves to connect the three subsequent sections and to highlight their relationships as a whole.

Our method proceeds as follows:  
First, we identify a finite-dimensional approximation space (FDAS) for $ PW_{\left[-\pi, \pi \right]}$. Then, we construct an appropriate random frame for the FDAS such that the frame coefficients of any function in the FDAS can be computed using the function samples. Finally, in the third step, we apply the canonical dual frame to reconstruct the original function and bound the reconstruction error.

\subsection{A suitable finite dimensional approximation space for $ PW_{[-\pi, \pi]} $}

Since we are working with functions in an infinite-dimensional Hilbert space, specifically the space of bandlimited functions, the first step is to project these functions onto a finite-dimensional approximation space (FDAS). This projection allows us to work with a finite number of samples, which in turn ensures that the sequence of quantized measurements $(q_\lambda)_{\lambda \in I}$ belongs to $\ell^2(I)$, thereby enabling us to operate within a Hilbert space framework. This framework is crucial for the techniques we will employ throughout this paper. In other words, the purpose of this first step of projection is to ensure that we do not lose the Hilbert space setting after the quantization of the samples.

\smallskip
Let $R > 1$ be a real number. Given random samples of a bandlimited function $f \in PW_{[-\pi, \pi]}$, our objective is to reconstruct $f$ on the interval $[-R, R]$. Clearly, one basic condition that the finite-dimensional approximation space (FDAS), say $V$, should satisfy is that the orthogonal projection of $f$ onto $V$ should be a very good approximation of $f$ on $[-R, R]$. This is because if the projection changes the value of the function in the region where we want to reconstruct the function then it no matter how well we reconstruct the projected function using the quantized sample is we can never do better than the projection error which itself 
could be significant on $[-R,R]$. To construct such an FDAS, we utilize the localization property outlined in Remark \ref{paper1localizationpropertyofmodifiedshannonformula}.

Fix $R_1 > R$ (see Figure \ref{fig:Fig6}). Let $V$ be defined in such a way that for any function 
\begin{equation}\label{paper1modifiedshanonformula}
            f = \frac{1}{\sqrt{2}}\sum_{k \in \mathbb{Z}}f\left(\frac{k}{2 }\right) g\left(\cdot-\frac{k}{2 }\right) \in PW_{[-\pi,\pi]} \subset V(g),
        \end{equation}
(see \eqref{paper14new} for the above representation of bandlimited functions) we have
\[
Pf = \frac{1}{\sqrt{2}} \sum_{k \in \left[2 R_1\right]} f\left(\frac{k}{2}\right) g\left(\cdot - \frac{k}{2}\right),
\]
where $P$ denotes the orthogonal projection from $L^2(\mathbb{R})$ onto $V$.

\smallskip
Note that to achieve the above formula for $Pf$, it is sufficient to define $V := V^{R_1}(g)$, where
\begin{equation}\label{definitionofV^{R_1}(g)}
   V^{R_1}(g) := \text{span} \left\{ g\left(\cdot - \frac{k}{2}\right) : k \in \left[ 2 R_1 \right] \right\}. 
\end{equation}
This holds because the translates of $g$, that is, $\left\{ g\left(\cdot - \frac{k}{2}\right) \right\}$, are orthonormal by construction (see \eqref{paper13}).
Given a function $f \in PW_{[-\pi, \pi]}$, it is clear from the above expression for $Pf$ that this projection is computed by zeroing out the samples of $f$ outside $\left[-R_1, R_1\right]$, that is, $\{ f\left(\frac{n}{2}\right) \}_{|\frac{n}{2}| \geq R_1}$, in \eqref{paper1modifiedshanonformula}. As the samples are locally weighted due to the decay of $g$, this replacement has a minimal effect on $f$ within the region $[-R, R]$, provided $R_1$ is sufficiently larger than $R$. As a result, $Pf$ will be a good approximation of $f$ on $[-R, R]$, though not necessarily outside this region, especially near the boundary points $-R_1$ and $R_1$, where significant discrepancies may occur between $f$ and $Pf$. We refer the reader to Section \ref{paper176}, where the error bounds between \( f \) and \( Pf \) are made explicit.
\begin{figure}
\begin{center}
    \begin{tikzpicture}
\draw[<->] (-6.5,0)--(6.5,0);
\draw (0,0.1)--(0,-0.1) node[below] {$0$};
\draw (4,0.1)--(4,-0.1) node[below] {$R$};
\draw (-4,0.1)--(-4,-0.1) node[below] {$-R$};
\draw (5,0.1)--(5,-0.1) node[below] {$R_1$};
\draw (-5,0.1)--(-5,-0.1) node[below] {$-R_1$};
\end{tikzpicture} 
\end{center}
\caption{Relative positions of $R_1$ and $R$}
    \label{fig:Fig6}
\end{figure}
\subsection{A random frame for the approximation space}
The next step is to identify an appropriate random frame for the FDAS $V^{R_1}(g)$. We start by introducing existing techniques and motivate our case based on them. 

\smallskip
\textbf{Existing theory and practise \cite{paper16}.}
Let $\mathcal{H}$ be a finite-dimensional Hilbert space, and let $\{f_i\}_ {i=1}^m$ be a collection of elements in $\mathcal{H}$. Define the (analysis) operator \cite{paper12} $E: \mathcal{H} \to \mathbb{C}^m$ as
\[
E(f) := \left\{ \langle f, f_i \rangle \right\}_{i=1}^m.
\]
Let $p$ be a positive integer that divides $m$, and let $V \in \mathbb{R}^{p \times m}$ represent either the $\Sigma\Delta$ operator \eqref{paper136} or the $\beta$ condensation operator \eqref{paper138}. Now consider the operator $VE := V \circ E: \mathcal{H} \to \mathbb{C}^p$. It would have the following form.
\begin{align*}
    \left(VE(f)\right)_j&= \left(V \left\{ \left\langle f, f_i \right\rangle \right\}_{i=1}^m\right)_j \hspace{0.2cm} \forall \hspace{0.1cm} j \in \{1, \ldots, p \}.\\
    &= \sum_{i=1}^m V_{ji}\left\langle f, f_i \right\rangle= \left\langle f,\sum_{i=1}^m V_{ji} f_i \right\rangle.
\end{align*}
Note that $VE$ is just the analysis operator associated with the collection $\left\{ \sum_{i=1}^m V_{ji} f_i \right\}_{j=1}^p$. Assume this collection forms a frame for $V$ (in which case the frame will have $p$ elements) and let $\widetilde{VE}$ denote the synthesis operator associated with its canonical frame. Then the following frame reconstruction formula holds:
\[
\widetilde{VE}VE(f) = f \hspace{0.2cm} \forall \hspace{0.1cm} f \in V.
\]
Now, suppose the measurement $y = Ef$ is quantized using a noise-shaping quantization scheme with transfer operator $H$. That is,  $y - q= Hu$. Then, the error between $f^\sharp := \widetilde{VE}Vq$ and $f$ can be calculated as
\begin{align*}
    \left\|f-f^\sharp\right\|&=\left\|\widetilde{VE}VEf-\widetilde{VE}Vq \right\|\\
    &=\left\|\widetilde{VE}V(y-q) \right\|\\
    &=\left\|\widetilde{VE}VHu \right\|\\
    \numberthis & \leq \left\|\widetilde{VE}\right\|\left\|VH\right\|_{\infty \rightarrow 2}\left\|u\right\|_{\infty}.\label{paper1VEbound}
\end{align*} 
Subsequently, the bounds in \eqref{paper140} and \eqref{paper141} can be used to obtain decay rates for $\Sigma \Delta$ and $\beta$ quantization schemes, respectively.

\smallskip
If the information provided to us is in the form of samples, then it is preferable to find a frame consisting of (linear combinations of) reproducing kernels. This is because in such a case the frame coefficients can be computed from the function samples. 

A common approach used for obtaining a frame of reproducing kernels is to assume that the signal space $\mathcal{H}$ is a reproducing kernel Hilbert space (RKHS) and demonstrate that the sampling set $\{x_i\}_{i=1}^m$ forms a stable set of sampling (with high probability, if the samples are random). In this case, there exist constants $A,B > 0$ such that:
\[
A \|f\|^2 \leq \sum_{i=1}^m \left| f(x_i) \right|^2 \leq B \|f\|^2 \quad \forall f \in \mathcal{H}.
\]
Since $\mathcal{H}$ is an RKHS, we have $f(x_i) = \langle f, k_{x_i} \rangle$ for all $i \in \{1, \ldots, m\}$. Consequently,
\[
A \|f\|^2 \leq \sum_{i=1}^m \left| \langle f, k_{x_i} \rangle \right|^2 \leq B \|f\|^2 \quad \forall f \in \mathcal{H}.
\]
That is, the collection $\{k_{x_i}\}_{i=1}^m$ forms a frame for $\mathcal{H}$, and the frame coefficients for any $f \in \mathcal{H}$ are given by the samples of $f$.

Inspired by this method, and considering the additional step of quantization, we aim to show that $\left\{ \sum_{i=1}^m V_{ji} k_{x_i} \right\}_{j=1}^p$ forms a frame for $\mathcal{H}$. Specifically, we wish to prove the existence of constants $A,B > 0$ such that:
\[
A \|f\|^2 \leq \sum_{j=1}^p \left| \left\langle f, \sum_{i=1}^m V_{ji} k_{x_i} \right\rangle \right|^2 \leq B \|f\|^2 \quad \forall f \in \mathcal{H}.
\]
In other words,
\begin{equation}\label{paper1modifiedsamplinginequalitywithV}
    A \|f\|^2 \leq \sum_{j=1}^p\left|\sum_{i=1}^m V_{ji} f(x_i)\right|^2 \leq B\|f\|^2 \hspace{0.2cm} \forall \hspace{0.1cm} f \in \mathcal{H} .
\end{equation}
        Further, by defining the operator $E$ as $Ef = \{\langle f,k_{x_i} \rangle\}_{i=1}^m=\{f(x_i)\}_{i=1}^m$ for all $f \in  \mathcal{H}$, \eqref{paper1modifiedsamplinginequalitywithV} can also be written as 
\begin{equation}\label{paper1newmodifiedsamplinginequalitywithV}
    A \|f\|^2 \leq \sum_{j=1}^p\left|(VE(f))_j\right|^2 \leq B\|f\|^2 \hspace{0.2cm} \forall \hspace{0.1cm} f \in \mathcal{H} .
\end{equation}
If we are able to prove this inequality \eqref{paper1modifiedsamplinginequalitywithV}, then by setting $E = \{k_{x_i}\}_{i=1}^m$, we can use the bound in \eqref{paper1VEbound} to control the quantization error.




Having established the concept of an appropriate frame in a general sense, we now focus on our specific case.

\smallskip
\textbf{Our technique.} First, recall that our approximation space is fixed as $V^{R_1}(g)$. Since we aim to work with a reasonable number of samples, we sample in a region larger than $\left[-R_1, R_1\right]$. We give a heuristic argument for this. If there are no sample points in an interval $[a, b] \subset [-R_1, R_1]$, then it is unreasonable to expect that a function concentrated in $[a, b]$ can be reconstructed using a feasible number of frame measurements, as no samples are taken from the region where the function is concentrated. Hence, it follows that sampling should occur over the interval $[-R_2, R_2]$, where $R_2 > R_1$ (see Figure \ref{fig:Fig7} for reference). Subsequently, \textbf{ assume that $\{x_i\}_{i=1}^m$ is a sequence of i.i.d. random variables uniformly distributed on $[-R_2, R_2]$}, where $R_2 > R_1$. \\
\begin{figure}
\begin{center}
    \begin{tikzpicture}
\draw[<->] (-6.5,0)--(6.5,0);
\draw (0,0.1)--(0,-0.1) node[below] {$0$};
\draw (4,0.1)--(4,-0.1) node[below] {$R$};
\draw (-4,0.1)--(-4,-0.1) node[below] {$-R$};
\draw (5,0.1)--(5,-0.1) node[below] {$R_1$};
\draw (-5,0.1)--(-5,-0.1) node[below] {$-R_1$};
\draw (6,0.1)--(6,-0.1) node[below] {$R_2$};
\draw (-6,0.1)--(-6,-0.1) node[below] {$-R_2$};
\end{tikzpicture} 
\end{center}
\caption{Relative positions of $R_2$,$R_1$ and $R$}
    \label{fig:Fig7}
\end{figure}
\smallskip
Let $f \in PW_{[-\pi, \pi]}$. Since we are working in the finite-dimensional space $V^{R_1}(g)$, we require the samples of the projected function, $\{Pf(x_i)\}_{i=1}^m$, to compute the frame measurements of $Pf$. However, the available samples are of the original function $f$, not the projected function $Pf$. Given that, by construction, $Pf$ closely approximates $f$ on the interval $[-R, R]$, we will use the original samples $\{f(x_i)\}_{i=1}^m$ instead of $\{Pf(x_i)\}_{i=1}^m$ to calculate frame measurements. It is important to note that some sampling points lie in the region $[-R_2, R_2] \setminus [-R, R]$, where $Pf$ may not approximate $f$ well.

\textbf{The goal is to devise a reconstruction process in which the error caused by this possibly inaccurate approximation of samples in $[-R_2,R_2]\setminus [-R,R]$ can be controlled and minimized.} First, we will explore how these inaccurately approximated samples could negatively impact the reconstruction process.

\smallskip
Consider the vector  $\left\{f(x_i)\right\}_{i=1}^m$. As we want to prove an inequality of the type \eqref{paper1newmodifiedsamplinginequalitywithV}, let us observe the action of the matrix $V$ on the  $\left\{f(x_i)\right\}_{i=1}^m$ for any given $f \in V^{R_1}(g)$.  From here on we use $\nu$ to denote both $\nu_{\Sigma \Delta}$ and $\nu_\beta$. We shall explicitly state what $\nu$ represents in cases where it is unclear from the context. Consider,
\[
V\left\{f(x_i)\right\}_{i=1}^m=
			\begin{pmatrix}
				{\nu} &  &  &  \\
				& {\nu}     &  &  \\ 	
				&   & \ddots &   \\
				&  &  & {\nu} 		\\
				&  &  & & {\nu} 
			\end{pmatrix} \begin{pmatrix}
				{f(x_1)}   \\
				{f(x_2)}  \\ 	
				\vdots &   \\
				{f(x_{m-1})} 		\\
				{f(x_m)} 
			\end{pmatrix}.
\]
Thus,
\begin{equation}\label{paper1orginalVf}
V\left\{f(x_i)\right\}_{i=1}^m=
                \begin{pmatrix}
				{\nu_1 f(x_1) + \cdots + \nu_{\frac{m}{p}} f\left(x_{\frac{m}{p}}\right)}   \\
				{\nu_{1} f\left(x_{\frac{m}{p}+1}\right) + \cdots + \nu_{\frac{m}{p}} f\left(x_{\frac{2m}{p}}\right)}    \\ 	
				\vdots &   \\
				{\nu_{1} f\left(x_{\frac{(p-2)m}{p}+1}\right) + \cdots + \nu_{\frac{m}{p}} f\left(x_{\frac{(p-1)m}{p}}\right)}\\
				{\nu_{1} f\left(x_{\frac{(p-1)m}{p}+1}\right) + \cdots + \nu_{\frac{m}{p}} f(x_m)} 
			\end{pmatrix}.
\end{equation}
That is, when we apply $V$, it essentially takes $\frac{m}{p}$ samples at a time and computes their linear combination. Since the sample points are i.i.d. and uniformly distributed, we cannot be certain whether all of the $\frac{m}{p}$ function samples taken at a time are accurately approximated, as some sample points may lie in $[-R_2, R_2] \setminus [-R, R]$. This leads to the intermixing of the accurately and inaccurately approximated samples which negatively effects our reconstruction process. Furthermore, we cannot reorder the random samples without altering their distribution and as our theory critically uses the fact that the random samples are uniformly distributed, this is not an option. Thus, our solution is to partition the sample points into three collections. We first describe the process roughly and then define rigorously.

\smallskip
The first step is to divide the interval $[-R_2, R_2]$ into three parts. Fix $\epsilon > 0$ such that $\epsilon R \geq 1$, and let $R_2 = (1 + 3\epsilon)R$. Consider the interval $[-(1 + 3\epsilon)R, (1 + 3\epsilon)R]$, partitioned as follows: $I_{1\epsilon} := (-(1 + \epsilon)R, (1 + \epsilon)R)$, $I_{2\epsilon} := (-(1 + 2\epsilon)R, -(1 + \epsilon)R] \cup [(1 + \epsilon)R, (1 + 2\epsilon)R)$ and $I_{3\epsilon} := [-(1 + 3\epsilon)R, -(1 + 2\epsilon)R] \cup [(1 + 2\epsilon)R, (1 + 3\epsilon)R]$ (see Figure \ref{fig:Fig8} for reference).

\smallskip
Additionally, let $R_1 = (1 + \frac{5}{2} \epsilon)R$. Now, choose $x_1$ uniformly distributed over $[-R_2, R_2] = [-(1 + 3\epsilon)R, (1 + 3\epsilon)R]$. If $x_1 \in I_{1\epsilon}$, define $y^1_1 := x_1$; if $x_1 \in I_{2\epsilon}$, define $y^2_1 := x_1$; if $x_1 \in I_{3\epsilon}$, define $y^3_1 := x_1$. Follow the same process for $x_2, \dots, x_m$. At the end, we have three collections: $\{y^1_i\}_{i=1}^{m_1}$, $\{y^2_i\}_{i=1}^{m_2}$, and $\{y^3_i\}_{i=1}^{m_3}$, where $m_1, m_2$, and $m_3$ represent the random number of points in $I_{1\epsilon}, I_{2\epsilon}$, and $I_{3\epsilon}$, respectively. These random variables can take values from $0$ to $m$. \\

\begin{figure}
\begin{tikzpicture}
\draw[<->] (-7.5,0)--(7.5,0);
\draw[red,-, line width=2pt] (-3,0)--(3,0);
\draw[blue,-, line width=2pt] (-5,0)--(-3,0);
\draw[blue,-, line width=2pt] (3,0)--(5,0);
\draw[green,-, line width=2pt] (5,0)--(7,0);
\draw[green,-, line width=2pt] (-7,0)--(-5,0);
\draw (0,0.1)--(0,-0.1) node[below] {$0$};
\draw (2,0.1)--(2,-0.1) node[below] {$R$};
\draw (-2,0.1)--(-2,-0.1) node[below] {$-R$};
\draw (3,0.1)--(3,-0.1) node[below] {$(1+\epsilon)R$};
\draw (-3,0.1)--(-3,-0.1) node[below] {$-(1+\epsilon)R$};
\draw (5,0.1)--(5,-0.1) node [below] {$(1+2\epsilon)R$};
\draw (-5,0.1)--(-5,-0.1) node[below] {$-(1+2\epsilon)R$};
\draw (6,0.1) -- (6,-0.1);
\draw (6,0.1)  node[above] {$\left(1+\frac{5}{2} \epsilon\right)R$};
\draw (-6,0.1) -- (-6,-0.1);
\draw (-6,0.1) node[above]{$-\left(1+\frac{5}{2}\epsilon\right)R$};
\draw (7,0.1)--(7,-0.1) node[below] {$(1+3\epsilon)R$};
\draw (-7,0.1)--(-7,-0.1) node[below] {$-(1+3\epsilon)R$};
\draw[fill=blue] (-3,0) circle (0.07);
\draw[fill=blue] (3,0) circle (0.07);
\draw[fill=green] (-5,0) circle (0.07);
\draw[fill=green] (5,0) circle (0.07);
\draw[fill=green] (-7,0) circle (0.07);
\draw[fill=green] (7,0) circle (0.07);
\end{tikzpicture} 
\caption{Partition of $\left[-\left(1+ 3 \epsilon\right)R, \left(1+ 3 \epsilon\right)R\right]$}
    \label{fig:Fig8}
\end{figure}
If $m_1, m_2$, or $m_3$ are not divisible by $\frac{m}{p}$, the next step is to discard a few samples towards the end  from each of the three collections of random variables, so that the cardinality of all the three collections is a multiple of $\frac{m}{p}$. This is done using the floor function.

\smallskip
We now formalize the process.

        \begin{definition} \label{paper1definitionofallrandomvariables}
Let $\left\{x_i\right
\}_{i=1}^m$ be a sequence of random variables uniformly distributed on $[-(1 + 3\epsilon)R, (1 + 3\epsilon)R]$. Define,
		\begin{enumerate}
        \item $m_i:=\left|\{t \in \{1, \ldots,m\}: x_t \in I_{i\epsilon}\}\right| \hspace{0.2cm} \forall \hspace{0.1cm} i \in \{1,2,3\}$.
			\item $p_i:= \sum_{j=1}^i\left\lfloor \frac{m_jp}{m}\right\rfloor \hspace{0.2cm} \forall \hspace{0.1cm} i \in \{1,2,3\}$.
			\item $\widetilde{m}_i:=\left\lfloor \frac{m_ip}{m}\right\rfloor \frac{m}{p} \hspace{0.2cm} \forall \hspace{0.1cm} i \in \{1,2,3\}$.
			\item $\widetilde{m}:=\frac{m}{p}p_3$. That is, if $p_3 \neq 0$, $\frac{\widetilde{m}}{p_3}=\frac{m}{p}$. 
   	\item For each $i \in \{1,2,3\}$, let
   \begin{align*}
       t^i_1&:= \inf\{t \in \{1, \ldots,m\}: x_t \in I_{i\epsilon}\}\\
       t^i_2&:=\inf\{ t \in \{t^i_1+1, \ldots, m\}: x_t \in I_{i\epsilon}\}\\
       &\hspace{2cm}\vdots\\
   t^i_{\widetilde{m}_i}&:=\inf\{ t \in \{t^i_{\widetilde{m}_i-1}+1, \ldots, m\}: x^i_t \in I_{i\epsilon}\}.
 \end{align*} 
Finally, let 
\begin{equation*}
  y^i_j:=x_{t^i_j} \hspace{0.2cm} \forall \hspace{0.1cm}  j \in \{1,\ldots, \widetilde{m}_i\}.
\end{equation*}Conditioned on the random variables $m_1,m_2$ and $m_3$, each of the random variables $\{y^1_i\}_{i=1}^{\widetilde{m}_1}$,$ \{y^2_i\}_{i=1}^{\widetilde{m}_2}$ and $\{y^3_i\}_{i=1}^{\widetilde{m}_3}$ will be i.i.d uniformly distributed on $I_{1\epsilon}$, $I_{2\epsilon}$ and $I_{3\epsilon}$ respectively.
			\item Define $\{\epsilon^1_i\}_{i=1}^{m}$,$\{\epsilon^2_i\}_{i=1}^{m}$ and $\{\epsilon^3_i\}_{i=1}^{m}$ to be sequences of $\pm 1$ Bernoulli independent random variables that are also independent from all the above defined random variables.
		\end{enumerate}
\end{definition}
Some remarks are in order regarding the random variables defined in the above definition.
\begin{remark}\label{paper1samplesnotordered}
   We emphasize that the sample points are not being reordered, as reordering $\{x_i\}_{i=1}^m$ would change their distribution. Instead, we partition the samples into the three collections $\{y^1_i\}_{i=1}^{\widetilde{m}_1}$, $\{y^2_i\}_{i=1}^{\widetilde{m}_2}$, and $\{y^3_i\}_{i=1}^{\widetilde{m}_3}$. Note that within the resulting collections $\{y^1_i\}_{i=1}^{\widetilde{m}_1}$, $\{y^2_i\}_{i=1}^{\widetilde{m}_2}$, and $\{y^3_i\}_{i=1}^{\widetilde{m}_3}$, the samples remain unordered. Further, we mention that the random variable $p_3$ defined above actually represents the number of elements in the random frame we define. This will become clearer in the subsequent sections. 
\end{remark}
The reason for doing this is because, we want to replace the vector $\{f(x_i)\}_{i=1}^m$ with  $\{f(y^1_i)\}_{i=1}^{\widetilde{m}_1}\frown\{f(y^2_i)\}_{i=1}^{\widetilde{m}_2}\frown\{f(y^3_i)\}_{i=1}^{\widetilde{m}_3}$.

\begin{equation}\label{paper1rearrangementofsamples}
 \begin{pmatrix}
				{f(y_1)}   \\ 
    {f(y_2)}  \\
				\vdots &   \\
         \vdots &   \\
               \vdots &      \\ 	
				\vdots &   \\	
				\vdots &   \\
    {f(y_{m-1})}\\  
                    {f(y_m)}
			\end{pmatrix} \longrightarrow  \begin{pmatrix}
				{f(y^1_1)}   \\ 	
				\vdots &   \\
         {f(y^1_{\widetilde{m}_1})}\vspace{0.1cm}\\ 
                {f(y^2_1)}   \\ 	
				\vdots &   \\
                {f(y^2_{\widetilde{m}_2})}\vspace{0.1cm}\\
				{f(y^3_1)}   \\ 	
				\vdots &   \\
                    {f(y^3_{\widetilde{m}_3})}
			\end{pmatrix}.
\end{equation}
Thus, as $\widetilde{m}_1$,$\widetilde{m}_2$ and $\widetilde{m}_3$ are multiples of $\frac{m}{p}$ (which is the cardinality of the vector $\nu$), we get
\[
V\begin{pmatrix}
				{f(y^1_1)}   \\ 	
				\vdots &   \\
{f(y^1_{\widetilde{m}_1})}\vspace{0.1cm}\\ 
                {f(y^2_1)}   \\ 	
				\vdots &   \\
                {f(y^2_{\widetilde{m}_2})}\vspace{0.1cm}\\
				{f(y^3_1)}   \\ 	
				\vdots &   \\
                    {f(y^3_{\widetilde{m}_3})}
			\end{pmatrix}=
                \begin{pmatrix}
				\nu_1 f(y^1_1) + \cdots + \nu_{\frac{m}{p}}
f\left(y^1_{\frac{m}{p}}\right)   \\
\vdots &   \\
{\nu_{1} f\left(y^1_{\frac{(p_1-1)m}{p}+1}\right) + \cdots + \nu_{\frac{m}{p}} f\left(y^1_{\frac{p_1m}{p}}\right)} \\
				{\nu_{1} f\left(y^2_{1}\right) + \cdots + \nu_{\frac{m}{p}} f\left(y^2_{\frac{m}{p}}\right)}\\
\vdots &   \\
{\nu_{1} f\left(y^2_{\frac{(p_2-p_1-1)m}{p}+1}\right) + \cdots + \nu_{\frac{m}{p}} f\left(y^2_{\frac{(p_2-p_1)m}{p}}\right)} \\
				{\nu_{1} f\left(y^3_{1}\right) + \cdots + \nu_{\frac{m}{p}} f\left(y^3_{\frac{m}{p}}\right)}\\
\vdots &   \\
{\nu_{1} f\left(y^3_{\frac{(p_3-p_2-1)m}{p}+1}\right) + \cdots + \nu_{\frac{m}{p}} f\left(y^3_{\frac{(p_3-p_2)m}{p}}\right)} 
			\end{pmatrix}.
\]
It can be observed from the above calculations that when $V$ is applied on the vector $\{f(y^1_i)\}_{i=1}^{\widetilde{m}_1}\frown\{f(y^2_i)\}_{i=1}^{\widetilde{m}_2}\frown\{f(y^3_i)\}_{i=1}^{\widetilde{m}_3}$, no intermixing of samples take place among the three different partitions and hence unlike the earlier case (see \eqref{paper1orginalVf}),  the inaccurately approximated samples and the accurately approximated samples do not mix.

\smallskip
This type of partitioning is based on the following fundamental idea: if the elements of a frame are localized within a certain region, then, under appropriate conditions, the elements of the canonical dual frame can also be shown to be localized within some region. Although substantial research (see \cite{paper1localbalaneressannams2006, paper1localbalanjfaa2006part2, paper1localbalanjfaa2006part1,paper1localcorderoacha2004,paper1localfornasierconstrapprox2005} and references therein) has been conducted on localized frames, the existing results are not directly applicable to our case. Thus, we develop our own bounds, though the core principle remains the same: localization properties of the frame are generally transferred to the canonical dual frame under suitable conditions. In order to leverage this property, we first start by proving that a linear combination of the reproducing kernels at this partitioned points will form a random frame for $V^{\left(1+\frac{5}{2}\epsilon\right)R}(g)$ with high probability. We now defer the reader to Section \ref{paper177}, where we define (see \eqref{paper13004}) and prove (see Lemma \ref{paper172}) that the collection $\{h_p\}_{p=1}^{p_3}$ will form a random frame for $V^{\left(1+\frac{5}{2}\epsilon\right)R}(g)$ with high probability. 

\subsection{Bounding the reconstruction error}\label{paper1boundingthereconstructionerror}
First we observe that the elements of the frame $\{h_p\}_{p=1}^{p_3}$ are localized in certain regions.
\begin{enumerate}
    \item For $j \in \{1,\dots,p_1\}$, $h_j=\sqrt{\frac{2(1+\epsilon)R}{p_1}}\sum_{i=1}^{\widetilde{m}_1}V_{ji}\epsilon^1_ik_{y^1_i}$ is localized in the region $I_{1 \epsilon}$ and its small neighborhood. This is can be concluded from two facts, firstly, for any $x \in \mathbb{R}$, the kernel $k_x$ is localized around $x$ (see Remark \ref{paper1localizationofkernel}), and secondly, $y^1_i \in I_{1\epsilon}$ for all $i \in \{1, \ldots, \widetilde{m}_1\}$. We include the neighborhood of $I_{1 \epsilon}$ in the localization region to account for points in $\{y^1_i\}_{i=1}^{\widetilde{m}_1}$ near the boundary of $I_{1 \epsilon}$.
    \item Similarly, for $j \in \{p_1+1,\dots,p_2\}$, $h_j=\sqrt{\frac{2\epsilon R}{p_2-p_1}}\sum_{i=1}^{\widetilde{m}_2}V_{j(\widetilde{m}_1+i)}\epsilon^2_ik_{y^2_i}$ is localized in the region $I_{2 \epsilon}$ and its neighborhood.
    \item For $j \in \{p_2+1,\dots,p_3\}$, $h_j=\sqrt{\frac{2\epsilon R}{p_3-p_2}}\sum_{i=1}^{\widetilde{m}_3}V_{j(\widetilde{m}_1+\widetilde{m}_2+i)}\epsilon^3_ik_{y^3_i}$ is similarly localized in $I_{3 \epsilon}$ and its surrounding neighborhood.
\end{enumerate}

We use the property that when $\{h_j\}_{j=1}^{p_3}$ forms a frame, then this localization property is naturally transferred to the canonical dual frame, denoted by $\{\widetilde{h}_j\}_{j=1}^{p_3}$. Although we do not explicitly prove this, we use it as a foundation for our error bound calculations in the Section \ref{paper178}.

Specifically, we use the fact that $\widetilde{h}_j$ is localized in $I_{1 \epsilon}$ and its neighborhood for all $j \in \{1, \dots, p_1\}$. Similarly, $h_j$ is localized in $I_{2 \epsilon}$ for $j \in \{p_1+1, \dots, p_2\}$, and in $I_{3 \epsilon}$ for $j \in \{p_2+1, \dots, p_3\}$.

The advantage of this localization in both the frame and the dual frame is that it limits the influence of inaccurate samples used to outside our region of interest $[-R, R]$, during reconstruction. The idea is as follows: when approximating the samples of the projected function $P_Vf$ using samples of the original function $ f $, the inaccurately approximated samples are primarily located in the region $I_{3 \epsilon}$. This implies that the frame coefficients $\left\{\langle f, h_j \rangle\right\}_{j=p_2+1}^{p_3}$ are not accurately computed. 

However, considering the frame reconstruction formula, for all $f \in V^{R_1}(g)$:
\[
f = \sum_{j=1}^{p_3} \left\langle f, h_j \right\rangle \widetilde{h}_j
= \sum_{j=1}^{p_2} \left\langle f, h_j \right\rangle \widetilde{h}_j + \sum_{j=p_2+1}^{p_3} \left\langle f, h_j \right\rangle \widetilde{h}_j,
\]
we observe that the inaccurately approximated frame coefficients are multiplied by their respective canonical dual frame vectors, which are themselves localized in the neighborhood of $I_{3 \epsilon}$. Hence, the errors in the approximated samples only affect the region $I_{2 \epsilon} \cup I_{3 \epsilon}$. This region excludes our area of interest, $[-R, R]$, which is strictly contained within $I_{1 \epsilon}$. Therefore, the reconstruction should lead to only a small error in the $L^2[-R, R]$ norm. This is exactly what is made rigourous in Section \ref{paper178} at the end of which we state and prove our main results Theorem \ref{paper16001} and \ref{paper12001}. 

 	\section{A suitable finite dimensional approximation space for $PW_{[-\pi, \pi]}$} \label{paper176}

        Let $g$ be as defined in Subsection \ref{paper1500}, and $r\geq 4$ be a positive integer. As $g$ is a Schwartz class function,  there exists a positive constant $C_r>1$ such that $|g(t)|<\frac{C_r}{(1+|t|)^r}\hspace{0.2cm} \forall \hspace{0.1cm} t \in \mathbb{R}.$ Let $R, \epsilon>0$ and partition the real line as in Figure \ref{fig:Fig9}.
\begin{figure}
\begin{tikzpicture}
\draw[<->] (-7.5,0)--(7.5,0);
\draw[red,-, line width=2pt] (-3,0)--(3,0);
\draw[blue,-, line width=2pt] (-5,0)--(-3,0);
\draw[blue,-, line width=2pt] (3,0)--(5,0);
\draw[green,-, line width=2pt] (5,0)--(7,0);
\draw[green,-, line width=2pt] (-7,0)--(-5,0);
\draw (0,0.1)--(0,-0.1) node[below] {$0$};
\draw (2,0.1)--(2,-0.1) node[below] {$R$};
\draw (-2,0.1)--(-2,-0.1) node[below] {$-R$};
\draw (3,0.1)--(3,-0.1) node[below] {$(1+\epsilon)R$};
\draw (-3,0.1)--(-3,-0.1) node[below] {$-(1+\epsilon)R$};
\draw (5,0.1)--(5,-0.1) node [below] {$(1+2\epsilon)R$};
\draw (-5,0.1)--(-5,-0.1) node[below] {$-(1+2\epsilon)R$};
\draw (6,0.1) -- (6,-0.1);
\draw (6,0.1)  node[above] {$\left(1+\frac{5}{2} \epsilon\right)R$};
\draw (-6,0.1) -- (-6,-0.1);
\draw (-6,0.1) node[above]{$-\left(1+\frac{5}{2}\epsilon\right)R$};
\draw (7,0.1)--(7,-0.1) node[below] {$(1+3\epsilon)R$};
\draw (-7,0.1)--(-7,-0.1) node[below] {$-(1+3\epsilon)R$};
\draw[fill=blue] (-3,0) circle (0.07);
\draw[fill=blue] (3,0) circle (0.07);
\draw[fill=green] (-5,0) circle (0.07);
\draw[fill=green] (5,0) circle (0.07);
\draw[fill=green] (-7,0) circle (0.07);
\draw[fill=green] (7,0) circle (0.07);
\end{tikzpicture} 
\caption{Partitions of the real line}
    \label{fig:Fig9}
\end{figure}
Recall (see \eqref{definitionofV^{R_1}(g)}), \[
V^{\left(1+\frac{5}{2}\epsilon\right)R}(g) := \text{span} \left\{ g\left(\cdot - \frac{k}{2}\right) : k \in \left[ 2 \left(1+\frac{5}{2}\epsilon\right)R \right] \right\}.
\]
			Next, we state three lemmas which will be repeatedly used by us throughout the rest of the paper. Their proofs consist mostly of straightforward computations. Thus, we highlight only the most important steps.
	\begin{lemma}\label{paper122}
		Let $k \in \mathbb{Z}$ and $R>0$ be a real number. Then
		\[\bigg\|g\left(\cdot- \frac{k}{2}\right)\bigg\|_{L^2[-R,R]} \leq
		\begin{cases}
				  \frac{C_r}{\sqrt{2r-1}\left(1-R+\frac{|k|}{2}\right)^{r- \frac{1}{2}}} & |k|>2 R\\
				  \frac{\sqrt{2}C_r}{\sqrt{2r-1}} & |k|\leq 2 R .
		\end{cases}
		\]
	\end{lemma}
        \begin{proof}
            Follows from elementary computations.
        \end{proof}
	\begin{lemma}\label{paper123}
		Let $x \in \mathbb{R}$. Then 
		\begin{equation}\label{paper110004}
			\sum_{k \in \mathbb{Z}}\frac{1}{\left(1+\big|x-\frac{k}{2 }\big|\right)^r} \leq 1+\frac{2}{r-1}\left(1+2^{r-1}\right) =:\Gamma_r.
		\end{equation}	
	\end{lemma}
    \begin{proof}
        The proof is moved to the appendix.
    \end{proof}
	\begin{lemma}\label{paper161}
		Given  $\gamma \in (0,1)$, if $\epsilon R\geq 2\left(\frac{2C_r}{\sqrt{(2r-1)(r-1)}\gamma}\right)^\frac{1}{r-1}$, then
		\begin{equation}
			\int_{-(1+3\epsilon)R}^{(1+3\epsilon)R}|f(x)|^2dx \geq (1-\gamma)\int_{\mathbb{R}}|f(x)|^2dx \hspace{0.2cm} \forall \hspace{0.1cm} f \in V^{\left(1+\frac{5}{2}\epsilon\right)R}(g).
		\end{equation} 
	\end{lemma}
     
    \begin{proof}
        The proof is moved to the appendix.
    \end{proof}

	\begin{lemma} \label{paper126}
		Let $f \in C_{[-\pi,\pi]}$ and $\mathcal{R}<\left(1+\frac{5}{2}\epsilon\right)R$ be a positive real number. Then the following bounds hold.
		\begin{align}
			&\label{paper17} \|f-Pf\|_{L^2[-\mathcal{R},\mathcal{R}]}\leq \frac{2\sqrt{2}C_r}{\sqrt{(2r-1)}(r- \frac{3}{2})\left(\left(1+\frac{5}{2}\epsilon\right)R-\mathcal{R}\right)^{r- \frac{3}{2}}}. \\
			&\label{paper18} \|f-Pf\|_{L^\infty[-\mathcal{R},\mathcal{R}]}\leq \frac{2\sqrt{2}C_r}{(r-1)\left(\left(1+\frac{5}{2}\epsilon\right)R-\mathcal{R}\right)^{r-1}}.\\
			&\label{paper153}  \|f-Pf\|_{L^\infty(\mathbb{R})}\leq \max\bigg\{\frac{\|g'\|_{L^1(\mathbb{R})}}{\sqrt{2}}+{\sqrt{2}}\|g\|_{L^1(\mathbb{R})},1\bigg\}:=C_{g}.
		\end{align}
	\end{lemma}
    
    \begin{proof}
        The proof is moved to the appendix.
    \end{proof}
	         In the following lemma, we prove certain upper bounds on the reproducing kernel of the space $V^{\left(1+\frac{5}{2}\epsilon\right)R}(g)$.  The central principle behind establishing these bounds is that for any point $x$, the reproducing kernel  $k_x \in V^{\left(1+\frac{5}{2}\epsilon\right)R}(g)$ is well localized around $x$.
	\begin{lemma}\label{paper125}
		The space $V^{\left(1+\frac{5}{2}\epsilon\right)R}(g)$ is a reproducing kernel Hilbert space with the reproducing kernel $K$ satisfying
		\begin{enumerate}
        \item 	\begin{equation*}
			K(x,y)=\sum_{ k \in \left[2 \left(1+\frac{5}{2}\epsilon\right)R\right]}\overline{g\left(y-\frac{k}{2}\right)}g\left(x-\frac{k}{2}\right) \hspace{0.2cm} \forall \hspace{0.1cm} x,y \in \mathbb{R}.
		\end{equation*}
			\item  \begin{equation}\label{paper169}
         |K(x,y)|\leq \frac{2C_r^2}{(1+|\frac{x-y}{2}|)^r}\Gamma_{r-\frac{1}{2}} \hspace{0.2cm} \forall \hspace{0.1cm} x,y \in \mathbb{R}.
   \end{equation}
   \item $\sup_{x \in \mathbb{R}}\sqrt{K(x,x)}<C:=\sqrt{3}$.
		\end{enumerate}
		 Moreover, for any real number $R_2>R$, we have 
		\begin{equation*} 
			\|k_x\|_{L^2[-R,R]} \leq  \frac{6C^2_r\Gamma_{r-\frac{1}{2}}}{\sqrt{2r-1}(\frac{R_2-R}{2})^{r-\frac{1}{2}}} \hspace{0.2cm} \forall \hspace{0.1cm} |x|\geq R_2,
		\end{equation*}
		where $k_x$ is the reproducing kernel at $x$ and $\Gamma_r$ is as defined in \eqref{paper110004}.
	\end{lemma}
	\begin{proof}
		As $\left\{g\left(\cdot- \frac{k}{2}\right): k \in \mathbb{Z}\right\}$ is an orthonormal system, by the definition of $V^{\left(1+\frac{5}{2}\epsilon\right)R}(g)$, the spanning set $\left\{g\left(\cdot-\frac{k}{2}\right): k \in \left[2 \left(1+\frac{5}{2}\epsilon\right)R\right]\right\}$ forms an orthonormal basis for $V^{\left(1+\frac{5}{2}\epsilon\right)R}(g)$. 
		This implies that for all $ f \in V^{\left(1+\frac{5}{2}\epsilon\right)R}(g)$ and $ x \in \mathbb{R}$, we have $ f(x)=\left\langle f, k_x \right\rangle , \text{ where }$
        \begin{equation} \label{paper13003}
            k_x(\cdot)=\sum_{ k \in \left[2 \left(1+\frac{5}{2}\epsilon\right)R\right]}\overline{g\left(x-\frac{k}{2}\right)}g\left(\cdot-\frac{k}{2}\right) .
        \end{equation}
		So, the space $V^{\left(1+\frac{5}{2}\epsilon\right)R}(g)$ is a reproducing kernel Hilbert space with reproducing kernel
		\begin{equation*}
			K(x,y)=k_y(x)=\sum_{ k \in \left[2 \left(1+\frac{5}{2}\epsilon\right)R\right]}\overline{g\left(y-\frac{k}{2}\right)}g\left(x-\frac{k}{2}\right).
		\end{equation*}
	Let $x,y \in \mathbb{R}$. Then
	\begin{align*}
		|K(x,y)|&\leq\sum_{k \in \left[2 \left(1+\frac{5}{2}\epsilon\right)R\right]}\left|g\left(y-\frac{k}{2}\right)\right|\left|g\left(x-\frac{k}{2}\right)\right|\\
		&\leq \sum_{k \in \left[2 \left(1+\frac{5}{2}\epsilon\right)R\right]}\frac{C_r^2}{\left(1+\left|y-\frac{k}{2}\right|\right)^r\left(1+\left|x-\frac{k}{2}\right|\right)^r}\\
		& \leq \frac{C^2_r}{(1+\left|\frac{x-y}{2}\right|)^r}\left(\sum_{k \in \mathbb{Z}}\frac{1}{\left(1+\left|y-\frac{k}{2}\right|\right)^r}+\sum_{k \in \mathbb{Z}}\frac{1}{\left(1+\left|x-\frac{k}{2}\right|\right)^r}\right)\\
		&\leq \frac{C_r^2}{(1+|\frac{x-y}{2}|)^r}\left(2\Gamma_{r-\frac{1}{2}}\right).
	\end{align*}
        The second last, and last inequality were obtained by using Lemma  \ref{paper116000} and \ref{paper123}, respectively.
	As $V^{\left(1+\frac{5}{2}\epsilon\right)R}(g) \subset PW_{[-3\pi, 3\pi]}$, $K(x,x)\leq \tilde{K}(x,x)$ for all $x \in \mathbb{R}$, where $\tilde{K}$ is the reproducing kernel for $PW_{[-3\pi, 3\pi]}$. That is, $\sup_{x \in \mathbb{R}} \sqrt{K(x,x)}\leq \sqrt{3}$.\\
	Let $x \in \mathbb{R}$ be such that $|x|\geq R_2.$ Then
	\begin{align*}
		\|k_x\|_{L^2[-R,R]}&=\left\|\sum_{ k \in \left[2 \left(1+\frac{5}{2}\epsilon\right)R\right]}\overline{g\left(x-\frac{k}{2}\right)}g\left(\cdot-\frac{k}{2}\right)\right\|_{L^2[-R,R]}\\
        \numberthis \label{paper12002} &\leq \sum_{ k \in [2 R]}\frac{C_r}{\left(1+\left|x-\frac{k}{2 }\right|\right)^r} \left\|g\left(\cdot-\frac{k}{2}\right)\right\|_{L^2[-R,R]}\\
        &+\sum_{ k \in \left[2 \left(1+\frac{5}{2}\epsilon\right)R\right],  k > \left\lfloor 2 R \right\rfloor}\frac{C_r}{\left(1+\left|x-\frac{k}{2 }\right|\right)^r} \left\|g\left(\cdot-\frac{k}{2}\right)\right\|_{L^2[-R,R]}\\
        &+\sum_{ k \in \left[2 \left(1+\frac{5}{2}\epsilon\right)R\right],  k < \left\lfloor 2 R \right\rfloor}\frac{C_r}{\left(1+\left|x-\frac{k}{2 }\right|\right)^r} \left\|g\left(\cdot-\frac{k}{2}\right)\right\|_{L^2[-R,R]} \\
		& \leq \frac{6C^2_r}{\sqrt{2r-1}\left(\frac{R_2-R}{2}\right)^{r-\frac{1}{2}}}\Gamma_{r-\frac{1}{2}}.
		\end{align*}
        The first term in \eqref{paper12002} is bounded using the fact that $\left|x-\frac{k}{2 }\right|\geq\left|\left|x|-|\frac{k}{2 }\right|\right| \geq R_2-R$ for all $k \in [2 R]$. The second and third terms in \eqref{paper12002} are bounded by using  Lemma \ref{paper122} to bound $\left\|g\left(\cdot-\frac{k}{2}\right)\right\|_{L^2[-R,R]}$.
	\end{proof}
 \begin{remark}\label{paper1localizationofkernel}
     From the previous lemma, we conclude that for all $y \in \mathbb{R}$,
\[\left|k_y(x)\right| \leq \frac{2C_r^2}{(1+|\frac{x-y}{2}|)^r}\Gamma_{r-\frac{1}{2}}.\]
This essentially means that kernel $k_y$ is concentrated at $y$ and decays very fast as we move further away from $y$. This property is critical to our method.
 \end{remark}
	\section{Finding a random frame for the approximation space} \label{paper177}
In this section, our aim is to find an appropriate frame consisting of a linear combination of reproducing kernels for $V^{\left(1+\frac{5}{2}\epsilon\right)R}(g)$. The frame inequality that we prove is motivated by the classical sampling inequality with the addition of the condensation operator and the random flips.
	\begin{definition}
			Assume that $\{x_i\}_{i=1}^m$ is a sequence of i.i.d. random variables uniformly distributed on $[-(1+3 \epsilon)R, (1+3 \epsilon)R]$. Let $\{p_i\}_{i=1}^3, \{\widetilde{m}_i\}_{i=1}^3,\{\epsilon^1_i\}_{i=1}^{m},\{\epsilon^2_i\}_{i=1}^{m}, \{\epsilon^3_i\}_{i=1}^{m},\{y^1_i\}_{i=1}^{\widetilde{m}_1}, \{y^2_i\}_{i=1}^{\widetilde{m}_2} $ and $\{y^3_i\}_{i=1}^{\widetilde{m}_3}$ be as defined in Definition \ref{paper1definitionofallrandomvariables}. Define the sampling operator $E$ from $V^{\left(1+\frac{5}{2}\epsilon\right)R}(g)$ to $\mathbb{C}^{\widetilde{m}}$ as
		\begin{align*}
			E(f)&=\{f(y^1_i)\}_{i=1}^{\widetilde{m}_1}\frown\{f(y^2_i)\}_{i=1}^{\widetilde{m}_2}\frown\{f(y^3_i)\}_{i=1}^{\widetilde{m}_3}\\
		\numberthis \label{paper1concatenatedvector}	&=\frown_{j=1}^3\{f(y^j_i)\}_{i=1}^{\widetilde{m}_j}.
		\end{align*}
	\end{definition}
	Let $V$ be either the $\Sigma\Delta$ \eqref{paper136} or the $\beta$ \eqref{paper138} condensation operator  with $p_3$ rows and $\widetilde{m}$ columns. Define the matrices $W_{p_3\times p_3}$ and $\Phi_{\widetilde{m} \times \widetilde{m}}$ as follows.
	\begin{equation}\label{paper13005}
	    (W)_{ij}=
	\begin{cases}
		\sqrt{\frac{2(1+\epsilon)R}{p_1}} & \hspace{-0.1cm}i=j, i \in \{1, \ldots, p_1\},\\
		\sqrt{\frac{2\epsilon R}{p_2-p_1}} & \hspace{-0.1cm}i=j, i \in \{p_1+1, \ldots, p_2\},\\
		\sqrt{\frac{2\epsilon R}{p_3-p_2}} & \hspace{-0.1cm}i=j, i \in \{p_2+1, \ldots, p_3\},\\
		0 & i \neq j.
	\end{cases}
	\end{equation}
 \[(\Phi)_{ij}=
\begin{cases}                
		\epsilon^1_i & \hspace{-0.1cm}i=j, i \in \{1, \ldots, \widetilde{m}_1\},\\
		\epsilon^2_i & \hspace{-0.1cm}i=j, i \in \{\widetilde{m}_1+1, \ldots,\widetilde{m}_1+\widetilde{m}_2\},\\
		\epsilon^3_i & \hspace{-0.1cm}i=j, i \in \{\widetilde{m}_1+\widetilde{m}_2+1, \ldots, \widetilde{m}\},\\
		0 & i \neq j.
	\end{cases}\]
 
 Next, we finding concentration inequalities for $p_1,p_2$ and $p_3$. For all  $i \in \{1, \ldots,m\}$, define the following random variables
	\[X_i:=\begin{cases}
		1 & \text{ if } x_i \in I_{1\epsilon},\\
		0 & \text{ if } x_i \in [-(1+3\epsilon)R,(1+3\epsilon)R]\setminus I_{1\epsilon}.
	\end{cases}\] 
	That is, 
	\[X_i=\begin{cases}
		1 & \text{ with probability } \frac{1+\epsilon}{1+3\epsilon}, \\
		0 & \text{ with probability } \frac{2\epsilon}{1+3\epsilon}.
	\end{cases}\] 
	Then, $m_1=\sum_{i=1}^mX_i$. As $\mathbb{E}m_1=\frac{m(1+\epsilon)}{(1+3\epsilon)}$, using the Chernoff's bound for small deviations (Theorem \ref{paper13001}) with $\delta=1/2$, we get 
	\begin{equation*}
		\mathbb{P}\left(\bigg|m_1-\frac{m(1+\epsilon)}{1+3\epsilon}\bigg|\geq \frac{m(1+\epsilon)}{2(1+3\epsilon)}\right)\leq 2 \exp\left(-\frac{m(1+\epsilon)}{12(1+3\epsilon)}\right).
	\end{equation*}
    That is, $\frac{m(1+\epsilon)}{2(1+3\epsilon)}\leq m_1\leq \frac{3m(1+\epsilon)}{2(1+3\epsilon)}$
    with probability greater than $1-2 \exp\left(-\frac{m(1+\epsilon)}{12(1+3\epsilon)}\right)$.
	Thus the event,
	\begin{equation} \label{paper14000}
	    B_1=\left\{\frac{p}{2(1+3\epsilon)R}-\frac{1}{(1+\epsilon)R}\leq\frac{p_1}{(1+\epsilon)R}\leq \frac{3p}{2(1+3\epsilon)R}\right\}
	\end{equation}
	has probability greater than $1-2\exp\left(-\frac{m(1+\epsilon)}{12(1+3\epsilon)}\right)$. Similarly, it can be shown that
 \begin{equation}\label{paper14001}
     B_i=\left\{\frac{p}{2(1+3\epsilon)R}-\frac{1}{\epsilon R}\leq \frac{p_i-p_{i-1}}{\epsilon R}\leq \frac{3p}{2(1+3\epsilon)R}\right\}
 \end{equation}
	has probability greater than $1-2\exp\left(-\frac{m\epsilon}{12(1+3\epsilon)}\right)$ for $i \in \{2,3\}$.
 From \eqref{paper13005}, it is clear that $\|W\|_{2 \rightarrow 2}\leq \max\left\{\sqrt{\frac{2(1+\epsilon)R}{p_1}},\sqrt{\frac{2\epsilon R}{p_2-p_1}}, \sqrt{\frac{2\epsilon R}{p_3-p_2}} \right\},$ thus the event $\cap_{i=1}^3B_i$ (see \eqref{paper14000},\eqref{paper14001}) is a subset of the event \[\|W\|_{2 \rightarrow 2}\leq \max\left\{\sqrt{\frac{2}{\frac{p}{2(1+\epsilon)R}-\frac{1}{(1+\epsilon)R}}},\sqrt{\frac{2}{\frac{p}{2(1+\epsilon)R}-\frac{1}{\epsilon R}}},\sqrt{\frac{2}{\frac{p}{2(1+\epsilon)R}-\frac{1}{\epsilon R}}}\right\},\]
		which in turn is a subset of the event 
  \begin{equation}\label{paper11001}
      \|W\|_{2 \rightarrow 2}\leq \sqrt{\frac{2}{\frac{p}{2(1+\epsilon)R}-\frac{1}{\epsilon R}}}.
  \end{equation}

Note that $WV \Phi E$ is an operator from $V^{\left(1+\frac{5}{2}\epsilon\right)R}(g)$ to $\mathbb{C}^{p_3}$, and for any $f \hspace{0.05 cm}\in V^{\left(1+\frac{5}{2}\epsilon\right)}(g)$ 
\begin{align}
   \label{paper1WVPhi=h} WV \Phi E(f)&= \{\left\langle f,h_j \right\rangle\}_{j=1}^{p_3},
\end{align}
 where
 \begin{equation} \label{paper13004}
     h_j:=\sum_{i=1}^{\widetilde{m}_1}(WV)_{ji}\epsilon^1_ik_{y^1_i}+\sum_{i=1}^{\widetilde{m}_2}(WV)_{j(\widetilde{m}_1+i)}\epsilon^2_ik_{y^2_i}+\sum_{i=1}^{\widetilde{m}_3}(WV)_{j(\widetilde{m}_1+\widetilde{m}_2+i)}\epsilon^3_ik_{y^3_i}
 \end{equation}
	 for all $j \in \{1, \ldots p_3\}.$ For any $x\in \mathbb{R}$, here $k_x$ refers to the reproducing kernel \eqref{paper13003} at $x$. Hence, $WV \Phi E$ is nothing but the analysis operator of the collection $\{h_j\}_{j=1}^{p_3}$. 

\smallskip
  Using the block diagonal structure of $V$, it can further be concluded that 
	\begin{enumerate}
		\item $h_j=\sqrt{\frac{2(1+\epsilon)R}{p_1}}\sum_{i=1}^{\widetilde{m}_1}V_{ji}\epsilon^1_ik_{y^1_i}$ for all $j \in \{1, \ldots ,p_1\}$,
		\item $h_j=\sqrt{\frac{2\epsilon R}{p_2-p_1}}\sum_{i=1}^{\widetilde{m}_2}V_{j(\widetilde{m}_1+i)}\epsilon^2_ik_{y^2_i}$ for all $j \in \{p_1+1, \ldots ,p_2\}$,
		\item $h_j=	\sqrt{\frac{2\epsilon R}{p_3-p_2}}\sum_{i=1}^{\widetilde{m}_3}V_{j(\widetilde{m}_1+\widetilde{m}_2+i)}\epsilon^3_ik_{y^3_i}$ for all $j \in \{p_2+1, \ldots ,p_3\}$.
	\end{enumerate}
  
        In Lemma \ref{paper172}, we show that $\{h_j\}_{j=1}^{p_3}$ forms a frame for  $V^{\left(1+\frac{5}{2}\epsilon\right)R}(g)$ with high probability when $m$ and $p$ are large. The lemma will be the consequence of  Lemma \ref{paper155}, which we prove now.
        \subsection{Preparation for the frame inequality}
        Our approach adapts the methods provided in  \cite{paper123} to the quantization setting.
        \begin{definition}
        Let $I$ be a subset of $\mathbb{R}$ whose Lebesgue measure $|I|$ satisfies $1\leq|I|<\infty$. We define the following two operators on $L^2(\mathbb{R})$,
		\[ T_If:=\mathbbm{1}_{I}f \text{ and } A_I:=PT_IP.\]
	\end{definition}
	Let $A^{'}_I$ be the restriction of the operator $A_I$ to $V^{\left(1+\frac{5}{2}\epsilon\right)R}(g)$. Then $A^{'}_I$ is a compact, positive semidefinite operator on $V^{\left(1+\frac{5}{2}\epsilon\right)R}(g)$ with norm less than or equal to one. Applying the spectral theorem for compact self-adjoint operators to $A^{'}_I$, we can find an orthonormal basis for $V^{\left(1+\frac{5}{2}\epsilon\right)R}(g)$ consisting of eigenvectors $\{\phi_k\}_{k \in \left[ 2 \left(1+\frac{5}{2}\epsilon\right)R\right]}$ of $A^{'}_I$ with corresponding eigenvalues $\{\lambda_k\}_{k \in \left[ 2 \left(1+\frac{5}{2}\epsilon\right)R\right]}$ satisfying $1\geq\lambda_k\geq0$ for all $k \in \left[ 2 \left(1+\frac{5}{2}\epsilon\right)R\right]$.
  
	Let $m_* \in \mathbb{N}$. Assume that $\{z_i\}_{i=1}^{m_*}$ is a sequence of i.i.d random variables that are uniformly distributed over $I$ and that $\{\epsilon_i\}_{i=1}^{m_*}$ is a sequence of $\pm 1$ Bernoulli independent random variables that are also independent from $\{z_i\}_{i=1}^{m_*}$. Let ${p_*}$ be a positive integer that divides $m_*$, and let the ${p_*} \times m_*$ matrix $V_*$ be either the $\Sigma\Delta$ or the $\beta$ condensation operator. For a complex sequence $u \in \mathbb{C}^{m_*}$, denote 
	$u \otimes u := u\cdot u^*=u\cdot \overline{u}^{\top}$.
 
	Define,
        \begin{enumerate}
            \item The vectors $\{u^j\}_{j=1}^{p_*}$ as \begin{equation}\label{paper1eq:12}
                u^j := \left(\overline{\sum_{i=1}^{m_*}{\left(V_*\right)}_{ji}\epsilon_i\phi_k(z_i)}\right)_{k \in \left[ 2 \left(1+\frac{5}{2}\epsilon\right)R\right]}  \hspace{0.2cm} \forall  \hspace{0.1cm} j \in \{1, ...,p_*\}.
            \end{equation}
        \item The matrix $\Lambda$ as
        \begin{equation} \label{paper1eq:13}
            \hspace{0.8cm} \Lambda :=  \mathbb{E}\left(u^j \otimes u^j\right).
        \end{equation}	
        \end{enumerate}
	\begin{lemma}\label{paper1boundonuandlambda}
		Let $u^j$ and $ \Lambda $ be as defined in \eqref{paper1eq:12}, \eqref{paper1eq:13} respectively. Then the following hold.
		\begin{equation} \label{paper127}
			\|u^j\|_2 \leq C \hspace{0.2cm} \forall  \hspace{0.1cm} j \in \{1, ...,p_*\} .
		\end{equation}
		\begin{equation} \label{paper128}
			\Lambda_{kl}=\frac{1}{|I|}\lambda_k\delta_{kl} \hspace{0.2cm} \forall \hspace{0.1cm} k,l \in \left[ 2 \left(1+\frac{5}{2}\epsilon\right)R\right].
		\end{equation}
             Here $C$ is as defined in Lemma \ref{paper125}, and $\delta_{kl}:=1$ if $k=l$ and $\delta_{kl}:=0$ for $k \neq l$.
	\end{lemma}
	\begin{proof}
		First note that $u^j= \sum_{i=1}^{m_*}u^j_i$ where 
		\[u^j_i:=\left(\overline{{\left(V_*\right)}_{ji}\epsilon_i\phi_k(z_i)}\right)_{k \in \left[ 2 \left(1+\frac{5}{2}\epsilon\right)R\right]}\hspace{0.2cm} \forall  \hspace{0.1cm} i \in \{1, ...,m_*\}.\]	
		Using the fact that $k_{z_i} \in {V}^{\left(1+\frac{5}{2}\epsilon\right)R}(g) \hspace{0.2cm} \forall  \hspace{0.1cm} i \in \{1, ...,m_* \}$ and that $\{\phi_k\}_{k \in \left[ 2 \left(1+\frac{5}{2}\epsilon\right)R\right]}$ is an orthonormal basis for ${V}^{\left(1+\frac{5}{2}\epsilon\right)R}(g)$, we obtain
		\begin{align*}
			\|u^j\|_2&\leq \sum_{i=1}^{m_*}\|u^j_i\|_2= \sum_{i=1}^{m_*}\left(\sum_{k \in \left[ 2 \left(1+\frac{5}{2}\epsilon\right)R\right]}\left|{\left(V_*\right)}_{ji}\epsilon_i\phi_k(z_i)\right|^2\right)^{\frac{1}{2}}\\
			&=\sum_{i=1}^{m_*}\left|{\left(V_*\right)}_{ji}\right|\left(\sum_{k \in \left[ 2 \left(1+\frac{5}{2}\epsilon\right)R\right]}\left|\phi_k(z_i)\right|^2\right)^{\frac{1}{2}}
			=\sum_{i=1}^{m_*}\left|{\left(V_*\right)}_{ji}\right|\left(\sum_{k \in \left[ 2 \left(1+\frac{5}{2}\epsilon\right)R\right]}\left|\left\langle \phi_k,k_{z_i}\right\rangle \right|^2\right)^{\frac{1}{2}}\\
			&=\sum_{i=1}^{m_*}\left|{\left(V_*\right)}_{ji}\right|\left\|k_{z_i}\right\| 
			\leq\sup_{z \in \mathbb{R}}\sqrt{K(z,z)}\sum_{i=1}^{m_*}\left|{\left(V_*\right)}_{ji}\right|\\
			&= \sup_{z \in \mathbb{R}}\sqrt{K(z,z)}< C.
		\end{align*}
		To get the last statement, we used the fact that rows in $V_*$ are $\ell_1$ normalized. Let $j \in \{1,2,...,p_*\}$ be arbitrary. Then  \[\left(u^j \otimes u^j\right)_{kl}=\overline{\left(\sum_{i=1}^{m_*}{\left(V_*\right)}_{ji}\epsilon_i\phi_k(z_i)\right)}\left(\sum_{r=1}^{m_*}{\left(V_*\right)}_{jr}\epsilon_r\phi_l(z_r)\right).\] Therefore, for all ${k,l \in \left[ 2 \left(1+\frac{5}{2}\epsilon\right)R\right]}$, we have
		\begin{align*}
			\mathbb{E}\left(u^j \otimes u^j\right)_{kl}&= \mathbb{E}_ z\mathbb{E}_\epsilon\left(\sum_{i=1,r=1}^{m_*}{\left(V_*\right)}_{ji}{\left(V_*\right)}_{jr}\epsilon_i\overline{\phi_k(z_i)}\epsilon_r\phi_l(z_r)\right)\\&=\mathbb{E}_z\sum_{i=1,r=1}^{m_*}{\left(V_*\right)}_{ji}{\left(V_*\right)}_{jr}\overline{\phi_k(z_i)}\phi_l(z_r)\mathbb{E}_\epsilon\epsilon_i\epsilon_r\\
            &=\mathbb{E}_z\sum_{i=1,r=1,i\neq r}^{m_*}{\left(V_*\right)}_{ji}{\left(V_*\right)}_{jr}\overline{\phi_k(z_i)}\phi_l(z_r)\mathbb{E}\epsilon_i\mathbb{E}\epsilon_r+\mathbb{E}_z\sum_{i=1}^{m_*}{\left(V_*\right)}_{ji}^2\overline{\phi_k(z_i)}\phi_l(z_i)\mathbb{E}\epsilon^2_i\\&=\mathbb{E}_z\sum_{i=1}^{m_*}{\left(V_*\right)}_{ji}^2\overline{\phi_k(z_i)}\phi_l(z_i)=\frac{\|\nu\|_2^2}{\|\nu\|_1^2|I|}\int_I\overline{\phi_k(z)}\phi_l(z)dz\\
		&=\frac{\|\nu\|_2^2}{\|\nu\|_1^2|I|}\overline{\left\langle A_I\phi_k,\phi_l \right\rangle}=\frac{\|\nu\|_2^2}{\|\nu\|_1^2|I|}\lambda_k\delta_{kl}.
			\end{align*}
	\end{proof}
	\begin{lemma} \label{paper155}
	Let $m_*,p_*,{V_*},\{z_i\}_{i=1}^{m_*}, I$  be as defined above and $t \in (0,1)$. Then
		\begin{align}
		\label{paper119} \frac{\|\nu\|^2_2}{\|\nu\|_1^2}(\|f\|^2_{2,I}-t\|f\|^2 ) &\leq \frac{|I|}{p_*}\sum_{j=1}^{p_*}\bigg|\sum_{i=1}^{m_*}{\left(V_*\right)}_{ji}\epsilon_if(z_i)\bigg|^2 \\
		\nonumber&\leq	\frac{\|\nu\|^2_2}{\|\nu\|_1^2}(\|f\|^2_{2,I}+t\|f\|^2 )  \hspace{0.4 cm}\forall  f \hspace{0.05 cm}\in {V}^{\left(1+\frac{5}{2}\epsilon\right)R}(g) 
		\end{align}
		$\text{with probability greater than } 1- 2^{\frac{3}{4}}p_*\exp\left(-\frac{t^2p_*}{21\tilde{C}^2|I|}\right),$ where $\tilde{C}$:=$\frac{\|\nu\|_1C}{\|\nu\|_2}$.
	\end{lemma}
	\begin{proof}
	\noindent Let $u^j, \Lambda $ be as defined in \eqref{paper1eq:12},\eqref{paper1eq:13} respectively. Then, using  \cite[Theorem 1.1]{paper14}, we obtain
	\begin{align}
		\label{paper1prop:1} \mathbb{P} \left(\bigg\|\frac{1}{p_*}\sum_{j=1}^{p_*}u^j\otimes u^j-\Lambda\bigg\|_{2\rightarrow 2}\leq \frac{\|\nu\|_2^2t}{\|\nu\|_1^2|I|} \right) \geq 1- 2^{\frac{3}{4}}p_*\exp\left(-\frac{t^2p_*\|\nu\|^2_2}{21\|\nu\|^2_1C^2|I|}\right).
	\end{align}
        Although $\{u^j\}_{j=1}^{p_*}$ is not an i.i.d sequence  as required by the conditions of  \cite[Theorem 1.1]{paper14},  the proof of  \cite[Theorem 1.1]{paper14} is still valid. Also, according to their theorem statement,  $\|\Lambda\|_{2 \rightarrow 2} \leq 1$ is sufficient. However, we have $\|\Lambda\|_{2 \rightarrow 2} \leq \frac{\|\nu\|_2^2}{\|\nu\|_1^2|I|}\leq 1$, which is stronger, hence, by making the necessary modifications to their proof, we obtain the above bound.
        
	Let $f=\sum_{k \in \left[ 2 \left(1+\frac{5}{2}\epsilon\right)R\right]}c_k\phi_k \in {V}^{\left(1+\frac{5}{2}\epsilon\right)R}(g)$ for some $c=\{ c_k \}_{k \in \left[ 2 \left(1+\frac{5}{2}\epsilon\right)R\right]}$. Then
	\begin{align}
 		\nonumber \left\langle\left(\frac{1}{p_*}\sum_{j=1}^{p_*}u^j\otimes u^j\right)c,c \right\rangle
		&=\frac{1}{p_*}\sum_{j=1}^{p_*}\left\langle u^j\otimes u^jc,c \right\rangle =\frac{1}{p_*}\sum_{j=1}^{p_*}\sum_{k \in \left[ 2 \left(1+\frac{5}{2}\epsilon\right)R\right]}(u^j\otimes u^j c)_k \overline{c}_{k}\\
		\nonumber &=\frac{1}{p_*}\sum_{j=1}^{p_*}\sum_{k \in \left[ 2 \left(1+\frac{5}{2}\epsilon\right)R\right]}\left(\sum_{l \in \left[ 2 \left(1+\frac{5}{2}\epsilon\right)R\right]}\left(u^j \otimes u^j\right)_{kl}c_l\right)\overline{c}_k\\
            \begin{split}
            \nonumber &=\frac{1}{p_*}\sum_{j=1}^{p_*}\sum_{k \in \left[ 2 \left(1+\frac{5}{2}\epsilon\right)R\right]}\left(\sum_{l \in \left[ 2 \left(1+\frac{5}{2}\epsilon\right)R\right]}\left(\overline{\sum_{i=1}^{m_*}{\left(V_*\right)}_{ji}\epsilon_i\phi_k(z_i})\right)\right.\\
 &  \left. \hspace{0.2cm}\cdot \left(\sum_{r=1}^{m_*}{\left(V_*\right)}_{jr}\epsilon_r\phi_l(z_r)\right)c_l\right)\overline{c}_k
\end{split}\\
\nonumber &= \frac{1}{p_*}\sum_{j=1}^{p_*}\left|\sum_{i=1}^{m_*}{\left(V_*\right)}_{ji}\epsilon_i\sum_{k \in \left[ 2 \left(1+\frac{5}{2}\epsilon\right)R\right]}c_k\phi_k(z_i)\right|^2\\
		\label{paper131}&=\frac{1}{p_*}\sum_{j=1}^{p_*}\left|\sum_{i=1}^{m_*}{\left(V_*\right)}_{ji}\epsilon_if(z_i)\right|^2.
	\end{align}
	Further,
	\begin{align}
	\nonumber \left\langle\Lambda c,c\right\rangle &=\sum_{k \in \left[ 2 \left(1+\frac{5}{2}\epsilon\right)R\right]}(\Lambda c)_k\overline{c}_k =\frac{1}{|I|}\sum_{k \in \left[ 2 \left(1+\frac{5}{2}\epsilon\right)R\right]}
	\lambda_k c_k\overline{c}_k=\frac{\left\langle A_If,f\right\rangle}{|I|}\\ &=\frac{\|\nu\|_2^2\|f\|^2_{2,I}}{\|\nu\|_1^2|I|}. \label{paper1eq:17}
	\end{align}	
	Using \eqref{paper131} and \eqref{paper1eq:17}, we get
	\begin{align}
		\nonumber \left\|\frac{1}{p_*}\sum_{i=1}^{p_*}u^j\otimes u^j-\Lambda\right\|_{2\rightarrow 2}&= \sup_{\|c\|_2=1}\left|\left\langle \left(\frac{1}{p_*}\sum_{i=1}^{p_*}u^j\otimes u^j\right)c,c \right\rangle- \left\langle \Lambda c,c \right\rangle\right|\\
		&=\sup_{\|f\| = 1, f \in  {V}^{\left(1+\frac{5}{2}\epsilon\right)R}(g)}\left|\frac{1}{p_*}\sum_{j=1}^{p_*}\left|\sum_{i=1}^{m_*}{\left(V_*\right)}_{ji}\epsilon_if(z_i)\right|^2- \frac{\|\nu\|_2^2\|f\|^2_{2,I}}{\|\nu\|_1^2|I|}\right|. \label{paper129}
	\end{align}
	Using \eqref{paper1prop:1} and  \eqref{paper129}, we get
	\[\mathbb{P}\left(\sup_{\|f\| = 1, f \in  {V}^{\left(1+\frac{5}{2}\epsilon\right)R}(g)}\left|\frac{1}{p_*}\sum_{j=1}^{p_*}\left|\sum_{i=1}^{m_*}{\left(V_*\right)}_{ji}\epsilon_if(z_i)\right|^2- \frac{\|\nu\|_2^2\|f\|^2_{2,I}}{\|\nu\|_1^2|I|}\right|\leq \frac{\|\nu\|_2^2t}{\|\nu\|_1^2|I|}\right)\]\[\geq 1- 2^{\frac{3}{4}}p_*\exp\left(-\frac{t^2p_*\|\nu\|^2_2}{21\|\nu\|^2_1C^2|I|}\right).\]
    That is,
\begin{align*}
	\frac{\|\nu\|^2_2}{\|\nu\|_1^2}\left(\frac{p_*}{|I|}\|f\|^2_{2,I}-\frac{tp_*}{|I|}\|f\|^2 \right) &\leq\sum_{j=1}^{p_*}\left|\sum_{i=1}^{m_*}{\left(V_*\right)}_{ji}\epsilon_if(z_i)\right|^2 \\&\leq	\frac{\|\nu\|^2_2}{\|\nu\|_1^2}\left(\frac{p_*}{|I|}\|f\|^2_{2,I}+\frac{tp_*}{|I|}\|f\|^2 \right)  \hspace{0.4 cm}\forall  f \hspace{0.05 cm}\in {V}^{\left(1+\frac{5}{2}\epsilon\right)R}(g),\|f\| = 1
\end{align*}
$\text{with probability greater than } 1- 2^{\frac{3}{4}}p_*\exp\left(-\frac{t^2p_*}{21\tilde{C}^2|I|}\right).$ Finally, using the fact that for any $f\in {V}^{\left(1+\frac{5}{2}\epsilon\right)R}(g)$, $\frac{f}{\|f\|}$ has norm one, we get \eqref{paper119}.
\end{proof}
\subsection{Frame inequality}
	\begin{lemma}\label{paper172}
	Let $\gamma,t \in (0,1)$ be such that $1-\gamma-3t>0$. Further, let  $W$ and $\{h_j\}_{j=1}^{p_3}$ be 
        as defined in \eqref{paper13005} and  \eqref{paper13004} respectively. If  $\epsilon R\geq 2\left(\frac{2C_r}{\sqrt{(2r-1)(r-1)}\gamma}\right)^\frac{1}{r-1}$, then the inequalities
		\begin{align} \label{paper154}
		\frac{\|\nu\|^2_2}{\|\nu\|_1^2}(1-\gamma-3t)\|f\|^2  & \leq \sum_{j=1}^{p_3}|\left\langle f, h_j \right\rangle|^2 \leq 
			\frac{\|\nu\|^2_2}{\|\nu\|_1^2}(1+3t)\|f\|^2 	  \hspace{0.4 cm}\forall  f \hspace{0.05 cm}\in V^{\left(1+\frac{5}{2}\epsilon\right)R}(g)
		\end{align}
		and 
		\begin{equation}\label{paper171}
			\|W\|_{2 \rightarrow 2} \leq \sqrt{\frac{2}{\frac{p}{2(1+\epsilon)R}-\frac{1}{\epsilon R}}}
		\end{equation}
		are true together with probability greater than 
  \begin{equation}\label{paper11005}
      1-6\exp\left(-\frac{m\epsilon}{12(1+3\epsilon)}\right)-5p\exp\left(-\frac{t^2}{42\tilde{C}^2}\left(\frac{p}{2(1+3\epsilon)R}-\frac{1}{\epsilon R}\right)\right).
  \end{equation}
	\end{lemma}
	\begin{proof}
		 Conditioning on $m_1,m_2$ and $m_3$, and applying Lemma \ref{paper155} to the random variables $\{y^1_i\}_{i=1}^{\widetilde{m}_1}$, $\{\epsilon^1_i\}_{i=1}^{\widetilde{m}_1}$, the $p_1\times \widetilde{m}_1$ matrix $ V_*=[V_{ij}]_{i\in \{1, \ldots,p_1\}, j \in \{1, \ldots,\widetilde{m}_1\}}$ and $I=I_{1\epsilon}$, we get that the event 
		\begin{align} \label{paper151}
			A_1=\bigg\{
			\frac{\|\nu\|^2_2}{\|\nu\|_1^2}\left(\|f\|^2_{2,I_{1\epsilon}}-t\|f\|^2 \right) &\leq\sum_{j=1}^{p_1}\frac{2(1+\epsilon)R}{p_1}\bigg|\sum_{i=1}^{\widetilde{m}_1}V_{ji}{\epsilon^1_i}f(y^1_i)\bigg|^2 \\
			\nonumber &\leq \frac{\|\nu\|^2_2}{\|\nu\|_1^2}\left(\|f\|^2_{2,I_{1\epsilon}}+t\|f\|^2 \right)	  \hspace{0.4 cm}\forall  f \hspace{0.05 cm}\in V^{\left(1+\frac{5}{2}\epsilon\right)R}(g) \bigg\}\\
			\nonumber=\bigg\{
			\frac{\|\nu\|^2_2}{\|\nu\|_1^2}\left(\|f\|^2_{2,I_{1\epsilon}}-t\|f\|^2 \right) &\leq\sum_{j=1}^{p_1}|\left\langle f, h_j \right\rangle|^2 \\
			\nonumber &\leq \frac{\|\nu\|^2_2}{\|\nu\|_1^2}\left(\|f\|^2_{2,I_{1\epsilon}}+t\|f\|^2 \right)	  \hspace{0.4 cm}\forall  f \hspace{0.05 cm}\in V^{\left(1+\frac{5}{2}\epsilon\right)R}(g) \bigg\}
		\end{align} 
		has probability greater than $ 1- 2^{\frac{3}{4}}p_1\exp\left(-\frac{t^2p_1}{42\tilde{C}^2(1+\epsilon)R}\right).$

  Again conditioning on $m_1,m_2$ and $m_3$, applying Lemma \ref{paper155} to the random variables $\{y^2_i\}_{i=1}^{\widetilde{m}_2}$,$\{\epsilon^2_i\}_{i=1}^{\widetilde{m}_2}$, the $p_2-p_1\times \widetilde{m}_2$ matrix $ V_*=[V_{ij}]_{i\in \{p_1+1, \ldots,p_2\}, j \in \{\widetilde{m}_1+1, \ldots,\widetilde{m}_1+\widetilde{m}_2\}}$ and $I=I_{2\epsilon}$, we get that the event 
		\begin{align}
			\nonumber A_2=\bigg\{
			\frac{\|\nu\|^2_2}{\|\nu\|_1^2}\left(\|f\|^2_{2,I_{2\epsilon}}-t\|f\|^2 \right) & \leq \sum_{j=p_1+1}^{p_2}\frac{2\epsilon R}{p_2-p_1}\bigg|\sum_{i=1}^{\widetilde{m}_2}V_{j\left(\widetilde{m}_1+i\right)}{\epsilon^2_i}f(y^2_i)\bigg|^2 \\
			\nonumber &\leq \frac{\|\nu\|^2_2}{\|\nu\|_1^2}\left(\|f\|^2_{2,I_{2\epsilon}}+t\|f\|^2 \right)	  \hspace{0.4 cm}\forall  f \hspace{0.05 cm}\in V^{\left(1+\frac{5}{2}\epsilon\right)R}(g) \bigg\}\\
			\nonumber=\bigg\{
			\frac{\|\nu\|^2_2}{\|\nu\|_1^2}\left(\|f\|^2_{2,I_{2\epsilon}}-t\|f\|^2 \right) &\leq\sum_{j=p_1+1}^{p_2}|\left\langle f, h_j \right\rangle|^2 \\
			\nonumber &\leq \frac{\|\nu\|^2_2}{\|\nu\|_1^2}\left(\|f\|^2_{2,I_{2\epsilon}}+t\|f\|^2 \right)	  \hspace{0.4 cm}\forall  f \hspace{0.05 cm}\in V^{\left(1+\frac{5}{2}\epsilon\right)R}(g) \bigg\}
		\end{align} 
		has probability greater than $ 1- 2^{\frac{3}{4}}(p_2-p_1)\exp\left(-\frac{t^2(p_2-p_1)}{42\tilde{C}^2\epsilon R}\right).$
		Similarly, we get that conditionally on $m_1,m_2$ and $m_3$, the event	
		\begin{align*}
			A_3=\bigg\{
			\frac{\|\nu\|^2_2}{\|\nu\|_1^2}\left(\|f\|^2_{2,I_{3\epsilon}}-t\|f\|^2 \right) &\leq\sum_{j=p_2+1}^{p_3}|\left\langle f, h_j \right\rangle|^2 \\
			\nonumber &\leq \frac{\|\nu\|^2_2}{\|\nu\|_1^2}\left(\|f\|^2_{2,I_{3\epsilon}}+t\|f\|^2 \right)	  \hspace{0.4 cm}\forall  f \hspace{0.05 cm}\in V^{\left(1+\frac{5}{2}\epsilon\right)R}(g) \bigg\}
		\end{align*}
		has probability greater than $ 1- 2^{\frac{3}{4}}(p_3-p_{2})\exp\left(-\frac{t^2(p_3-p_{2})}{42\tilde{C}^2\epsilon R}\right).$ Note that we require $|I_{2\epsilon}|=|I_{3\epsilon}|=2\epsilon R\geq 1$ in order to apply Lemma \ref{paper155}, but this holds true since we initially assumed that $\epsilon R\geq1$.
  
		From the union bound, we have
  \begin{equation} \label{paper11003}
      \mathbb{P}\left(\cap_{i=1}^3\left(A_i \cap B_i\right)\right) \geq 1-\mathbb{P}\left(\left(A_1 \cap B_1\right)^{\complement}\right)-\mathbb{P}\left(\left(A_2 \cap B_2\right)^{\complement}\right)-\mathbb{P}\left(\left(A_3 \cap B_3\right)^{\complement}\right).
  \end{equation}

        Further, using \eqref{paper14000} and \eqref{paper151},
        \begin{align*}
           \mathbb{P}\left(A_1 \cap B_1\right)&=\mathbb{E}\left(\mathbbm{1}_{A_1 \cap B_1}\right)=\mathbb{E}\left(\mathbbm{1}_{A_1}\mathbbm{1}_{B_1}\right)=\mathbb{E}\left(\mathbb{E}\left(\mathbbm{1}_{A_1}\mathbbm{1}_{B_1}|m_1,m_2,m_3 \right) \right)\\
&=\mathbb{E}\left(\mathbbm{1}_{B_1}\mathbb{E}\left(\mathbbm{1}_{A_1}|m_1,m_2,m_3\right) \right)=\mathbb{E}\left(\mathbbm{1}_{B_1}\mathbb{P}\left({A_1}|m_1,m_2,m_3\right) \right)\\
&\geq \mathbb{E}\left(\mathbbm{1}_{B_1}\left(1- 2^{\frac{3}{4}}p_1\exp\left(-\frac{t^2p_1}{42\tilde{C}^2(1+\epsilon)R}\right) \right)\right)\\
&\geq \left(1-2^{\frac{3}{4}}\frac{3p\left(1+\epsilon\right)}{2\left(1+3\epsilon\right)}\exp\left(-\frac{t^2}{42\tilde{C}^2}\left(\frac{p}{2\left(1+3\epsilon\right)R}-\frac{1}{\left(1+\epsilon\right) R}\right)\right)\right)\mathbb{P}(B_1).
\end{align*}
That is, if $1-2^{\frac{3}{4}}\frac{3p\left(1+\epsilon\right)}{2\left(1+3\epsilon\right)}\exp\left(-\frac{t^2}{42\tilde{C}^2}\left(\frac{p}{2\left(1+3\epsilon\right)R}-\frac{1}{\left(1+\epsilon\right) R}\right)\right)\geq0$ (we need not consider the case where this term is negative as this would imply that the bound in \eqref{paper11005} is negative), then
\begin{align*}
\mathbb{P}\left(A_1 \cap B_1\right)
&\geq \left(1-2^{\frac{3}{4}}\frac{3p\left(1+\epsilon\right)}{2\left(1+3\epsilon\right)}\exp\left(-\frac{t^2}{42\tilde{C}^2}\left(\frac{p}{2\left(1+3\epsilon\right)R}-\frac{1}{\left(1+\epsilon\right) R}\right)\right)\right)\\
& \hspace{0.2cm}\cdot \left(1-2\exp\left(-\frac{m\left(1+\epsilon\right)}{12\left(1+3\epsilon\right)}\right)\right)\\
& \hspace{-0.8cm}\geq 1-2\exp\left(-\frac{m\left(1+\epsilon\right)}{12\left(1+3\epsilon\right)}\right)-2^{\frac{3}{4}}\frac{3p\left(1+\epsilon\right)}{2\left(1+3\epsilon\right)}\exp\left(-\frac{t^2}{42\tilde{C}^2}\left(\frac{p}{2\left(1+3\epsilon\right)R}-\frac{1}{\left(1+\epsilon\right) R}\right)\right).
\end{align*}
Bounding $\mathbb{P}\left(A_2 \cap B_2\right)$ and $\mathbb{P}\left(A_3 \cap B_3\right)$ in a similar way, and then using the bounds in \eqref{paper11003}, we obtain that with probability greater than \[1-2\exp\left(-\frac{m\left(1+\epsilon\right)}{12\left(1+3\epsilon\right)}\right)-2^{\frac{3}{4}}\frac{3p\left(1+\epsilon\right)}{2\left(1+3\epsilon\right)}\exp\left(-\frac{t^2}{42\tilde{C}^2}\left(\frac{p}{2\left(1+3\epsilon\right)R}-\frac{1}{\left(1+\epsilon\right) R}\right)\right)\]\[-2\cdot2\exp\left(-\frac{m\epsilon}{12\left(1+3\epsilon\right)}\right)-2\cdot 2^{\frac{3}{4}}\frac{3p\epsilon}{2\left(1+3\epsilon\right)}\exp\left(-\frac{t^2}{42\tilde{C}^2}\left(\frac{p}{2\left(1+3\epsilon\right)R}-\frac{1}{\epsilon R}\right)\right)\]
        the inequalities
		\begin{align}\label{paper15000}
		\frac{\|\nu\|^2_2}{\|\nu\|_1^2}\left(\|f\|^2_{2,[-(1+3\epsilon)R,(1+3\epsilon)R]}-3t\|f\|^2 \right)&\leq\sum_{j=1}^{p_3}|\left\langle f, h_j \right\rangle|^2\\
  \nonumber &\hspace{-2cm}\leq\frac{\|\nu\|^2_2}{\|\nu\|_1^2}\left(\|f\|^2_{2,[-(1+3\epsilon)R,(1+3\epsilon)R]}+3t\|f\|^2 \right) \hspace{0.2cm} \forall \hspace{0.1cm} f \in V^{\left(1+\frac{5}{2}\epsilon\right)R}(g)
		\end{align}
		and (see \eqref{paper11001})
		\begin{equation*}
			\|W\|_{2 \rightarrow 2} \leq \sqrt{\frac{2}{\frac{p}{2(1+\epsilon)R}-\frac{1}{\epsilon R}}}
		\end{equation*}
		are true together. The inequality \eqref{paper15000} is attained by adding the inequalities in $A_1, A_2$ and $A_3$. Simplifying the probability bound and using Lemma \ref{paper161} to bound the inequality  \eqref{paper15000} from below, we conclude the proof.
	\end{proof}

\smallskip
         Using Lemma \ref{paper172} , we conclude that with probability greater than $1-6\exp\left(-\frac{m\epsilon}{12(1+3\epsilon)}\right)-5p\exp\left(-\frac{t^2}{42\tilde{C}^2}\left(\frac{p}{2(1+3\epsilon)R}-\frac{1}{\epsilon R}\right)\right)$, the following hold.
	   \begin{enumerate}
	       \item The collection $\{h_j\}_{j=1}^{p_3}$ forms a frame for $ V^{\left(1+\frac{5}{2}\epsilon\right)R}(g)$.
             \item The frame operator corresponding to $\{h_j\}_{j=1}^{p_3}$ given by 
             \begin{equation}
		Sf=\sum_{j=1}^{p_3}\left\langle f, h_j \right\rangle h_j \hspace{0.2cm} \forall \hspace{0.1cm} f \in  V^{\left(1+\frac{5}{2}\epsilon\right)R}(g)
	\end{equation}
	is invertible. 
        \item Let 
        $G_{WV}$ denote the synthesis operator  \cite{paper12} associated with the canonical dual frame $\{S^{-1}h_j\}_{j=1}^{p_3}$, that is,
	\begin{align*}
		&G_{WV}(\{c_j\}_{j=1}^{p_3})=\sum_{j=1}^{p_3} c_jS^{-1}h_j \hspace{0.2cm} \forall \hspace{0.1cm} \{c_j\}_{j=1}^{p_3} \in \mathbb{C}^{p_3}.
	\end{align*}
        Then, as $WV \Phi E$ is the analysis operator corresponding to the frame $\{h_j\}_{j=1}^{p_3}$ (see \eqref{paper1WVPhi=h}),  by the frame reconstruction formula, we get that for all $f \in  V^{\left(1+\frac{5}{2}\epsilon\right)R}(g)$,
        \begin{align*}
            G_{WV}  WV\Phi Ef=\sum_{j=1}^{p_3} \left\langle f,h_j \right\rangle S^{-1}h_j=f
        \end{align*}
        Thus the operator
	\begin{equation} \label{paper145} 
		F_{WV}:= G_{WV} WV
	\end{equation} satisfies $F_{WV}\Phi E=I$, where $I$ is the identity operator on $V^{\left(1+\frac{5}{2}\epsilon\right)R}(g)$. Thus it is a left inverse of the operator $\Phi E$.
        \item  It can be shown that $S^{-1}=G_{WV}G^*_{WV}$ and that ${\|S^{-1}\|=\|G_{WV}\|^2}$(see  \cite{paper12}). Hence using \eqref{paper154} and  \cite[Lemma 5.1.5]{paper12} we conclude that  
	\begin{enumerate}
		\item \begin{equation} \label{paper110001}
		    \|G_{WV}\|=\sqrt{\|S^{-1}\|}\leq \frac{\|\nu\|_1}{\|\nu\|_2\sqrt{(1-\gamma-3t)}}.
		\end{equation}
		\item 
		\begin{equation} \label{paper110002}
			\|I-B^{-1}S\|\leq \frac{B-A}{B} = \frac{6t+\gamma}{1+3t}:=\alpha,
		\end{equation}
	\end{enumerate}
		  where $A=\frac{\|\nu\|^2_2}{\|\nu\|_1^2}\left(1-3t-\delta\right)$ and $B=\frac{\|\nu\|^2_2}{\|\nu\|_1^2}\left(1+3t\right)$. As $\alpha<1$, we further have the following expansion for $S^{-1}$,
	\begin{align*}
		S^{-1}&=\frac{1}{B}\sum_{r=0}^{\infty}(I-B^{-1}S)^r.
	\end{align*}
	\end{enumerate}
\begin{remark}\label{paper1reasonforflips}
Now, we remark on the importance of the random flips in our method.
 
\smallskip
Suppose we remove the random sign flips $\left\{\epsilon_i\right\}_{i=1}^m$. That is,
	we define, the vectors $\{u^j\}_{j=1}^{p_*}$ (see \eqref{paper1eq:12}) as \begin{equation}
                u^j := \left(\overline{\sum_{i=1}^{m_*}{\left(V_*\right)}_{ji}\phi_k(z_i)}\right)_{k \in \left[ 2 \left(1+\frac{5}{2}\epsilon\right)R\right]}  \hspace{0.2cm} \forall  \hspace{0.1cm} j \in \{1, ...,p_*\}.
            \end{equation}

Then, going through the calculations of Lemma \ref{paper1boundonuandlambda}, we can observe that the bound $\|u_j\|_2 \leq C$ for all $j \in \{1, \ldots, p^{*}\}$  remains unchanged. However, $\Lambda$ changes. We calculate it below.
\begin{align*}
			\Lambda_{kl}=\mathbb{E}\left(u^j \otimes u^j\right)_{kl}&= \mathbb{E}_ z\left(\sum_{i=1,r=1}^{m_*}{\left(V_*\right)}_{ji}{\left(V_*\right)}_{jr}\overline{\phi_k(z_i)}\phi_l(z_r)\right)\\&=\mathbb{E}_z\sum_{i=1,r=1}^{m_*}{\left(V_*\right)}_{ji}{\left(V_*\right)}_{jr}\overline{\phi_k(z_i)}\phi_l(z_r)\\
            &=\mathbb{E}_z\sum_{i=1,r=1,i\neq r}^{m_*}{\left(V_*\right)}_{ji}{\left(V_*\right)}_{jr}\overline{\phi_k(z_i)}\phi_l(z_r)+\mathbb{E}_z\sum_{i=1}^{m_*}{\left(V_*\right)}_{ji}^2\overline{\phi_k(z_i)}\phi_l(z_i)\\&=\frac{1}{|I|^2}\overline{\int_I\phi_k(z)dz}\int_I\phi_l(z)dz+\frac{\|\nu\|_2^2}{\|\nu\|_1^2|I|}\int_I\overline{\phi_k(z)}\phi_l(z)dz\\
		&=\frac{1}{|I|^2}\overline{\int_I\phi_k(z)dz}\int_I\phi_l(z)dz+\frac{\|\nu\|_2^2}{\|\nu\|_1^2|I|}\overline{\left\langle A_I\phi_k,\phi_l \right\rangle}\\&=\frac{1}{|I|^2}\overline{\int_I\phi_k(z)dz}\int_I\phi_l(z)dz+\frac{\|\nu\|_2^2}{\|\nu\|_1^2|I|}\lambda_k\delta_{kl}.
			\end{align*}
Now, using this new $\Lambda$ in the computations done in the proof of Lemma \ref{paper155}, we obtain the following inequality instead of \eqref{paper119}.
\begin{align}
		\label{paper1newwrongframeineq}\frac{1}{|I|}\left|\int_If(x)dx\right|^2+\frac{\|\nu\|^2_2}{\|\nu\|_1^2}(\|f\|^2_{2,I}-t\|f\|^2 ) &\leq \frac{|I|}{p_*}\sum_{j=1}^{p_*}\bigg|\sum_{i=1}^{m_*}{\left(V_*\right)}_{ji}\epsilon_if(z_i)\bigg|^2 \\
		\nonumber&\hspace{-2cm}\leq	\frac{\|\nu\|^2_2}{\|\nu\|_1^2}(\|f\|^2_{2,I}+t\|f\|^2 )+\frac{1}{|I|}\left|\int_If(x)dx\right|^2  \hspace{0.2 cm}\forall  f \hspace{0.1 cm}\in {V}^{\left(1+\frac{5}{2}\epsilon\right)R}(g) 
\end{align}
To prove the frame inequality, it is clear that the additional term $\frac{1}{|I|^2}\left|\int_If(x)dx\right|^2$ must be eliminated. One way to do this is to completely ignore the presence of this term in the left-hand side inequality and use the Holder inequality to bound this term on the right hand side. Hence, we get 
\begin{align}
		\label{paper1withoutflipsineq} \frac{\|\nu\|^2_2}{\|\nu\|_1^2}(\|f\|^2_{2,I}-t\|f\|^2 ) &\leq \frac{|I|}{p_*}\sum_{j=1}^{p_*}\bigg|\sum_{i=1}^{m_*}{\left(V_*\right)}_{ji}\epsilon_if(z_i)\bigg|^2 \\
		\nonumber&\hspace{-2cm}\leq	\frac{\|\nu\|^2_2}{\|\nu\|_1^2}(2\|f\|^2_{2,I}+t\|f\|^2 )  \hspace{0.2 cm}\forall  f \hspace{0.1 cm}\in {V}^{\left(1+\frac{5}{2}\epsilon\right)R}(g) 
\end{align}
Notice that, in the next section (see \eqref{paper166}), we bound the reconstruction error in terms of $\alpha$ (defined in \eqref{paper110002}). In case, the random flips are included $\alpha$ can be made arbitrarily small by choosing an appropriate value of $t$ and $\gamma$. However, if instead the random flips are excluded and we work with the inequality \eqref{paper1withoutflipsineq}. Then, continuing the along the same steps, we arrive at the following value for $\alpha$ in \eqref{paper110002}. 
\[\alpha=\frac{1+6t+\gamma}{2+3t}.\]
The above $\alpha$, however unlike the one in \eqref{paper110002}, cannot be made arbitrarily small by choosing small values of $t$ and $\gamma$. Hence, the error bounds that we prove in the next section no longer give any kind of decay. This is why we have included the random flips in our method, to obtain an \( \alpha \) that can be made arbitrarily small. Notice that in the limiting case, that is when \( \alpha = 0 \), this essentially means we have a tight frame. That is, the dual frame of \(\{h_j\}_{j=1}^{p_3}\) would be a constant multiple of it (see \cite{paper12}) and, therefore, would satisfy all the localization properties that it satisfies. Consequently, this would help in the localization of the error, as explained in Subsection \ref{paper1boundingthereconstructionerror}.
\end{remark}
 \section{Bounding the reconstruction error} \label{paper178}
 In this section, we use the random frame from the previous section to bound the error between the original function and the reconstructed function. We prove our two main results towards the end of the section. 
	For any $f \in C_{[-\pi,\pi]}$, define
        \begin{enumerate}
            \item $\tilde{f}:=Pf$,
            \item $\tilde{y}:= \Phi E\tilde{f}=\frown_{j=1}^3\{\epsilon^j_i\tilde{f}(y^j_i)\}_{i=1}^{\widetilde{m}_j}$,
            \item 
            \begin{equation} \label{paper164}
                y:=\frown_{j=1}^3\{\epsilon^j_if(y^j_i)\}_{i=1}^{\widetilde{m}_j},
            \end{equation}
             \item $e=\frown_{j=1}^3\{e^j_i\}_{i=1}^{\widetilde{m}_j}:= y- \tilde{y}.$\\
        \end{enumerate}
        The first step is to establish a crucial lemma. From the lemma, our two main results follow as corollaries. However, before that we remark the following.
	\begin{remark} \label{paper162}
        The block diagonal structure of $V$ implies
	\begin{enumerate}
		\item $(Ve)_j=\sum_{i=1}^{\frac{m}{p}}V_{j((j-1)\frac{m}{p}+i)}e_{y^1_{(j-1)\frac{m}{p}+i}} \hspace{0.2cm} \forall \hspace{0.1cm} j \in \{1, \ldots,p_1\}.$
		\item $(Ve)_j=\sum_{i=1}^{\frac{m}{p}}V_{j((j-1)\frac{m}{p}+i)}e_{y^2_{(j-p_1-1)\frac{m}{p}+i}} \hspace{0.2cm} \forall \hspace{0.1cm} j \in \{p_1+1, \ldots,p_2\}.$
		\item $(Ve)_j=\sum_{i=1}^{\frac{m}{p}}V_{j((j-1)\frac{m}{p}+i)}e_{y^3_{(j-p_2-1)\frac{m}{p}+i}} \hspace{0.2cm} \forall \hspace{0.1cm} j \in \{p_2+1, \ldots,p_3\}.$
	\end{enumerate}	
	Now, by definition, $y^1_{(j-1)\frac{m}{p}+i}\in I_{1\epsilon} \hspace{0.2cm} \forall \hspace{0.1cm} i \in \{1, \ldots,\frac{m}{p}\}$ and $ j \in \{1, \ldots,p_1\}$, $y^2_{(j-p_1-1)\frac{m}{p}+i}\in I_{2\epsilon} \hspace{0.2cm} \forall \hspace{0.1cm} i \in \{1, \ldots,\frac{m}{p}\}$ and $ j \in \{p_1+1, \ldots,p_2\}$, and $y^3_{(j-p_2-1)\frac{m}{p}+i}\in I_{3\epsilon} \hspace{0.2cm} \forall \hspace{0.1cm} i \in \{1, \ldots,\frac{m}{p}\}$ and $ j \in \{p_2+1, \ldots,p_3\}$. Therefore, using Lemmas \ref{paper126} and \ref{paper125}, we can conclude the following.
	\begin{enumerate}
		\item For $j \in \{1, \ldots,p_1\}$,
	\begin{enumerate}
		\item   $\begin{aligned}[t]
                 \left|(Ve)_j\right|=&\left|\sum_{i=1}^{\frac{m}{p}}V_{j((j-1)\frac{m}{p}+i)}e_{y^1_{(j-1)\frac{m}{p}+i}}\right|
                 \leq \|f-\tilde{f}\|_{L^\infty(I_{1\epsilon})}\sum_{i=1}^{\frac{m}{p}}V_{j((j-1)\frac{m}{p}+i)}\\
                 &=\|f-\tilde{f}\|_{L^\infty(I_{1\epsilon})}\leq \frac{2\sqrt{2}C_r}{(r-1)(\frac{3\epsilon R}{2})^{r-1}} .
            \end{aligned}$
		\item $\begin{aligned}[t]\|h_j\|&= \left\|\sqrt{\frac{2(1+\epsilon)R}{p_1}}\sum_{i=1}^{\widetilde{m}_1}V_{ji}\epsilon^1_ik_{y^1_i}\right\| \leq \sqrt{\frac{2(1+\epsilon)R}{p_1}}\sup_{y \in \mathbb{R}}\|k_y\|\sum_{i=1}^{\widetilde{m}_1}V_{ji}\\
  &=\sqrt{\frac{2(1+\epsilon)R}{p_1}}\sup_{y \in \mathbb{R}}\|k_y\|\leq \sqrt{\frac{2(1+\epsilon)R}{p_1}}C. \end{aligned}$\\
	\end{enumerate}
	\item Similarly, for all $j \in \{p_1+1, \ldots,p_2\}$,
	\begin{enumerate}
	\item $\left|(Ve)_j\right|\leq \left\|f-\tilde{f}\right\|_{L^\infty(I_{2\epsilon})} \leq\frac{2\sqrt{2}C_r}{(r-1)(\frac{\epsilon R}{2})^{r-1}}$.
	\item $\|h_j\|_{L^2[-R,R]} \leq  \sqrt{\frac{2\epsilon R}{p_2-p_1}}\sup_{x \in I_{2\epsilon}}\|k_x\|_{L^2[-R,R]} \leq \sqrt{\frac{2\epsilon R}{p_2-p_1}}\frac{6C^2_r}{\sqrt{2r-1}(\frac{\epsilon R}{2})^{r-\frac{1}{2}}}\Gamma_{r-\frac{1}{2}}$.
	\item $\|h_j\| \leq \sqrt{\frac{2\epsilon R}{p_2-p_1}}\sup_{x \in \mathbb{R}}\|k_x\|\leq \sqrt{\frac{2\epsilon R}{p_2-p_1}}C$.\\
	\end{enumerate}
	\item  Likewise, for all $j \in \{p_2+1, \cdots,p_3\}$,
	\begin{enumerate}
		\item $|(Ve)_j| \leq \left\|f-\tilde{f}\right\|_{L^\infty(\mathbb{R})}\leq C_{g}$.
		\item $\|h_j\|_{L^2[-R,R]} \leq  \sqrt{\frac{2\epsilon R}{p_3-p_2}}\sup_{x \in I_{3\epsilon}}\left\|k_x\right\|_{L^2[-R,R]} \leq \sqrt{\frac{2\epsilon R}{p_3-p_2}}\frac{6C^2_r}{\sqrt{2r-1}(\epsilon R)^{r-\frac{1}{2}}}\Gamma_{r-\frac{1}{2}}$.
		\item $\|h_j\| \leq \sqrt{\frac{2\epsilon R}{p_3-p_2}}\sup_{x \in \mathbb{R}}\|k_x\|\leq \sqrt{\frac{2\epsilon R}{p_3-p_2}}C$.
	\end{enumerate}
	\end{enumerate}
	\end{remark}
As mentioned earlier, we now prove the following important lemma.
    \begin{lemma}\label{finalerrorboundlemma}
Assume that the conditions in Lemma \ref{paper172} hold. Then, with probability greater than $1-6\exp\left(-\frac{m\epsilon}{12(1+3\epsilon)}\right)-5p\exp\left(-\frac{t^2}{42\tilde{C}^2}\left(\frac{p}{2(1+3\epsilon)R}-\frac{1}{\epsilon R}\right)\right)$, we have
	\begin{align}
	   \label{paper170} \|f-F_{WV}q\|_{L^2[-R,R]}&\leq\frac{2\sqrt{2}C_r}{\sqrt{2r-1}(r-\frac{3}{2})(\frac{5}{2}\epsilon R)^{r-\frac{3}{2}}}\\
		\nonumber&+ \frac{\|\nu\|_1}{\|\nu\|_2\sqrt{1-\gamma-3t}}\sqrt{\frac{2}{\frac{p}{2(1+\epsilon)R}-\frac{1}{\epsilon R}}}\|VH\|_{\infty \rightarrow 2}\|u\|_{\infty}\\
		\nonumber&+ \left(\frac{\|\nu\|_1}{\|\nu\|_2}\right)^4\frac{C_r^\sharp}{(\frac{\epsilon R}{2})^{r-\frac{5}{2}}}\left(\frac{(1+\epsilon)R}{(\frac{\epsilon R}{2})^{\frac{3}{2}}}+\frac{12}{\sqrt{2r-1}}+\frac{1}{\sqrt{2r-1}\epsilon R}\right)\\
		\nonumber&+\left(\frac{\|\nu\|_1}{\|\nu\|_2}\right)^22\epsilon R C C_g\frac{\alpha^2}{1-\alpha} \hspace{0.2cm} \forall \hspace{0.1cm} f \in C_{[-\pi,\pi]}.
	\end{align} 
    Here $F_{WV}$ is as defined in \eqref{paper145}, $y$ is as defined in \eqref{paper164} and $q=Q(y)$, where $Q$ is any noise-shaping quantizer with transfer operator $H$.
    \end{lemma}
    \begin{proof}
As $y$ is quantized using a noise-shaping quantizer with transfer operator $H$, we have $y-q=Hu$. Consequently, using Lemma \ref{paper172}, we get that with probability greater than $1-6\exp\left(-\frac{m\epsilon}{12(1+3\epsilon)}\right)-5p\exp\left(-\frac{t^2}{42\tilde{C}^2}\left(\frac{p}{2(1+3\epsilon)R}-\frac{1}{\epsilon R}\right)\right)$, the following holds. First,
\begin{align}
\label{paper1projectionerror} \|f-F_{WV}q\|_{L^2[-R,R]}&\leq \|f- \tilde{f}\|_{L^2[-R,R]} +\|\tilde{f}-F_{WV}q\|_{L^2[-R,R]} \hspace{0.2cm} \forall \hspace{0.1cm} f \in C_{[-\pi,\pi]}.
\end{align}
Further, using \eqref{paper171} and \eqref{paper110001}, the following hold uniformly for all $f \in C_{[-\pi,\pi]}$ with the same probability.
\begin{align}
	\nonumber	\left\|\tilde{f}- F_{WV}q\right\|_{L^2[-R,R]}& = \left\|F_{WV}\Phi E\tilde{f}-F_{WV}q\right\|_{L^2[-R,R]}\\
	\nonumber	&= \left\|F_{WV}\tilde{y}-F_{WV}q\right\|_{L^2[-R,R]}\\
	\nonumber	&= \left\|F_{WV}(y-e-q)\right\|_{L^2[-R,R]}\\
	\nonumber	&=\|F_{WV}(Hu-e)\|_{L^2[-R,R]}\\
	\nonumber	&\leq \|F_{WV}Hu\|_{L^2[-R,R]}+ \|F_{VW}e\|_{L^2[-R,R]}\\
     \nonumber &\leq \|G_{WV}WVHu\|+\|G_{WV}WVe\|_{L^2[-R,R]}\\
	\nonumber	& \leq \|G_{WV}\|\|W\|_{2 \rightarrow 2}\|VH\|_{\infty \rightarrow 2}\|u\|_{\infty}+\|G_{WV}WVe\|_{L^2[-R,R]}\\
	\nonumber& \leq \frac{\|\nu\|_1}{\|\nu\|_2\sqrt{1-\gamma-3t}}\sqrt{\frac{2}{\frac{p}{2(1+\epsilon)R}-\frac{1}{\epsilon R}}}\|VH\|_{\infty \rightarrow 2}\|u\|_{\infty}\\
 \label{paper120}&+\|G_{WV}WVe\|_{L^2[-R,R]}.
	\end{align}
	Next, we bound the second term in \eqref{paper120}.
	\[\left\|G_{WV}WVe\right\|_{L^2[-R,R]}=\left\|\sum_{j=1}^{p_3}(WVe)_jS^{-1}h_j\right\|_{L^2[-R,R]}=\left\|S^{-1}\left(\sum_{j=1}^{p_3}(WVe)_jh_j\right)\right\|_{L^2[-R,R]}.\]
	Let $\tilde{h}:=\sum_{j=1}^{p_3}(WVe)_jh_j$. Then 
	\begin{align}
	\nonumber	\left\|G_{WV}WVe\right\|_{L^2[-R,R]}&=\left\|S^{-1}\tilde{h}\right\|_{L^2[-R,R]}=\left\|\frac{1}{B}\sum_{r=0}^{\infty}(I-B^{-1}S)^r\tilde{h}\right\|_{L^2[-R,R]}\\
	\nonumber&=\frac{1}{B}\left\|\tilde{h}+\tilde{h}-B^{-1}S\tilde{h}+\sum_{r=2}^{\infty}(I-B^{-1}S)^r\tilde{h}\right\|_{L^2[-R,R]}\\
	\nonumber	& \leq \frac{2}{B}\left\|\tilde{h}\right\|_{L^2[-R,R]}+B^{-2}\left\|S\tilde{h}\right\|_{L^2[-R,R]}+\frac{1}{B}\left\|\sum_{r=2}^{\infty}(I-B^{-1}S)^r\tilde{h}\right\| \\
	\nonumber	&\leq \frac{2}{B}\left\|\tilde{h}\right\|_{L^2[-R,R]}+B^{-2}\left\|S\tilde{h}\right\|_{L^2[-R,R]}+\frac{1}{B}\sum_{r=2}^{\infty}\|I-B^{-1}S\|^r\|\tilde{h}\| \\
	\label{paper166} & \leq \frac{2}{B}\left\|\tilde{h}\right\|_{L^2[-R,R]}+B^{-2}\left\|S\tilde{h}\right\|_{L^2[-R,R]}+ \frac{1}{B}\frac{\alpha^2}{1-\alpha}\|\tilde{h}\|,
	\end{align} 
 where we used \eqref{paper110002} in the last step. Now, we separately bound each of the three terms in \eqref{paper166}.

\smallskip
        Using the bounds in Remark \ref{paper162} and the definition of the matrix $W$ (see \eqref{paper13005}), $\|\tilde{h}\|_{L^2[-R,R]}$ from the first term in \eqref{paper166} can be bound as follows.
	\begin{align}
		\nonumber\|\tilde{h}\|_{L^2[-R,R]}&=\left\|\sum_{j=1}^{p_3}(WVe)_jh_j\right\|_{L^2[-R,R]}\\ 
		\nonumber & \hspace{-0.5cm} \leq \left\|\sum_{j=1}^{p_1}(WVe)_jh_j\right\| +\left\|\sum_{j=p_1+1}^{p_2}(WVe)_jh_j\right\| +\left\|\sum_{j=p_2+1}^{p_3}(WVe)_jh_j\right\|_{L^2[-R,R]}\\
  \nonumber & \hspace{-0.5cm} \leq \sum_{j=1}^{p_1}W_{jj}|(Ve)_j|\|h_j\| +\sum_{j=p_1+1}^{p_2}W_{jj}|(Ve)_j|\|h_j\| +\sum_{j=p_2+1}^{p_3}W_{jj}|(Ve)_j|\|h_j\|_{L^2[-R,R]}\\
		\label{paper158} & \hspace{-0.5cm}  \leq \frac{4\sqrt{2}C C_r}{(r-1)}\left(\frac{(1+\epsilon)R}{(\frac{3\epsilon R}{2})^{r-1}}+\frac{\epsilon R}{(\frac{\epsilon R}{2})^{r-1}}\right)+ \frac{12\epsilon RC^2_rC_{g}\Gamma_{r-\frac{1}{2}}}{\sqrt{2r-1}(\epsilon R)^{r-\frac{1}{2}}}.
	\end{align}
	Using the fact that $1-3t-\gamma>0$ implies $\|S\|\leq \frac{\|\nu\|^2_2}{\|\nu\|_1^2}(1+3t)<2$, $\left\|S\tilde{h}\right\|_{L^2[-R,R]}$ from the second term in \eqref{paper166} can be bound in the following way.
	\begin{align}	\nonumber\left\|S\tilde{h}\right\|_{L^2[-R,R]}&=\left\|\sum_{j=1}^{p_3}(WVe)_jSh_j\right\|_{L^2[-R,R]}\\ 
		\nonumber & \hspace{-2cm}\leq \sum_{j=1}^{p_2}W_{jj}|(Ve)_j|\|S\|\|h_j\| +\sum_{j=p_2+1}^{p_3}W_{jj}|(Ve)_j|\|Sh_j\|_{L^2[-R,R]}\\
		\label{paper111004}& \hspace{-2cm} \leq \frac{8\sqrt{2}C C_r}{(r-1)}\left(\frac{(1+\epsilon)R}{(\frac{3\epsilon R}{2})^{r-1}}+\frac{\epsilon R}{(\frac{\epsilon R}{2})^{r-1}}\right)+\sqrt{\frac{2\epsilon R}{p_3-p_2}}\sum_{j=p_2+1}^{p_3}C_{g}\|Sh_{j}\|_{L^2[-R,R]},
	\end{align}
    Now, for $j \in \{p_2+1, \ldots,p_3\}$,
	\begin{align}
		\nonumber \|Sh_j\|_{L^2[-R,R]}&\leq\sum_{l=1}^{p_3}|\left\langle h_j, h_l \right\rangle |\|h_l\|_{L^2[-R,R]}\\
		 \label{paper111001}&\leq \sum_{l=1}^{p_1}|\left\langle h_j,h_l \right\rangle|\|h_l\| +\sum_{l=p_1+1}^{p_3}\|h_j\| \|h_l\| \|h_l\|_{L^2[-R,R]}.
	\end{align}
	The first term in \eqref{paper111001} can be bound in the following way.
	\begin{align*}
	\sum_{l=1}^{p_1}|\left\langle h_j,h_l \right\rangle|\|h_l\|  & \leq \sum_{l=1}^{p_1}\left|\left\langle \sqrt{\frac{2\epsilon R}{p_3-p_2}}\sum_{i=1}^{\widetilde{m}_3}V_{j(\widetilde{m}_1+\widetilde{m}_2+i)}\epsilon^3_ik_{y^3_i}, \sqrt{\frac{2(1+\epsilon)R}{p_1}}\sum_{n=1}^{\widetilde{m}_1}V_{ln}\epsilon^1_nk_{y^1_n} \right\rangle\right|\|h_l\| \\
		&\leq \sqrt{\frac{2\epsilon R}{p_3-p_2}}\frac{2C(1+\epsilon)R}{p_1}\sum_{l=1}^{p_1}\sum_{i=1}^{\widetilde{m}_3}\sum_{n=1}^{\widetilde{m}_1}V_{j(\widetilde{m}_1+\widetilde{m}_2+i)}V_{ln}\left|K(y_{n}^1,y_i^3)\right|\\
		\numberthis \label{paper111002} & \hspace{0cm} \leq \frac{4C(1+\epsilon)RC^2_r\Gamma_{r-\frac{1}{2}}}{(\frac{\epsilon R}{2})^r}\sqrt{\frac{2\epsilon R}{p_3-p_2}},
	\end{align*}
        where in order to obtain the last inequality, we used \eqref{paper169}. The second term in the equation \eqref{paper111001} can be bound in the following way.
        \begin{align*}
		\sum_{l=p_1+1}^{p_3}&\|h_j\|\|h_l\|\|h_l\|_{L^2[-R,R]}\\
		&\leq \sqrt{\frac{2 \epsilon R}{p_3-p_2}}C\left(\sum_{l=p_1+1}^{p_2}\|h_l\|\|h_l\|_{L^2[-R,R]}+\sum_{l=p_2+1}^{p_3}\|h_l\|\|h_l\|_{L^2[-R,R]}\right)\\
        \begin{split}
            &\leq \sqrt{\frac{2 \epsilon R}{p_3-p_2}}C\left(\sum_{l=p_1+1}^{p_2}\frac{12\epsilon RCC_r^2\Gamma_{r-\frac{1}{2}}}{(p_2-p_1)\sqrt{2r-1}(\frac{\epsilon R}{2})^{r-\frac{1}{2}}} \right. 
		\left. +\sum_{l=p_2+1}^{p_3}\frac{12\epsilon RCC_r^2\Gamma_{r-\frac{1}{2}}}{(p_2-p_1)\sqrt{2r-1}(\epsilon R)^{r-\frac{1}{2}}}\right)
        \end{split}\\
		\numberthis \label{paper111003}& =\sqrt{\frac{2 \epsilon R}{p_3-p_2}}\frac{12\epsilon RC^2C_r^2\Gamma_{r-\frac{1}{2}}}{\sqrt{2r-1}(\frac{\epsilon R}{2})^{r-\frac{1}{2}}}\left(1+\frac{1}{2^{r-\frac{1}{2}}}\right).
	\end{align*}
        Therefore, using \eqref{paper111001},\eqref{paper111002} and \eqref{paper111003} in \eqref{paper111004}, we get
	\begin{align} 
		\nonumber \|S\tilde{h}\|_{L^2[-R,R]} &\leq \frac{8\sqrt{2}C C_r}{(r-1)}\left(\frac{(1+\epsilon)R}{(\frac{3\epsilon R}{2})^{r-1}}+\frac{\epsilon R}{(\frac{\epsilon R}{2})^{r-1}}\right)\\
		\nonumber & \hspace{-1cm}+2\epsilon R C_{g}\left(\frac{4C(1+\epsilon)RC^2_r\Gamma_{r-\frac{1}{2}}}{(\frac{\epsilon R}{2})^r}+\frac{12 \epsilon R C^2C_r^2\Gamma_{r-\frac{1}{2}}}{\sqrt{2r-1}(\frac{\epsilon R}{2})^{r-\frac{1}{2}}}\left(1+\frac{1}{2^{r-\frac{1}{2}}}\right)\right)\\
        \label{paper163} & \hspace{-1cm}\leq \frac{8\sqrt{2}C C_r}{(r-1)}\left(\frac{(1+\epsilon)R}{(\frac{3\epsilon R}{2})^{r-1}}+\frac{\epsilon R}{(\frac{\epsilon R}{2})^{r-1}}\right)\\\nonumber & \hspace{-1cm}+ 16C_gCC^2_r\Gamma_{r-\frac{1}{2}}\left(\frac{(1+\epsilon)R}{\left(\frac{\epsilon R}{2}\right)^{r-1}}+\frac{12C}{\sqrt{2r-1}\left(\frac{\epsilon R}{2}\right)^{r-\frac{5}{2}}}\right).
	\end{align}
	Lastly, $\left\|\tilde{h}\right\|$ from the third term in \eqref{paper166} can be bound as follows.
	\begin{align}
		\nonumber\left\|\tilde{h}\right\| =\left\|\sum_{j=1}^{p_3}(WVe)_jh_j\right\| &\leq \sum_{j=1}^{p_2}W_{jj}|(Ve)_j|\|h_j\| +\sum_{j=p_2+1}^{p_3}W_{jj}|(Ve)_j|\|h_j\|\\
		\label{paper159} & \leq \frac{4\sqrt{2}C C_r}{(r-1)}\left(\frac{(1+\epsilon)R}{(\frac{3\epsilon R}{2})^{r-1}}+\frac{\epsilon R}{(\frac{\epsilon R}{2})^{r-1}}\right) + 2\epsilon RCC_{g}.
	\end{align}
	If we assume that $\frac{\alpha^2}{1-\alpha}<1$, then using \eqref{paper158},\eqref{paper163} and \eqref{paper159} in \eqref{paper166}, we obtain
    \begin{align}
        \begin{split}
           \nonumber\|G_{WV}WVe\|_{L^2[-R,R]}\leq& \left(\frac{\|\nu\|_1}{\|\nu\|_2}\right)^4\left(\frac{8C C_r\sqrt{\lambda }}{(r-1)}\left(\frac{(1+\epsilon)R}{(\frac{3\epsilon R}{2})^{r-1}}+\frac{\epsilon R}{(\frac{\epsilon R}{2})^{r-1}}\right)+ \frac{24\epsilon RC^2_rC_{g}\Gamma_{r-\frac{1}{2}}}{\sqrt{2r-1}(\epsilon R)^{r-\frac{1}{2}}}\right.\\
 &  \left. +\frac{8C C_r\sqrt{\lambda }}{(r-1)}\left(\frac{(1+\epsilon)R}{(\frac{3\epsilon R}{2})^{r-1}}+\frac{\epsilon R}{(\frac{\epsilon R}{2})^{r-1}}\right)\right.\\
 &  \left.+ 16C_gCC^2_r\Gamma_{r-\frac{1}{2}}\left(\frac{(1+\epsilon)R}{\left(\frac{\epsilon R}{2}\right)^{r-1}}+\frac{12C}{\sqrt{2r-1}\left(\frac{\epsilon R}{2}\right)^{r-\frac{5}{2}}}\right)
       \right.\\
 &  \left.+\frac{4\sqrt{2}C C_r}{(r-1)}\left(\frac{(1+\epsilon)R}{(\frac{3\epsilon R}{2})^{r-1}}+\frac{\epsilon R}{(\frac{\epsilon R}{2})^{r-1}}\right)\right)  + 2\epsilon RCC_{g}\left(\frac{\|\nu\|_1}{\|\nu\|_2}\right)^2\frac{\alpha^2}{1-\alpha}
\end{split}\\
\begin{split} \label{paper150000}
    &\leq \left(\frac{\|\nu\|_1}{\|\nu\|_2}\right)^4\left(36\sqrt{2}CC_gC^2_r\Gamma_{r-\frac{1}{2}}\left(\frac{(1+\epsilon)R}{(\frac{\epsilon R}{2})^{r-1}}+\frac{12C}{\sqrt{2r-1}\left(\frac{\epsilon R}{2}\right)^{r-\frac{5}{2}}}\right)\right.\\
 &  \left.+\frac{24C^2_rC_{g}\Gamma_{r-\frac{1}{2}}}{\sqrt{2r-1}(\epsilon R)^{r-\frac{3}{2}}}\right)+2\epsilon R C C_g\left(\frac{\|\nu\|_1}{\|\nu\|_2}\right)^2\frac{\alpha^2}{1-\alpha}
\end{split}\\
\nonumber& \leq \left(\frac{\|\nu\|_1}{\|\nu\|_2}\right)^4\frac{C_r^\sharp}{(\frac{\epsilon R}{2})^{r-\frac{5}{2}}}\left(\frac{(1+\epsilon)R}{(\frac{\epsilon R}{2})^{\frac{3}{2}}}+\frac{12}{\sqrt{2r-1}}+\frac{1}{\sqrt{2r-1}\epsilon R}\right)\\
			\nonumber&+\left(\frac{\|\nu\|_1}{\|\nu\|_2}\right)^22\epsilon R C C_g\frac{\alpha^2}{1-\alpha},
 		\end{align}
 	where $C^\sharp_{r}=36\sqrt{2}C_gC^2C_r^2\Gamma_{r-\frac{1}{2}}$. To obtain \eqref{paper150000}, we used that $\frac{1}{r-1}\leq \frac{1}{\sqrt{2r-1}} \hspace{0.2cm} \forall \hspace{0.1cm} r\geq4$.
  Hence, by using the inequalities \eqref{paper1projectionerror} and \eqref{paper120}, we obtain the stated result.
\end{proof}
 
	We now consider the two quantization schemes separately.
	\subsection{$\Sigma\Delta$ quantization}
		Let $Q^7_{\Sigma\Delta}$ denote the stable seventh-order $\Sigma \Delta$  scheme from Proposition \ref{paper139}. Let $f \in C_{-\pi, \pi}$. Then, $\|y\|_{\infty}\leq 1$. Further, let $q:=Q^7_{\Sigma\Delta}(y)$, $H$ be the $\widetilde{m}\times \widetilde{m}$ matrix 
 $D^7$ and $V$ be the $p_3\times \widetilde{m}$ matrix $\tilde{V}_{\Sigma\Delta}$, then using Proposition $\ref{paper139}$ and \eqref{paper140} we get
		\begin{equation*}
			\|VH\|_{\infty \rightarrow 2}\|u\|_{\infty} \leq \sqrt{p_3}(8\cdot7)^{7+1}C(7)\left(\frac{\widetilde{m}}{p_3}\right)^{-7}\delta \leq \sqrt{p}(56)^{8}C(7)\left(\frac{m}{p}\right)^{-7}\delta.
		\end{equation*}
		In this case, $\nu= \nu_{\Sigma \Delta}$, and since $ \nu_{\Sigma \Delta} \in \mathbb{R}^{\frac{m}{p}}$, $\frac{\|\nu\|_1}{\|\nu\|_2} \leq \sqrt{\frac{m}{p}}$. Thus, uniformly for all $f \in C_{[-\pi,\pi]}$, 
		\begin{align}
			\label{paper173}\|f-F_{W\tilde{V}_{\Sigma\Delta}}q\|_{L^2[-R,R]}&\leq\frac{2\sqrt{2}C_r}{\sqrt{2r-1}(r-\frac{3}{2})(\frac{5}{2}\epsilon R)^{r-\frac{3}{2}}}\\
			\nonumber&+ \sqrt{\frac{m}{p}}\frac{1}{\sqrt{1-\gamma-3t}}\sqrt{\frac{2}{\frac{p}{2(1+\epsilon)R}-\frac{1}{\epsilon R}}}\sqrt{p}(56)^{8}C(7)\left(\frac{m}{p}\right)^{-7}\delta\\
			\nonumber&+ \left(\frac{m}{p}\right)^2\frac{C_r^\sharp}{(\frac{\epsilon R}{2})^{r-\frac{5}{2}}}\left(\frac{(1+\epsilon)R}{(\frac{\epsilon R}{2})^{\frac{3}{2}}}+\frac{12}{\sqrt{2r-1}}+\frac{1}{\sqrt{2r-1}\epsilon R}\right)\\
			\nonumber&+2\epsilon R C C_g\left(\frac{m}{p}\right)\frac{\alpha^2}{1-\alpha}
		\end{align}
		with probability greater than
		$1-6\exp\left(-\frac{m\epsilon}{12(1+3\epsilon)R}\right)-5p\exp\left(-\frac{t^2p}{42mC^2}\left(\frac{p}{2(1+3\epsilon)R}-\frac{1}{\epsilon R}\right)\right)$.
        \begin{theorem} \label{paper16001}
		Let $R>1,\delta>0$ be real numbers, and $L$ be a positive integer. Assume that $\{x_i\}_{i=1}^m$ is a sequence of i.i.d random variables that are uniformly distributed on $[-R-3m^{\frac{1}{16}},R+3m^{\frac{1}{16}}]$. Let $\mathcal{Q}^7_{\Sigma\Delta}$ be the stable seventh-order $\Sigma\Delta$ quantization scheme from Proposition \ref{paper139}. If $m^{\frac{15}{16}}$  is an integer such that it divides $m$, $7$ divides $m^{\frac{1}{16}}-1$  and $m$ is sufficiently large, then with probability greater than 
		\[1-17m^{\frac{15}{16}}\exp\left(-\frac{m^{\frac{5}{16}}}{d_2R}\right)\]
		we have 
		\begin{equation}
			\|f-F_{W\tilde{V}_{\Sigma \Delta}}q\|_{L^2[-R,R]}\leq \frac{d_1R}{m^{\frac{3}{8}}} \hspace{0.2cm} \forall \hspace{0.1cm} f \in C_{[-\pi,\pi]},
		\end{equation}
		where $F_{W\tilde{V}_{\Sigma \Delta}}$ is as defined in \eqref{paper145}, $q:=Q^7_{\Sigma\Delta}(y)$, $y$ is as defined in \eqref{paper164}, and $d_1$ and $d_2$ are positive  constants given in the proof. 
	\end{theorem}
        \begin{proof}
        Take $\epsilon=\frac{m^{\frac{1}{16}}}{R},r=11, t=\frac{1}{12}\left(\frac{1}{m}\right)^{\frac{1}{4}}$ and $\gamma=\frac{1}{8}\left(\frac{1}{m}\right)^{\frac{1}{4}}$. Choose $p=\frac{m}{m^{\frac{1}{16}}}=m^{\frac{15}{16}}$, this causes no issues as we have already assumed all of the necessary conditions for $m$ in the theorem statement. With the above-mentioned choices, we can conclude the following. 
   \begin{equation}\label{paper112001} 
       \frac{\alpha^2}{1-\alpha}< \frac{(6t+\gamma)^2}{1-(3t+\gamma)}\leq \frac{5}{8}\left(\frac{1}{m}\right)^{\frac{1}{2}},\frac{1}{\sqrt{1-3t-\gamma}}\leq \sqrt{\frac{8}{5}} \text{ and }  C=\sqrt{3}.
   \end{equation}
   \begin{equation} \label{paper160001}
       1+3\epsilon=1+\frac{3m^{\frac{1}{16}}}{R}\leq 4m^{\frac{1}{16}} \text{ and similarly } 1+\epsilon \leq 2m^{\frac{1}{16}}.
   \end{equation}
    If $m^{\frac{15}{16}}\geq \frac{4R}{m^{\frac{1}{16}}}+12$, then $p\geq \frac{4}{\epsilon}+12,$ which in turn implies that
   \begin{equation}\label{paper112002} 
       \sqrt{\frac{1}{\frac{p}{(1+\epsilon)R}-\frac{1}{\epsilon R}}}\leq \sqrt{\frac{4(1+3\epsilon)R}{p}}\leq 2m^{\frac{1}{32}}\sqrt{\frac{R}{p}}.
   \end{equation}
   \begin{equation}\label{paper112003}
          m\geq 2^{\frac{80}{3}}\left(\frac{16C_{11}}{\sqrt{210}}\right)^{\frac{8}{3}} \text{ implies } \epsilon R\geq 2\left (\frac{2C_r}{\sqrt{(2r-1)(r-1)}\gamma}\right)^{\frac{1}{r-1}}.
   \end{equation}
		Thus, using \eqref{paper173}, we get that if $m\geq 2^{\frac{80}{3}}\left(\frac{16C_{11}}{\sqrt{210}}\right)^{\frac{8}{3}}$ and $m^{\frac{15}{16}}\geq \frac{4R}{m^{\frac{1}{16}}}+12$, then, uniformly over all $f\in C_{[-\pi,\pi]}$, 
		\begin{align*}
			\|f-F_{W\tilde{V}_{\Sigma\Delta}}q\|_{L^2[-R,R]}&\leq \frac{C_{11}}{m^{\frac{19}{32}}}+\frac{16}{\sqrt{5}}\frac{(56)^{8}C(7)\sqrt{R}}{m^\frac{6}{16}}\delta\\&+ \frac{2^{\frac{17}{2}}C_{11}^\sharp}{m^{\frac{13}{32}}}\left(\frac{4\sqrt{2}R}{m^{\frac{1}{32}}}+\frac{12}{\sqrt{21}}+\frac{1}{\sqrt{21}m^{\frac{1}{16}}}\right)+\frac{5\sqrt{3}C_g}{4m^{\frac{3}{8}}}
		\end{align*}
		with probability $1-6\exp{\left(-\frac{m^{\frac{17}{16}}}{12R\left(1+3m^{\frac{1}{16}}\right)}\right)}-5em^{\frac{15}{16}}\exp\left(-\frac{m^{\frac{5}{16}}}{145152R}\right)$. That is, for sufficiently large $m$, \[\|f-F_{W\tilde{V}_{\Sigma \Delta}}q\|_{L^2[-R,R]}\leq \frac{d_1R}{m^{\frac{3}{8}}} \] uniformly over all $f\in C_{[-\pi,\pi]}$
 with probability greater than $1-17m^{\frac{15}{16}}\exp\left(-\frac{m^{\frac{5}{16}}}{d_2R}\right),$ where 
  $d_2=145152$ and \[d_1=4\cdot\max\bigg\{C_{11},\frac{16}{\sqrt{5}}(56)^{8}C(7)\delta,\left(2^{11}+\frac{2^{\frac{17}{2}}13}{\sqrt{21}}\right)C_{11}^\sharp ,\frac{5\sqrt{3}C_g}{4}\bigg\} .\]
	\end{proof}
 \begin{remark}\label{paper110003}
     If we assume that $2^7+\frac{1}{\delta}\leq 2L$, then selecting $H=D^7$  in Lemma \ref{paper150}  also yields a stable $\Sigma \Delta$ quantizer of order $7$.  The preceding theorem holds true for this quantizer as well; only the constant $d_1$ changes.
 \end{remark}
	\subsection{Distributed noise-shaping quantization}
	Let $L$ be a positive integer and $\delta>0$. Fix $\beta \in (1,2L-\frac{1}{\delta})$. Let $f \in C_{[-\pi,\pi]}$. Then $\|y\|_{\infty}\leq 1$. Let $H$ be the $\widetilde{m}\times \widetilde{m}$ matrix 
$H_\beta$ and $V$ be the $p_3\times \widetilde{m}$ matrix $\tilde{V}_\beta$. Further, let $Q_\beta$ be the distributed noise-shaping quantizer from  Lemma \ref{paper150} and $q:=Q_\beta(y)$. Then, using \eqref{paper141} and Lemma \ref{paper150}, we get
	\begin{align*}
		\|VH\|_{\infty \rightarrow 2}\|u\|_{\infty}\leq\sqrt{p_3}\beta^{-\frac{\widetilde{m}}{p_3}+1}\delta\leq \sqrt{p}\beta^{-\frac{m}{p}+1}\delta.
	\end{align*}
	As in this case $\nu=\nu_\beta=[\beta^{-1} \beta^{-2}\cdots\beta^{-\frac{m}{p}}]$, it can be calculated that $\frac{\|\nu\|_1}{\|\nu\|_2}\leq \sqrt{\frac{\beta+1}{\beta-1}}$. Thus, uniformly over all $f\in C_{[-\pi,\pi]}$, 
	\begin{align}
		\label{paper174}\|f-F_{W\tilde{V}_{\beta}}q\|_{L^2[-R,R]}&\leq\frac{2\sqrt{2}C_r}{\sqrt{2r-1}(r-\frac{3}{2})(\frac{5}{2}\epsilon R)^{r-\frac{3}{2}}}\\
		\nonumber&+ \sqrt{\frac{\beta+1}{\beta-1}}\frac{1}{\sqrt{1-\gamma-3t}}\sqrt{\frac{2}{\frac{p}{2(1+\epsilon)R}-\frac{1}{\epsilon R}}}\sqrt{p}\beta^{-\frac{m}{p}+1}\delta\\
		\nonumber&+ \left(\frac{\beta+1}{\beta-1}\right)^2\frac{C_r^\sharp}{(\frac{\epsilon R}{2})^{r-\frac{5}{2}}}\left(\frac{(1+\epsilon)R}{(\frac{\epsilon R}{2})^{\frac{3}{2}}}+\frac{12}{\sqrt{2r-1}}+\frac{1}{\sqrt{2r-1}\epsilon R}\right)\\
		\nonumber&+2\epsilon R C C_g\left(\frac{\beta+1}{\beta-1}\right)\frac{\alpha^2}{1-\alpha}
	\end{align}
	with probability greater than
	$1-6\exp\left(-\frac{m\epsilon}{12(1+3\epsilon)R}\right)-5p\exp\left(-\frac{t^2(\beta-1)}{42(\beta+1)C^2}\left(\frac{p}{2(1+3\epsilon)R}-\frac{1}{\epsilon R}\right)\right)$.

 \begin{remark}\label{paper1alltypesoferrors}
        
        As can be seen, the error  \(\|f - F_{W\tilde{V}_{\beta}} q\|_{L^2[-R,R]}\) for distributed noise-shaping quantization, see \eqref{paper174} (and \(\|f - F_{W\tilde{V}_{\Sigma\Delta}} q\|_{L^2[-R,R]}\) for $\Sigma \Delta$ quantization, see \eqref{paper173}), is bounded by four terms:  
\begin{enumerate}
\item The first term, \(\|f - \tilde{f}\|_{L^2[-R,R]}\), quantifies the approximation error in the finite-dimensional approximation subspace. This decays polynomially in \(\epsilon R\) and hence decreases rapidly for large \(\epsilon R\).This happens when the sampling interval is significantly larger than the reconstruction interval.  

\item The second term decays exponentially (polynomially for $\Sigma \Delta$ quantization) in the oversampling rate relative to the number of frame vectors, that is, in \(\frac{m}{p}\).  

\item The third and fourth terms (in the case of distributed noise-shaping quantization as well as in the case of $\Sigma \Delta$ quantization) account for errors introduced when approximating projected function samples with function samples. 

The third term can be significantly reduced by increasing \(\epsilon R\). 

The decay of the fourth term is governed by \(\alpha\), which depends on the tightness of the frame bounds of \(\{h_j\}_{j=1}^{p_3}\) (see \eqref{paper1WVPhi=h}). Reducing \(\alpha\) (as defined in \eqref{paper110002}) requires tightening the frame bounds. However, since \(\alpha := \frac{6t+\gamma}{1+3t}\) and the bound on \(\|f - F_{W\tilde{V}_{\Sigma\Delta}} q\|_{L^2[-R,R]}\) holds with probability greater than  

   \[
   1 - 6\exp\left(-\frac{m\epsilon}{12(1+3\epsilon)R}\right) - 5p\exp\left(-\frac{t^2p}{42mC^2} \left(\frac{p}{2(1+3\epsilon)R} - \frac{1}{\epsilon R}\right)\right),
   \]  reducing \(\alpha\) by tightening the frame bounds (that is, decreasing \(t\)) requires an exponentially larger number of samples in terms of the oversampling rate.   
\end{enumerate}
Taking these limitations into account, we carefully choose constants in the subsequent theorem (and in Theorem \ref{paper16001} for the case of $\Sigma \Delta$ quantization) to balance the decay rates of all four error terms. 
            
        \end{remark}
        
        \begin{theorem}\label{paper12001}
		Let $R>1, \delta>0$ be real numbers, and $L$ be a positive integer. Fix $\beta \in (1,2L-\frac{1}{\delta})$. Assume that $\{x_i\}_{i=1}^m$ is a sequence of i.i.d random variables that are uniformly distributed on $[-R-3m^{\frac{1}{16}},R+3m^{\frac{1}{16}}]$. Let $\mathcal{Q}_\beta$ be the quantization scheme from Lemma \ref{paper150}. If $m^{\frac{15}{16}}$  is an integer such that it divides $m$ and $m$ is sufficiently large, then with probability greater than 
		\[1-17m^{\frac{15}{16}}\exp\left(-\frac{m^{\frac{3}{8}}}{d_2R}\right)\]
		we have 
		\begin{equation}
			\|f-F_{W\tilde{V}_{\beta}}q\|_{L^2[-R,R]}\leq \frac{d_1R}{m^{\frac{7}{16}}} \hspace{0.2cm} \forall \hspace{0.1cm} f \in C_{[-\pi,\pi]},
		\end{equation}
		where $F_{W\tilde{V}_{\beta}}$ is as defined in \eqref{paper145}, $q:=Q_{\beta}(y)$, $y$ is as defined in \eqref{paper164}, and $d_1$ and $d_2$ are positive constants given in the proof.
	\end{theorem}
	\begin{proof}
		Take $\epsilon, t$,$\gamma$, $p$ and $r$ as in the proof of Theorem \ref{paper16001}. Then \eqref{paper112001},\eqref{paper160001},\eqref{paper112002} and \eqref{paper112003} are still true and in addition we have that if $m\geq (\frac{15}{32\ln\beta}\ln m)^{16}$, then $\beta^{-m^{\frac{1}{16}}}\leq \frac{1}{m^{\frac{15}{32}}}$.
  
		Thus, using \eqref{paper174}, we get that for sufficiently large $m$, \[\|f-F_{W\tilde{V}_\beta}q\|_{L^2[-R,R]}\leq \frac{d_1R}{m^{\frac{7}{16}}}\] uniformly over all $f\in C_{[-\pi,\pi]}$ with probability greater than $1-17m^{\frac{15}{16}}\exp\left(-\frac{m^{\frac{3}{8}}}{d_2R}\right),$ where $d_2=145152\left(\frac{\beta+1}{\beta-1}\right)$ and \[d_1=4\cdot\max\bigg\{C_{11},\sqrt{\frac{\beta+1}{\beta-1}}\frac{16}{\sqrt{5}}\beta\delta,\left(\frac{\beta+1}{\beta-1}\right)^2\left(2^{11}+\frac{2^{\frac{17}{2}}13}{\sqrt{21}}\right)C_{11}^\sharp,\left(\frac{\beta+1}{\beta-1}\right)\frac{5\sqrt{3}C_g}{4}\bigg\}.\] 
	\end{proof}
        \section{Numerical experiments}\label{paper113001}
In order to test the accuracy of our main results, we run numerical experiments. Let $w$ be defined as 
\[ w(\xi)=\begin{cases} 
      e^{-\frac{1}{\xi}} & \xi> 0, \\
      0 & \xi\leq0.
   \end{cases}\] 
For $g$ we chose the following function defined via the Fourier transform \[ \hat{g}(\xi)=\begin{cases} 
      \frac{1}{\sqrt{2\lambda \pi}} & |\xi|\leq \pi, \\
      \frac{1}{\sqrt{2\lambda \pi}}\cos\left(\frac{\pi}{2}\nu\left(\frac{{|\xi|}-\pi}{(2\lambda -2)\pi}\right)\right)  & \pi < |\xi|\leq (2\lambda-1)\pi, \\
      0 & (2\lambda-1)\pi< |\xi|. 
   \end{cases}\] 
   Here $\nu(\xi):=\frac{w(\xi)}{w(\xi)+w(1-\xi)} \hspace{0.2cm} \forall \hspace{0.1cm} \xi \in \mathbb{R}$ and $\lambda=2$. Further, we fix the parameters $\epsilon=\frac{1}{2}$ and $R=5$. As a result: the sampling interval  $[-(1+3\epsilon)R,(1+3\epsilon)R]$ reduces to $\left[-\frac{25}{2},\frac{25}{2}\right]$, each sample $x_i$ is sampled according to the uniform distribution on $\left[-\frac{25}{2},\frac{25}{2}\right]$ and the finite dimensional approximation space is $V^{\left(1+\frac{5}{2}\epsilon\right)R}(g)=V^{\frac{45}{4}}(g)$ having a dimension of $45$ by  definition and hence the effective oversampling rate is $\frac{m}{45}$. 
\begin{figure}
    \centering
    \includegraphics[width=17cm, height=10cm]{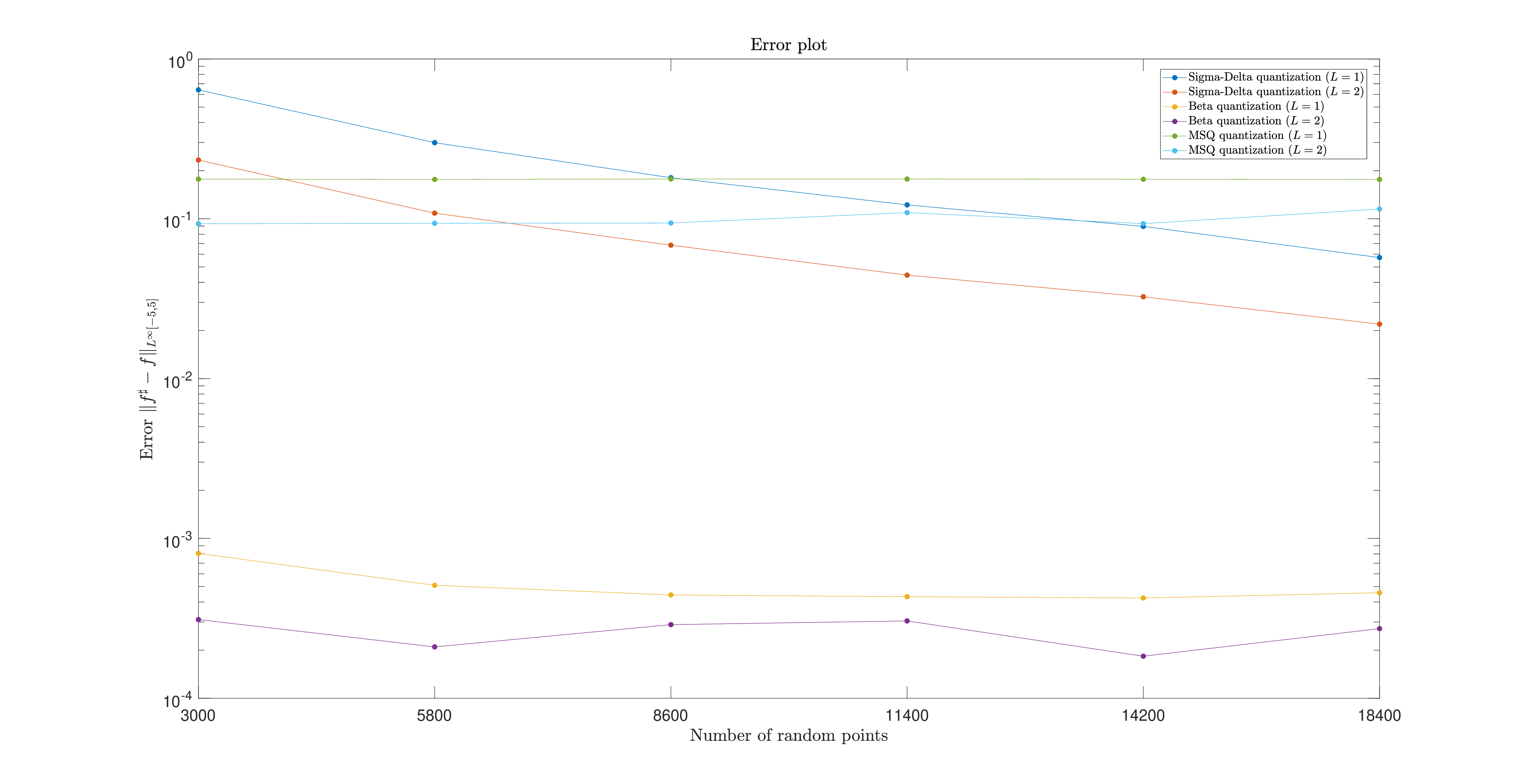}
\caption{ We plot the reconstruction error along with the sample size $m$ for $\Sigma\Delta$, $\beta$ and MSQ schemes. The value of $p$ is fixed as $200$.}
    \label{fig:Fig2}
\end{figure}

\smallskip
 The set of numerical experiments is conducted twice: once with an alphabet of cardinality 2 (that is, \( L = 1 \)) and once with an alphabet of cardinality 4 (that is, \( L = 2 \)). In each case, we randomly select 10 functions from the class \( PW_{[-\pi, \pi]} \), with a maximum amplitude of 0.35. For each randomly selected function \( f \), we apply our reconstruction method. Specifically, we sample \( f \) over the interval \( \left[-\frac{25}{2}, \frac{25}{2}\right] \), quantize the (randomly flipped) samples, and then reconstruct the function from its quantized measurements using the formula \( f^\sharp = F_{WV} q \). This process of sampling, quantization, and reconstruction is repeated 20 times for each function \( f \), and the average reconstruction error across these 20 iterations is calculated. Finally, we compute the average of the reconstruction errors across all 10 randomly chosen functions. The results, displayed in Figure \ref{fig:Fig2}, show the average error \( \left\| f - f^{\sharp} \right\|_{L^\infty[-5,5]} \) over all 200 iterations.

\smallskip

For the \( \Sigma \Delta \) quantization, the quantization alphabet is selected as \( \mathcal{A}^{1}_{1} \) (defined in equation \eqref{paper120001}) for \( L = 1 \) and \( \mathcal{A}^{\frac{1}{3}}_{2} \) for \( L = 2 \). The greedy quantizer, as outlined in Lemma \ref{paper150}, is then used to quantize the randomly flipped samples \( y \) (see equation \eqref{paper164}). The noise transfer operator is taken as the matrix \( H: u \mapsto h \ast u \) for all \( u \in \mathbb{R}^m \), where \( h \) arises from applying a rounding procedure on the seventh-order filter described by Deift et al. in \cite{paper17}.

\smallskip
For \( \beta \)-quantization, the greedy quantizer defined in Lemma \ref{paper150} is applied to quantize the randomly flipped samples. The noise transfer operator is chosen as \( H_\beta = H_{1.64} \), corresponding to \( \beta = 1.64 \) for both \( L = 1 \) and \( L = 2 \). The quantization alphabet is selected as \( \mathcal{A}^{1}_{1} \) for \( L = 1 \) and \( \mathcal{A}^{\frac{1}{3}}_{2} \) for \( L = 2 \).

\smallskip
Unlike the noise-shaping quantization approaches discussed in this study, where the samples must be partitioned into three collections due to the presence of the block diagonal matrix \( V \), no such partitioning is necessary in the MSQ scenario, as \( p = m \) (which implies that \( V \) is the identity matrix). Furthermore, the collection of symmetric Bernoulli random variables is not required. 

For MSQ, we define the quantization alphabet as follows:
\[
\mathcal{A}_L = \left\{ \left( -1 + \frac{2n+1}{2L} \right) \|y\|_{\infty}: n \in \{0, \dots, 2L-1\} \right\},
\]
which, for the specific case of \( \|y\|_{\infty} = 0.35 \), becomes
\begin{equation}\label{paper130001}
\mathcal{A}_L = \left\{ \left( -1 + \frac{2n+1}{2L} \right) 0.35 : n \in \{0, \dots, 2L-1\} \right\}.
\end{equation}
For each sample \( x_i \), the quantized value \( q^{MSQ}_i \) is then determined as
\[
q^{MSQ}_i := \arg \min_{r \in \mathcal{A}_L} |f(x_i) - r| \quad \forall \, i \in \{1, \dots, m\}.
\]
As outlined in the introduction, we apply a linear method for MSQ data: the function samples are first quantized using the memoryless scalar quantizer, and then the canonical dual frame is applied to these quantized samples for reconstruction.

The quantization alphabet for MSQ is chosen as \( \mathcal{A}_1 \) and \( \mathcal{A}_2 \) (defined in \eqref{paper130001}) for \( L = 1 \) and \( L = 2 \), respectively.

\smallskip
The following conclusions can be drawn from Figure \ref{fig:Fig2}. Consistent with our theoretical results, the \( \beta \)-scheme performs the best, followed by the \( \Sigma\Delta \)-scheme in second place, and the MSQ scheme exhibits the worst performance. As expected, the error for the MSQ scheme does not decrease with an increasing sample size. The \( \beta \)-scheme demonstrates a behavior in which the error does not continue to decay beyond a certain point as the number of samples increases. One possible explanation for this lack of continued error decay is that the theorem requires an increase in the sampling interval as the sample size grows. However, due to computational difficulties, we do not implement this adjustment, which may limit the potential for further error reduction. There may be inherent limitations arising from the fact that the projection error \( \|f - Pf\|_{L^2[-5, 5]} \) does not decrease with an increasing number of samples. 

It is also possible that the error would continue to decrease, albeit more slowly, with an increasing sample size, but computational constraints prevent us from using a larger number of samples. 

Additionally, the \( \beta \)-scheme performs significantly better in terms of computational speed when compared to both the MSQ and \( \Sigma\Delta \) schemes. This improved performance can be attributed to the fact that in the MSQ scheme, much larger frames are used, requiring more computations to calculate the dual frame. In the \( \Sigma\Delta \) scheme, the function samples must be quantized using a seventh-order \( \Sigma\Delta \) quantizer, which is computationally intensive.
\section{Concluding remarks} \label{paper140000}
       In this paper, we present the first error analysis for the reconstruction of bandlimited signals from coarsely quantized random samples with random sign flips, considering both the $\Sigma \Delta$ and distributed noise-shaping quantization approaches. Our theoretical findings suggest an advantage for the latter approach, which was also observed in our numerical experiments. Furthermore, in our numerical experiments, we investigated the performance of the MSQ scheme for our method and found that it does not perform well, as the reconstruction error does not decrease with an increasing number of samples. Hence, in this paper, we have discussed the performance of our proposed method for all three popular quantization schemes in the literature. Next, we discuss some important remarks regarding our method.
        \begin{itemize}
        \item  In real-life scenarios, it is very unlikely that measurements of a signal can be taken on a grid spread throughout the real line (which is the case considered in \cite{paper110}). Hence, our paper assumes the more realistic scenario in which samples can be taken over a bounded interval. The issue, however, is that the boundedness of the interval, along with the fact that the samples are uniformly distributed, creates a specific problem related to inaccurate samples from the boundaries (as explained in Section \ref{paper1SQRpipeline}), a problem that does not arise in the case considered by the authors in \cite{paper110} due their more benign sampling approach. Our entire paper primarily focuses on how to deal with these inaccuracies during the reconstruction process. \\
         \item 
            As can be seen from \eqref{paper120}, the error between the projected function and the reconstructed function from the quantized measurements can be bounded by two terms. The first term is the error from the quantization process, and the second term is the error that arises from projecting the bandlimited function into the approximation space and the subsequent error caused by the approximation of the samples. As observed, the second term will generally be much larger than the first term. The small magnitude of the quantization term is due to the presence of the matrix \( V \), but this also introduces \( V \) into the second term. As noted earlier, this effectively takes a linear combination of the samples \( \frac{m}{p} \) at a time. This destroys the localization property of the frame elements, as they are now linear combinations of the reproducing kernels. Without \( V \), the frame elements would simply be the reproducing kernels at the sample points, which are highly localized due to the decay properties (see Lemma \ref{paper125}) satisfied by our kernel. The entire process of partitioning the samples into three collections is done to address the issue caused by the addition of \( V \). Therefore, it is clear that there is a trade-off between the two terms in the error. The addition of \( V \) helps reduce the first term but increases the second term by destroying its localization properties. Hence, the parameters (such as the order of quantization in the \( \Sigma \Delta \) scheme) are chosen so that the errors in both terms match.\\

\item  
The problem of determining lower bounds for the reconstruction error is significantly more complex than finding upper bounds, and has only been addressed in a few studies. The first lower bounds for uniform sampling and coarse quantization of bandlimited signals were introduced in \cite{paper1krahmeracha2012}. Additionally, \cite{paper1chouanha2015} discusses lower bounds on analysis distortion for a fixed frame and given alphabet length. In both cases, the lower bounds exhibit exponential dependence on the oversampling rate. However, unlike these studies, our work must address the unique challenge of dealing with inaccurate samples. Consequently, a direct comparison with the aforementioned lower bounds is not feasible.

The key consideration in our work is that we operate with a finite number of uniform random samples over an interval while dealing with functions in an infinite-dimensional space. The first step in our approach, thus,  involves projection onto a finite-dimensional space, which inevitably introduces projection error, a factor that can only be reduced by increasing the sampling interval. The second point is that, in the projected space, we can attempt to recover the projected function with increasing accuracy by increasing the number of samples. However, note that as we take more samples, the inaccuracy in the sample approximations leads to the inclusion of more error in our calculations, which must be carefully managed.

Let \( f \in PW_{[-\pi,\pi]} \). The error between the projected function and the reconstructed function is given by:
\[
\| \tilde{f} - F_{WV} q \|_{L^2[-R,R]} = \| F_{WV} H u + F_{VW} e \|_{L^2[-R,R]}.
\]
As discussed in previous remarks, the first term, which arises from quantization, can decay rapidly, and it can be made significantly smaller than the second term, which arises from sample inaccuracy. Therefore, in addressing the question of the lower bound, it is appropriate to focus on bounding \( \| F_{VW} e \|_{L^2[-R,R]} \). Since \( F_{VW} e := \sum_{j=1}^{p_3} (WVe)_j S^{-1} h_j \) is a function of random variables, determining a lower bound for \( \| F_{VW} e \|_{L^2[-R,R]} \) requires understanding its concentration behavior as \( p \) and \( m \) increase. However, this is a challenging task due to the lack of an explicit formula for \( \{ S^{-1} h_j \}_{j=1}^{p_3} \), and thus this problem remains unresolved in this work and is left for future investigation.

Furthermore, as noted earlier, we expect that increasing the number of partitions in the sampling interval will further reduce \( \| F_{VW} e \|_{L^2[-R,R]} \), thereby improving the lower bound. As such, we defer a detailed analysis of the lower bound to future work, once we refine the reconstruction error by increasing the number of partitions.\\

\item  
Our method of partitioning the samples into three bins prevents the \( \Sigma \Delta \) scheme from operating in an online manner. This occurs because, in our method, after the samples are partitioned into three parts, they are concatenated into a single vector (see \eqref{paper1rearrangementofsamples}) by placing them sequentially. Consequently, the quantization scheme processes the concatenated vector from the first to the last sample. As a result, there is a non-zero probability that the first sample in the second partition may have been collected before the last sample in the first partition, meaning the last sample in the first partition will be quantized before the first sample in the second partition. Therefore, this process is not strictly online. However, we note that the process is approximately online. This approximation holds because, if \( I_{2\epsilon} \) and \( I_{3\epsilon} \) are small compared to \( I_{1\epsilon} \), as would typically be the case in practical applications, then the first partition contains most of the samples. Since the samples from the first partition are placed first in the concatenated vector (see \eqref{paper1concatenatedvector}), they are quantized first. Therefore, the non-online aspect of the quantization procedure is largely confined to the second and third partitions, which, in such cases, will be very small. Nevertheless, since the process is not strictly online, this is one of the reasons why we suggest that the \( \Sigma \Delta \) quantization scheme should not be the preferred method for our approach.

\smallskip

However, this issue does not arise with the \( \beta \) quantization scheme. Unlike the \( \Sigma \Delta \) quantization scheme, the \( \beta \) scheme can be viewed as running \( p_3 \) parallel quantizers simultaneously, allowing the samples in the different partitions to be quantized in parallel. Thus, although the quantization procedure is described using a concatenated vector that includes all three partitions, the structure of the noise-transfer operator \( H_{\beta} \) ensures that the quantization process is carried out separately on each partition without any interconnection between them. Let \( f \in C_{[-\pi,\pi]} \) and \( y \) be as defined in \eqref{paper164}. Then, from \( y - q = Hu \), we get:
\begin{equation} 
		\begin{pmatrix}
				{\epsilon^1_1f(y^1_1)}   \\ 	
				\vdots &   \\
         {\epsilon^1_{\widetilde{m}_1}f(y^1_{\widetilde{m}_1})}\vspace{0.1cm}\\ 
                {\epsilon^2_1f(y^2_1)}   \\ 	
				\vdots &   \\
                {\epsilon^2_{\widetilde{m}_2}f(y^2_{\widetilde{m}_2})}\vspace{0.1cm}\\
				{\epsilon^3_1f(y^3_1)}   \\ 	
				\vdots &   \\
                    {\epsilon^3_{\widetilde{m}_3}f(y^3_{\widetilde{m}_3})}
			\end{pmatrix}-  \begin{pmatrix}
				{q_1}   \\ 
    {q_2}  \\
				\vdots &   \\
         \vdots &   \\
               \vdots &      \\ 	
				\vdots &   \\	
				\vdots &   \\
    {q_{\widetilde{m}-1}}\\  
                    {q_{\widetilde{m}}}
			\end{pmatrix}=\begin{pmatrix}
			\tilde{H}_\beta & & &  &  & & & &\\
			& \tilde{H}_\beta  &  &  & & & & &\\
			&   & \tilde{H}_\beta &  & & & & &\\
			&  &  & \tilde{H}_\beta	& & & & &\\
			&  &  & & \ddots & & & &\\
                &  &  & &  &\tilde{H}_\beta & & &\\
& &  &  &  & & \tilde{H}_\beta & &\\
& &  &  &  & &  & \tilde{H}_\beta &\\
& &  &  &  & & & & \tilde{H}_\beta \\ 
		\end{pmatrix}\begin{pmatrix}
				{u_1}   \\ 
    {u_2}  \\
				\vdots &   \\
         \vdots &   \\
               \vdots &      \\ 	
				\vdots &   \\	
				\vdots &   \\
    {u_{\widetilde{m}-1}}\\  
                    {u_{\widetilde{m}}}
			\end{pmatrix}.
	\end{equation}

The above matrix equation is equivalent to the following three matrix equations because of the structure of $H_\beta$.
\begin{equation*} 
		\begin{pmatrix}
				{\epsilon^1_1f(y^1_1)}   \\ 	
				\vdots &   \\
         {\epsilon^1_{\widetilde{m}_1}f(y^1_{\widetilde{m}_1})}
			\end{pmatrix}-  \begin{pmatrix}
				{q_1}   \\ 
				\vdots &   \\
            {q_{\widetilde{m}_1}}
			\end{pmatrix}=\begin{pmatrix}
			\tilde{H}_\beta & &\\
			& \ddots  & \\
			&   & \tilde{H}_\beta 
		\end{pmatrix}\begin{pmatrix}
				{u_1}   \\
				\vdots &   \\
                    {u_{\widetilde{m}_1}}
			\end{pmatrix},
	\end{equation*}
\begin{equation*} 
		\begin{pmatrix}
				{\epsilon^2_1f(y^2_1)}   \\ 	
				\vdots &   \\
         {\epsilon^2_{\widetilde{m}_2}f(y^2_{\widetilde{m}_2})}
			\end{pmatrix}-  \begin{pmatrix}
				q_{{\widetilde{m}_1+1}}   \\ 
				\vdots &   \\
            {q_{\widetilde{m}_1+\widetilde{m}_2}}
			\end{pmatrix}=\begin{pmatrix}
			\tilde{H}_\beta & &\\
			& \ddots  & \\
			&   & \tilde{H}_\beta 
		\end{pmatrix}\begin{pmatrix}
				{u_{\widetilde{m}_1}+1}   \\
				\vdots &   \\
                    {u_{\widetilde{m}_1+\widetilde{m}_2}}
			\end{pmatrix},
	\end{equation*} 
and 
\begin{equation*} 
		\begin{pmatrix}
				{\epsilon^3_1f(y^3_1)}   \\ 	
				\vdots &   \\
         {\epsilon^3_{\widetilde{m}_3}f(y^3_{\widetilde{m}_3})}
			\end{pmatrix}-  \begin{pmatrix}
				q_{\widetilde{m}_1+\widetilde{m}_2+1}   \\ 
				\vdots &   \\
            {q_{\widetilde{m}}}
			\end{pmatrix}=\begin{pmatrix}
			\tilde{H}_\beta & &\\
			& \ddots  & \\
			&   & \tilde{H}_\beta 
		\end{pmatrix}\begin{pmatrix}
				{u_{\widetilde{m}_1+\widetilde{m}_2}+1}   \\
				\vdots &   \\
                    {u_{\widetilde{m}}}
			\end{pmatrix}.
	\end{equation*} 
This basically means that as soon as a random sample is added to a  partition it can be quantized with respect to the samples already in that partition. That is, we don't have to wait to quantize the samples in the first partition before  moving on to quantizing the second partition and then the third partition. This makes are quantization procedure online. Hence, $\beta$ quantization is specially suited to our method because both involve partitioning the samples.
\end{itemize}
\bibliographystyle{amsplain}
\bibliography{references}
\section{Appendix}\label{appendix}	
  \begin{proof}[Proof of Lemma \ref{paper123}]
                Let $y>0$ and $p>1$. Then, using the integral test we have the following bound for the $p$ series.
            \begin{equation} \label{paper11000}
                \sum_{n \in \mathbb{Z}, n \geq 1}\frac{1}{(y+n)^p}\leq \frac{1}{(p-1)y^{p-1}}
            \end{equation}
            Consequently,
            \begin{align*}
                \sum_{k \in \mathbb{Z}}\frac{1}{\left(1+\big|x-\frac{k}{2 }\big|\right)^r}
                &=\sum_{k \in \mathbb{Z}}\frac{2^r}{\left(2 +|2 x-k|\right)^r}\\
                &=\left(\frac{2}{2+\{2 x\}}\right)^r +\sum_{k\geq1, k \in \mathbb{Z}}\left(\frac{2}{2+\{2 x\}+k}\right)^r+\sum_{k\geq1, k \in \mathbb{Z}}\left(\frac{2}{2+k-\{2 x\}}\right)^r\\
                &\leq 1+\frac{2}{r-1}+\frac{2}{r-1}\left(2\right)^{r-1}.
            \end{align*}
        Here  $\{2 x\}:=2 x - \left \lfloor 2 x \right \rfloor $ denotes the fractional part of $2 x$.
        \end{proof}
        Given a $\gamma \in (0,1)$, the following lemma states that if $\epsilon R$ is sufficiently large, then all functions $f \in V^{\left(1+\frac{5}{2}\epsilon\right)R}(g)$ are concentrated in the interval $[-(1+3\epsilon)R,(1+3\epsilon)R]$ up to a factor of $1-\gamma$. 
 \begin{proof}[Proof of Lemma \ref{paper161}]
            It is enough to show that
            \begin{equation*}
                \int_{[-(1+3\epsilon)R,(1+3\epsilon)R]^{{\mathsf{c}}}}|f(x)|^2dx\leq  \gamma \hspace{0.2cm} \forall \hspace{0.1cm} f \in V^{\left(1+\frac{5}{2}\epsilon\right)R}(g) \text{ satisfying } \|f\|=1.
            \end{equation*}
        Let $f \in V^{\left(1+\frac{5}{2}\epsilon\right)R}(g)$ be such that $\|f\|=1$. Then $f= \sum_{k \in \left[2  \left(1+\frac{5}{2}\epsilon\right)R\right]}c_k g\left(\cdot-\frac{k}{2 }\right)$ for some $\{c_k\}_{k \in \left[2  \left(1+\frac{5}{2}\epsilon\right)R\right]}$ with $\left(\sum_{k \in \left[2 \left(1+\frac{5}{2}\epsilon\right)R\right]}|c_k|^2\right)^{\frac{1}{2}}=1$.

        Hence,
	\begin{align*}
		&\|f\|_{L^2[-(1+3\epsilon)R,(1+3\epsilon)R]^{{\mathsf{c}}}}\\&\hspace{1cm}\leq \left(\sum_{k \in \left[2 \left(1+\frac{5}{2}\epsilon\right)R\right]}|c_k|^2\right)^{\frac{1}{2}}\left(\sum_{k \in \left[2 \left(1+\frac{5}{2}\epsilon\right)R\right]}\left\|g\left(\cdot-\frac{k}{2}\right)\right\|^2_{L^2[-(1+3\epsilon)R,(1+3\epsilon)R]^{{\mathsf{c}}}}\right)^{\frac{1}{2}}\\
		&\hspace{1cm}= \left(\sum_{k \in \left[2 \left(1+\frac{5}{2}\epsilon\right)R\right]}\left\|g\left(\cdot-\frac{k}{2}\right)\right\|^2_{L^2[-(1+3\epsilon)R,(1+3\epsilon)R]^{{\mathsf{c}}}}\right)^{\frac{1}{2}}.
	\end{align*}
 
	Let $k \in \left[2 \left(1+\frac{5}{2}\epsilon\right)R\right]$. Then 
	\begin{align*}
		&\left\|g\left(\cdot-\frac{k}{2}\right)\right\|^2_{L^2[-(1+3\epsilon)R,(1+3\epsilon)R]^{{\mathsf{c}}}}\\&\hspace{1cm}= \int_{-\infty}^{-(1+3\epsilon)R}\frac{C^2_r}{\left(1+|x-\frac{k}{2}|\right)^{2r}}dx + \int_{(1+3\epsilon)R}^{\infty}\frac{C^2_r}{\left(1+|x-\frac{k}{2}|\right)^{2r}}dx\\
		&\hspace{1cm}=\frac{C^2_r}{(2r-1)\left(1+\frac{k}{2}+(1+3\epsilon)R\right)^{2r-1}}+\frac{C^2_r}{(2r-1)\left(1-\frac{k}{2}+(1+3\epsilon)R\right)^{2r-1}}.
	\end{align*}
 
	Further, one can show that
	\begin{align*}
			\left\|g\left(\cdot-\frac{k}{2}\right)\right\|^2_{L^2[-(1+3\epsilon)R,(1+3\epsilon)R]^{{\mathsf{c}}}} &\leq 
		 \frac{4C^2_r}{(2r-1)(\frac{\epsilon R}{2})^{2r-2}(r-1)},
	\end{align*}
	from which the assertion follows.
        \end{proof}
        \begin{proof}[Proof of Lemma \ref{paper126}]
		Let $f \in C_{[-\pi,\pi]}$. Then 
		$f-Pf=\sum_{k \in \mathbb{Z} \setminus \left[2 \left(1+\frac{5}{2}\epsilon\right)R\right]}\frac{1}{\sqrt{2 }}f\left(\frac{k}{2 }\right)g\left(\cdot- \frac{k}{2 }\right).$
	Now, using the estimates in	Lemmas \ref{paper122} and \ref{paper123}, the proof of       \eqref{paper17} and \eqref{paper18} follows. It is well known (see \cite{paper120}) that if $\psi \in L^1(\mathbb{R}) \cap C^{1}(\mathbb{R})$ with $\psi' \in L^1(\mathbb{R})$, then for all $t \in \mathbb{R}$,
         \begin{equation*}
             \sum_{k \in \mathbb{Z}}\left|\psi\left(t-\frac{k}{2}\right)\right|\leq 2\|\psi\|_{L^1(\mathbb{R})} + \|\psi'\|_{L^1(\mathbb{R})}.
         \end{equation*}
         Using the above fact, one can show \eqref{paper153}.
	\end{proof}
\end{document}